\documentclass[useAMS,usenatbib]{mn2e}
\usepackage{graphicx}
\usepackage{times}
\usepackage{amssymb} %%% for math symbols as $\lesssim$ and/or $\gtrsim$
\usepackage{txfonts}
\usepackage{appendix}
\usepackage{natbib}
\usepackage{longtable, portland}
\usepackage{supertabular}
\usepackage{rotating}
\usepackage{array}
\usepackage{caption}
\usepackage{gensymb}

\title[Morphological classification of galaxies: impact from survey depth and resolution]{The impact from survey depth and resolution on the morphological classification of galaxies\\}
\author[Povi\'c et al.]{M. Povi\'c$^{1}$\thanks{E-mail: mpovic@iaa.es}, I. M\'arquez$^{1}$, J. Masegosa$^{1}$, J. Perea$^{1}$, A. del Olmo$^{1}$, C. Simpson$^{2}$, J. A. L. Aguerri$^{3}$, \newauthor B. Ascaso$^{4, 1}$, Y. Jim\'enez-Teja$^{5}$, C. L\'opez-Sanjuan$^{6}$, A. Molino$^{7, 1}$, A. M. P\'erez-Garc\'ia$^{3, 8}$, \newauthor K. Viironen$^{6}$, C. Husillos$^{1}$, D. Crist\'obal-Hornillos$^{6}$, C. Caldwell$^{2}$, N. Ben\'itez$^{1}$, E. Alfaro$^{1}$, \newauthor T. Aparicio-Villegas$^{5, 1}$, T. Broadhurst$^{9, 10}$, J. Cabrera-Ca\~no$^{11}$, F. J. Castander$^{12}$, J. Cepa$^{8, 3}$, \newauthor M. Cervi\~no$^{1, 3}$, A. Fern\'andez-Soto$^{13, 14}$, R. M. Gonz\'alez Delgado$^{1}$, L. Infante$^{15}$, \newauthor V. J. Mart\'inez$^{16, 14}$, M. Moles$^{6, 1}$, F. Prada$^{1}$, and J. M. Quintana$^{1}$\\
$^{1}$Instituto de Astrof\'isica de Andaluc\'ia (IAA-CSIC), Granada, Spain \\
$^{2}$Astrophysics Research Institute, Liverpool John Moores University, Liverpool, UK\\ 
$^{3}$Instituto de Astrof\'isica de Canarias (IAC), La Laguna, Tenerife, Spain\\ 
$^{4}$GEPI, Paris Observatory, Paris, France\\ 
$^{5}$Observat\'orio Nacional, Rio de Janeiro, Brazil\\
$^{6}$Centro de Estudios de F\'isica del Cosmos de Arag\'on (CEFCA), Teruel, Spain\\
$^{7}$Instituto de Astronom\'ia, Geof\'isica e Ci\'encias Atmosf\'ericas, Universidade de S\~ao Paulo, S\~ao Paulo, Brazil\\ 
$^{8}$Departamento de Astrof\'isica, Facultad de F\'isica, Universidad de la Laguna, La Laguna, Spain \\ 
$^{9}$Department of Theoretical Physics, University of Basque Country, Bilbao, Spain\\
$^{10}$IKERBASQUE, Basque Foundation for Science, Bilbao, Spain\\
$^{11}$Facultad de F\'isica. Departamento de F\'isica At\'omica, Molecular y Nuclear, Universidad de Sevilla, Sevilla, Spain\\
$^{12}$Institut de Ci\`encies de l'Espai, IEEC/CSIC, Barcelona, Spain\\ 
$^{13}$Instituto de F\'isica de Cantabria (CSIC-UC), Santander, Spain\\ 
$^{14}$Unidad Asociada Observatori Astron\`omic (IFCA - UV), Valencia, Spain\\
$^{15}$Departamento de Astronom\'ia, Pontificia Universidad Cat\'olica, Santiago de Chile, Chile\\ 
$^{16}$Observatori Astron\`omic de la Universitat de Val\`encia, Valencia, Spain\\ 
}
\begin{document}

\date{Accepted ??. Received ??; in original form ??}

\pagerange{\pageref{firstpage}--\pageref{lastpage}} \pubyear{2014}

\maketitle

\label{firstpage}

\begin{abstract}
We consistently analyse for the first time the impact of survey depth and spatial resolution on the most used morphological parameters for classifying galaxies through non-parametric methods: Abraham and Conselice-Bershady concentration indices, Gini, M20 moment of light, asymmetry, and smoothness. Three different non-local datasets are used, ALHAMBRA and SXDS (examples of deep ground-based surveys), and COSMOS (deep space-based survey). We used a sample of 3000 local, visually classified galaxies, measuring their morphological parameters at their real redshifts (z\,$\sim$\,0). Then we simulated them to match the redshift and magnitude distributions of galaxies in the non-local surveys. The comparisons of the two sets allow to put constraints on the use of each parameter for morphological classification and evaluate the effectiveness of the commonly used morphological diagnostic diagrams. All analysed parameters suffer from biases related to spatial resolution and depth, the impact of the former being much stronger. When including asymmetry and smoothness in classification diagrams, the noise effects must be taken into account carefully, especially for ground-based surveys. M20 is significantly affected, changing both the shape and range of its distribution at all brightness levels. We suggest that diagnostic diagrams based on 2\,-\,3 parameters should be avoided when classifying galaxies in ground-based surveys, independently of their brightness; for COSMOS they should be avoided for galaxies fainter than F814\,=\,23.0. These results can be applied directly to surveys similar to ALHAMBRA, SXDS and COSMOS, and also can serve as an upper/ lower limit for shallower/deeper ones.
\end{abstract}


\section{Introduction}
\label{sec_intro}

\indent \indent Morphology is one of the main characteristics of galaxies, and the morphological classification has been central to many advances in the picture of galaxy formation and evolution. Different correlations between morphology and other galaxy properties have been studied, including the relation with stellar mass \citep[e.g.,][]{deng13}, colour \citep[e.g.,][]{strateva01,hogg03,bell03,baldry04,weiner05,cirasuolo05,melbourne07,cassata07,povic13}, luminosity \citep[e.g.,][]{blanton03,kelm05}, environment \citep[e.g.,][]{cassata07}, black hole mass \citep[e.g.,][]{kormendy95, mclure00, graham01a, marconi03}, nuclear activity \citep[e.g.,][]{adams77, heckman78, ho95, kauffman03,choi09,pierce07,povic12,gabor09}, and X-ray properties \citep[e.g.,][]{hickox09, povic09a, povic09b}.\\   
\indent The methods of morphological classification of galaxies can be separated into three groups: i) visual \citep[e.g.][]{lintott08,nair10,baillard11,lintott11,willett13}, for classifying the nearby and well resolved galaxies, ii) parametric, based on the galaxy physical \citep[e.g.][]{vaucouleurs48,sersic63,peng02,peng10,simard02,simard11,souza04,barden12} and mathematical parameters, where assuming an analytic model for fitting the galaxy \citep[e.g.][]{kelly04,kelly05,ngan09,andrae11a,andrae11b,jimenez12}, they perform the decomposition of well resolved galaxies and provide the properties of different structures, and finally, iii) non-parametric, which does not assume any particular analytic model.\\ 
\indent The non-parametric methods are based on measuring the different galaxy quantities that correlate with the morphological types, i.e colours \citep[e.g.][]{strateva01}, spectral properties \citep[e.g.][]{humason31,morgan57,baldwin81,folkes96,sanchez10}, or light distribution \citep[e.g.][]{doi93,abraham94,abraham96,abraham03,bershady00,conselice00,graham01b,lotz04,yamauchi05}. The non-parametric methods are less time-consuming in comparison with other methods, and can provide an easy and fast separation between regular and irregular (or disturbed) sources and/or early- and late-type galaxies, down to intermediate redshifts ($\sim$\,1.5), or higher if dealing with space based, or good seeing ground-based data. Over the past years different morphological diagrams have been applied, relating galaxy light concentration and asymmetry or smoothness parameters either to classify galaxies \citep[][]{abraham94,conselice00,bershady00,graham01b,conselice06,cassata07,povic09a,deng13}, or to select merger candidates \citep[][]{lotz10a,lotz10b,urrutia08,villforth14}. However, some of the previous works showed that these parameters can depend significantly on e.g. the aperture definition for measuring the galaxy flux/radius \citep{strateva01,graham01b,lisker08}, and/or on the resolution and signal-to-noise ratio \citep[hereafter S/N;][]{conselice00,graham01b,lisker08,huertas09,andrae11a, andrae11b,povic12,cibinel13,carollo13,petty14}. Different trends of these parameters can be observed in relation to the data quality, spatial resolution and depth. Since a reliable morphological classification is essential for galaxy formation and evolution studies, it is crucial to disentangle how strong is this impact\footnote{Hereafter, when using the expression \textbf{'observational bias'} we will refer to the impact of spatial resolution and data depth on analysed morphological parameters.} on each morphological parameter when dealing with different data sets. In most of previous works this observational bias was analysed either for a particular parameter, either for a particular observational condition (survey), or for a particular sample of galaxies. In this work we present, for the first time, a systematic study, where in a consistent way the analysis was carried out for all parameters, on large sample of galaxies, and in relation with different observational conditions.\\
\indent We analysed the observational bias that might affect each of the six commonly used morphological parameters: Abraham concentration index, Gini coefficient, Conselice-Bershady concentration index, M20 moment of light, asymmetry index, and smoothness. We studied how this bias depends on the spatial resolution and magnitude/redshift distributions of galaxies, and how it affects the diagnostic diagrams used to classify galaxies. We used a visually classified sample of local galaxies and simulated them to map the observational conditions of three different ground- and space-based deep surveys: the Advanced Large Homogeneous Area Medium Band Redshift Astronomical survey \citep[][ALHAMBRA]{moles08}, the Subaru/XMM-\textit{Newton} Deep Survey \citep[][SXDS]{furusawa08}, and the Cosmos Evolution Survey \citep[][COSMOS]{scoville07}. With this analysis our main goal is to measure a set of morphological parameters in the real (local) conditions and in the simulated conditions of non-local surveys in order to study how the spatial resolution and data depth affect each parameter and the commonly used morphological diagrams. We emphasize that is out of our aim to classify the galaxies in the three selected non-local surveys, neither reclassify local galaxies once they were scaled to map the conditions of ALHAMBRA, SXDS, and COSMOS. Comparing the parameters of local galaxies measured before and after shifting them in redshift and magnitude, we were able to put constraints on the main diagrams, observing how the position and the shape of the regions typical of early- and late-type galaxies change in the local and non-local conditions. Moreover, analysing the morphological diagrams in ground-based and space-based surveys, and at different magnitude cuts, we quantified how strong is the impact from spatial resolution and survey depth, respectively. The results obtained in this work can be applied to surveys similar to ALHAMBRA, SXDS, and COSMOS.\\
\indent The paper is organized as follows: in Sec.~\ref{sec_data} we described the used local and high-redshift samples and the corresponding surveys. The applied methodology is described in Sec.~\ref{sec_method}. The results obtained for local and simulated galaxies are showed in Sec.~\ref{sec_results}. In Sec.~\ref{sec_discussion} we discussed how the impact from data depth and spatial resolution affects the morphological parameters and commonly used diagnostic diagrams for galaxy classification. Finally, we summarised our results in Sec.~\ref{sec_summary_conclusions}.\\  
\indent We assume the following cosmological parameters throughout the paper: $\Omega_{\Lambda}$\,=\,0.7, $\Omega_{M}$\,=\,0.3, and H$_0$\,=\,70 km s$^{-1}$ Mpc$^{-1}$. Unless otherwise specified, all magnitudes are given in the AB system \citep{oke83}.

\section[]{The data}
\label{sec_data}

\subsection[]{Local sample} 
\label{sec_data_locsample}

\begin{figure}
\centering
\begin{minipage}[c]{.49\textwidth}
\includegraphics[width=7.0cm,angle=0]{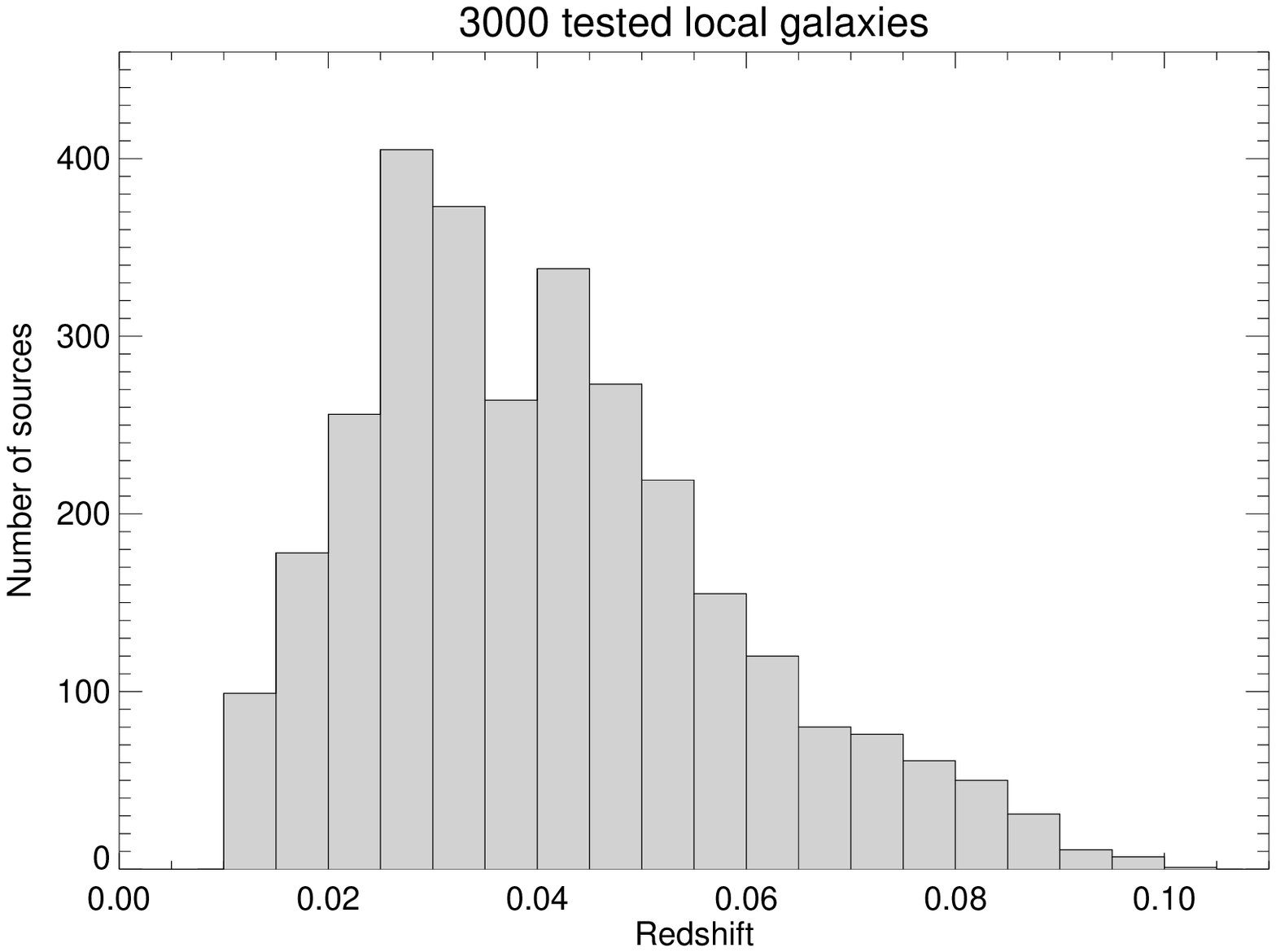}
\end{minipage}
\begin{minipage}[c]{.49\textwidth}
\includegraphics[width=7.0cm,angle=0]{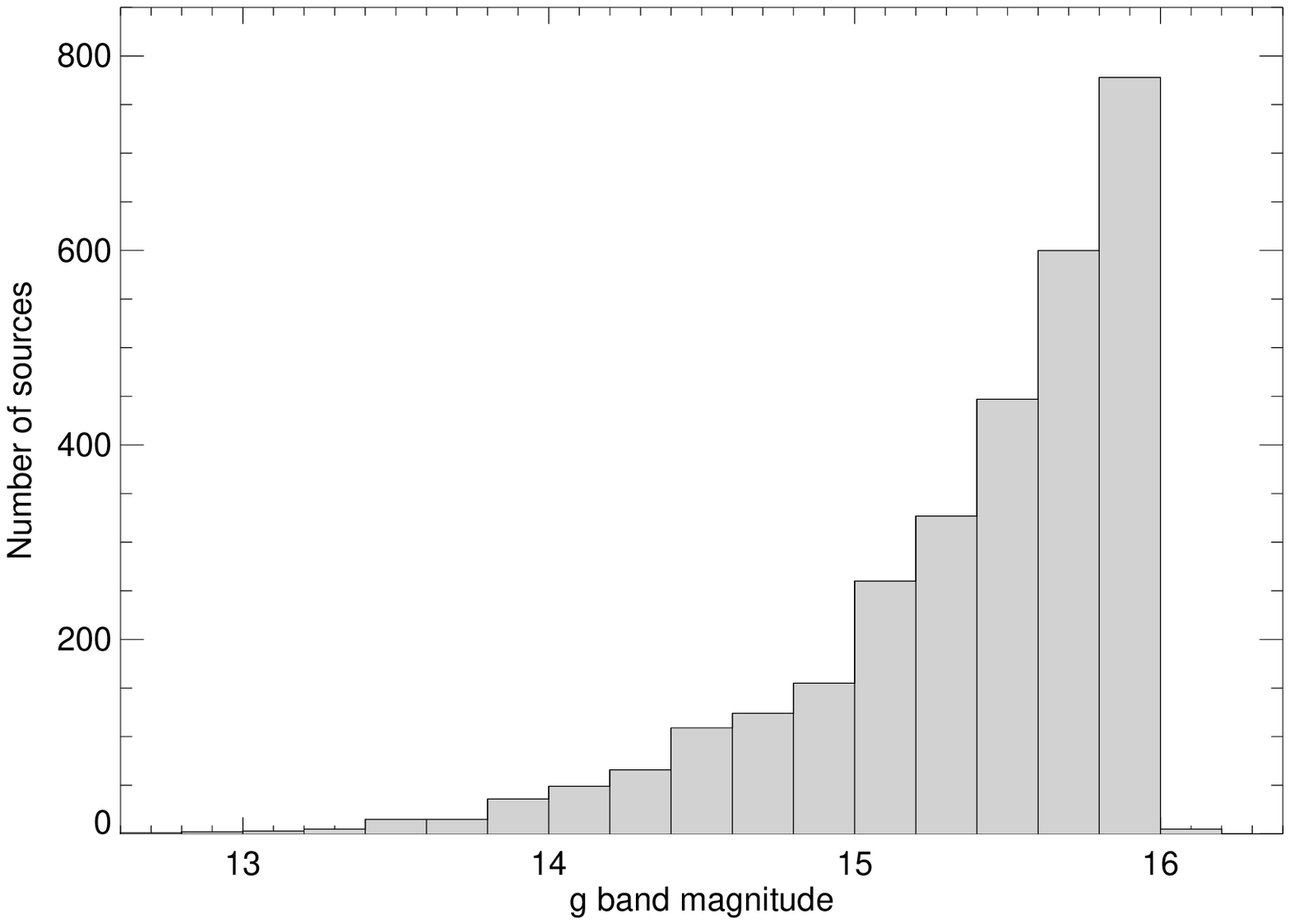}
\end{minipage}
\begin{minipage}[c]{.49\textwidth}
\includegraphics[width=7.0cm,angle=0]{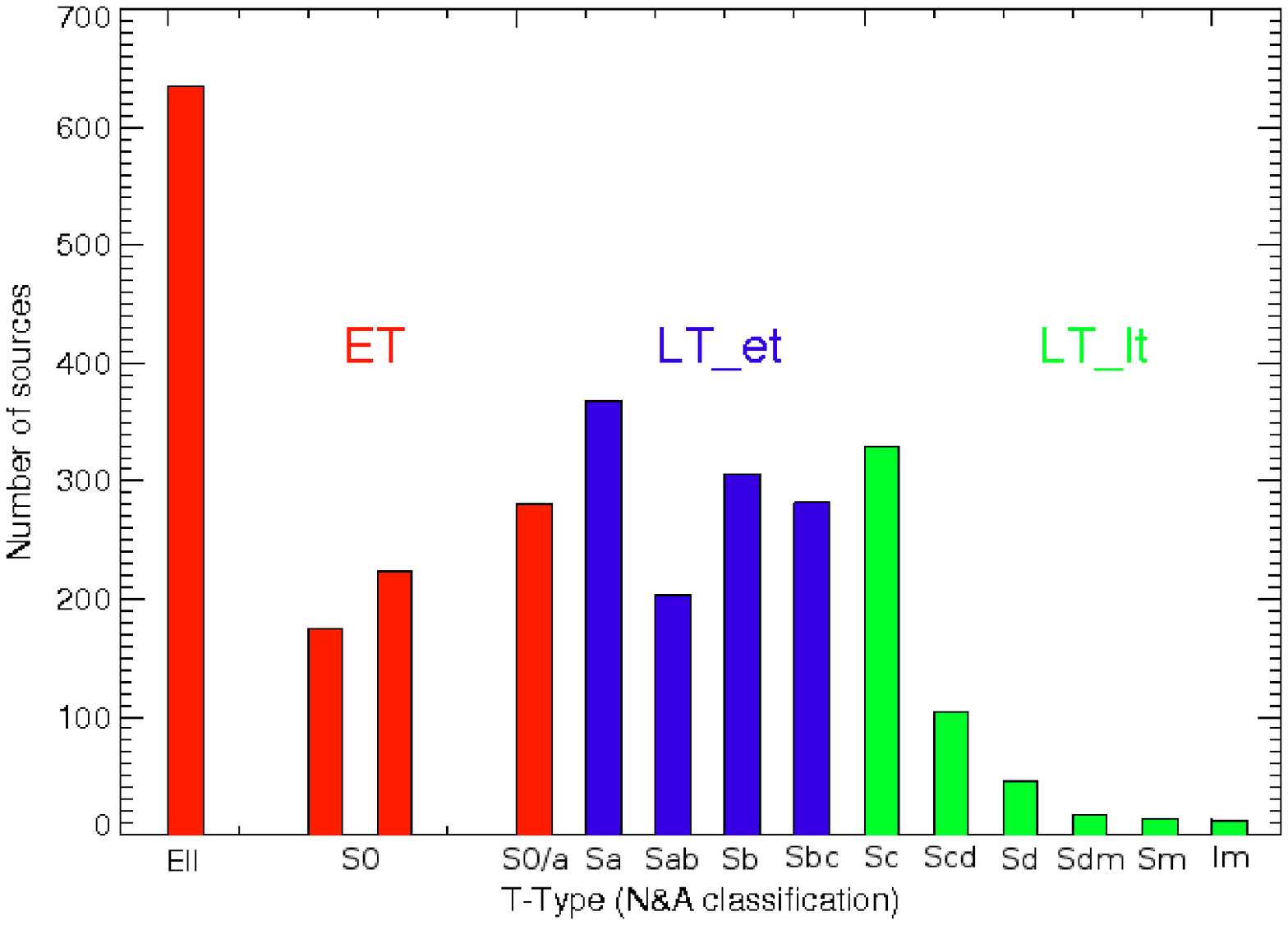}
\end{minipage}
\caption[ ]{Redshift (top), g band magnitude (middle), and morphological type (bottom) distributions of the local sample used in this work.\\\\(A colour version of this figure is available in the on-line journal)}
\label{fig_local_distgeneral}
\end{figure}

\indent To test how the observational bias affects the analysed morphological parameters, we used a sample of 3000 local galaxies at 0.01\,$\le$\,z\,$\le$\,0.1 (with a mean redshift of 0.04), observed in the Sloan Digital Sky Survey (SDSS) Data Release 4 (DR4) down to an apparent extinction-corrected magnitude of $g$\,$<$\,16, and visually classified by Nair \& Abraham (2010; hereafter N\&A), using the $g$ and $r$ bands. The number of local galaxies is selected as a compromise between the computing time and the classification accuracy, since the computing time to train the Support Vector Machine (SVM) through the galSVM code used in this work (see Sec.~\ref{sec_method}) is totally dependent and very sensitive to the size of the training data set \citep{huertas08,huertas09}. The galaxies were selected randomly out of $\sim$\,14,000 sources contained in the N\&A catalogue, making sure that the selected sub-sample is representative in terms of the general properties of the whole data set \citep[see][and their 
Fig. 3]{povic13}: $g$ band magnitude, redshift, $g$\,-\,$r$ colour, morphological classification, and inclination in the case of the late-type galaxies. On the other hand, Figure~\ref{fig_local_distgeneral} shows the redshift, g-band magnitude, and N\&A morphological classification (T-Type in their work) distributions of the selected local sample. As can be seen from the T-Type histogram, the selected sample occupies all range of morphologies, from elliptical to irregular galaxies. In this work, all analyses were performed dividing galaxies into three morphological groups:\\
\indent  \textbf{1) early-type galaxies (hereafter ET)} - with T-Type\,$\le$\,0, including elliptical (c0, E0, E+), lenticular (S0-, S0, S0+) and S0/a galaxies from the N\&A classification,\\
\indent  \textbf{2) early spirals (hereafter LT\_et)} - with 0\,$<$\,T-Type\,$\le$\,4, including Sa, Sab, Sb, and Sbc, and \\
\indent  \textbf{3) late spirals and irregular galaxies (hereafter LT\_lt)} - with T-Type\,$>$\,4, including Sc, Scd, Sd, Sdm, Sm, and Im galaxies. \\ 
\indent We stress that throughout this work, we do not re-classify local galaxies and every time we specify the morphological type, it refers to that from N\&A classification. Here we want to measure a set of the most used morphological parameters in real conditions of the local sample and in the simulated conditions of non-local surveys in order to
evaluate how the spatial resolution and data depth affect each parameter. 
ET, LT\_et, and LT\_lt galaxies represent 45\%, 38\%, and 17\% of the selected local sample, respectively, completely consistent with the whole N\&A sample. \\
\indent Irregular galaxies constitute only $<$\,1\% of all sources in the N\&A catalogue. With such a small population, we were not able to provide reliable statistical analyses in comparison to the other two broader groups, Ell/S0 and spirals. This is the reason why we do not study irregulars separately, and divide instead all late-type galaxies into two subgroups: earlier and later. In our LT\_lt defined group, galaxies classified as Irr by N\&A form only 2\% of the whole population, being therefore insignificant for affecting the results represented in this paper. \\

\subsection[]{Non-local samples} 
\label{sec_data_highz}

In this Section we describe the properties of non-local surveys used for studying the impact from spatial resolution and data depth on the morphological classification of galaxies. Since our work is statistical, to select the surveys we considered the following criteria:\\
\indent - a large number of detected galaxies, either a deep survey covering a smaller area, and/or a shallower survey covering a large area,\\
\indent - surveys with accurate photometry as well as accurate photometric redshift measurements available, and \\
\indent - surveys with different properties in terms of spatial resolution and depth.\\
\indent Finally, as a result of these criteria, we selected three non-local surveys for our analysis: deep ground-based surveys ALHAMBRA and SXDS (ALHAMBRA being shallower than SXDS, but with a large covered area and accurate photometric and photometric redshift measurements, and SXDS as an example of the deepest available ground-based data), and COSMOS as an example of deep space-based surveys. We had to take into account the number of selected surveys, since for each of them the applied computational procedure is time consuming. In the following sections we describe each of these surveys. Their basic properties are summarised in Table~\ref{tab_surveys}. We can not compare directly the results obtained between the three analysed non-local surveys: first, because they have different depth, so the analysed magnitude cuts are different (see below), and secondly they have different photometric information available, so the bands that we used are different as well. We tested each survey separately, and tried to represent the obtained results depending on the properties of each one. 

\begin{table*}
\begin{center}
\caption{Summary of the ground and space-based high-redshift data selected for studying the observational bias of morphological parameters. 
\label{tab_surveys}}
\begin{tabular}{| c | c | c | c | c | c | c |}
\hline
\textbf{Survey}&\textbf{tot. A/used A [deg$^2$]}&\textbf{Sp. Res. [arcsec]}&\textbf{Lim\_mag\,/\,Sel. sam. (band)}&\textbf{Num of gal}&\textbf{mag\_sim}&\textbf{z\_sim}\\
\hline
\textbf{ALHAMBRA}&4.0\,/\,$\sim$\,3.0&1.1&25.0\,/\,23.0 ($F613W$)&$>$\,43,000&$\le$\,20.0, $\le$\,21.5, $\le$\,23.0&$\sim$\,0.45, $\sim$\,0.62, $\sim$\,1.0\\
\textbf{SXDS}&1.22\,/\,$\sim$\,0.65&0.8&27.7\,/\,24.5 ($i'$)&$>$\,68,000&$\le$\,21.0, $\le$\,23.0, $\le$\,24.5&$\sim$\,0.65, $\sim$\,1.2, $\sim$\,2.2\\
\textbf{COSMOS}&2.0\,/$\sim$\,0.5&0.09&26.5\,/\,24.0 ($F814W$)&$>$\,107,000&$\le$\,21.0, $\le$\,23.0, $\le$\,24.0&$\sim$\,0.67, $\sim$\,1.1, $\sim$\,1.6\\
\hline
\end{tabular}
\end{center}
\begin{flushleft}
{\textbf{* Column description:} \textbf{Survey} - selected high-redshift surveys; \textbf{tot. A/used A} - the total covered area of each survey, and the area that covers the data used in this work; \textbf{Sp. Res.} - the average survey spatial resolution; \textbf{Lim. mag / Sel. sam. (band)} - limiting survey magnitude, and limiting magnitude of the selected sample in the correspondent band; \textbf{Num. of gal} - the final number of selected galaxies used in this work to obtain the magnitude and redshift distributions in each survey, and move later the local galaxies to these conditions (see Sec.~\ref{sec_method} and ~\ref{sec_results}); \textbf{mag\_sim} and \textbf{z\_sim} - magnitude cuts and the corresponding redshifts (for 95\% of the sample) used to simulate the local galaxies;}
\end{flushleft}
\end{table*}

\subsubsection[]{ALHAMBRA} 
\label{sec_data_highz_alh}

\indent The Advanced Large Homogeneous Area Medium Band Redshift Astronomical survey \citep[ALHAMBRA;][]{moles08} is a photometric survey that imaged $\sim$\,3.5\,deg$^2$ of the sky, in eight different fields, through 20 optical and 3 near-infrared (NIR) filters. The average spatial resolution of the ALHAMBRA images is $\sim$\,1\,arcsec \citep[not higher than 1.6, and ranging mainly between 0.8 and 1.2\,arcsec in the F613W band; see Table 1 in][for more details]{povic13}. More than half million sources were detected down to the magnitude limit $r$\,=\,25.0. \\
\indent The Bayesian Photometric Redshift code (\texttt{BPZ2.0}) was used to measure the photometric redshifts \citep{benitez00,molino14}, obtaining the accuracy for galaxies brighter than 22.5 and with magnitudes 22.5\,-\,24.5 in the constructed F814W band of $\sim$\,$\delta$z\,/\,(1\,+\,z)\,=\,0.011 and $\sim$\,$\delta$z\,/\,(1\,+\,z)\,=\,0.014, respectively. All images and catalogues are available through the ALHAMBRA webpage\footnote{http://alhambrasurvey.com}.

\subparagraph {\textit{ALHAMBRA sample selection\\}}
In this work we used the same sample as described in \cite{povic13}, whereby we will provide here only a brief summary. We selected only sources with the BPZ quality parameter ODDS\,$>$\,0.2, expecting no more than 3\% of photometric redshift outliers \citep{molino14}. For details about the galaxy-star separation see \cite{povic13}. And finally, we selected only objects with magnitudes $\le$\,23.0 in the F613W band, taking into account that above this magnitude limit S/N ratio decreases significantly, leading to less accurate photometric redshift estimations and geometrical galaxy-star classifications. The F613W photometric band was selected due to its higher signal-to-noise (S/N) ratio in comparison with other filters \citep{aparicio10}. The final selected sample has more than 43,000 galaxies.

\subsubsection[]{SXDS} 
\label{sec_data_highz_sxds}

\indent The Subaru/XMM-\textit{Newton} Deep Survey \citep[SXDS;][]{sekiguchi04} is a large, multiwavelength, and one of the deepest optical ground based surveys, cobvering five continuous, rectangle subfields, and a total area of 1.22\,deg$^2$. All subfields were observed in five broadband filters $BVR{_c}i'z'$, detecting in each more than 800,000 sources down to the limiting magnitudes $B$\,=\,28.4, $V$\,=\,27.8, $R_c$\,=\,27.7, $i'$\,=\,27.7, and $z'$\,=\,26.6 at 3$\sigma$ and a typical seeing of 0.8. All images and photometric catalogues are described in \cite{furusawa08}, and can be downloaded from the SXDS webpage\footnote{http://www.naoj.org/Science/SubaruProject/SXDS/index.html}.\\
\indent The photometric redshift distribution of each sample of galaxies was determined using multi-band photometry of objects in the K-band-selected catalogue of \cite{hartley13}, having a depth of $K$\,=\,24.3. In addition to the $JHK$ photometry, photometric measurements were obtained for the $uBVR_{c}i'z'$ and \textit{Spitzer}/IRAC channels 1 and 2 using the method of \cite{simpson12}. Photometric redshifts and object classifications were derived using the code EAzY \citep{brammer08}, using a development of the method described in \cite{simpson13} and Caldwell et al. (in prep). Measured photometric redshifts have accuracy of $\delta$z\,/\,(1\,+\,z)\,=\,0.031. 

\subparagraph {\textit{SXDS sample selection\\}}
The sample we selected in the SXDS field is based on the available photometric redshifts. Therefore, we used only those galaxies that overlap with the K-band survey, instead of the entire SXDS field. The survey covers 0.77\,deg$^2$, where the overlap region with the SXDS field is $\sim$\,0.63\,deg$^2$, covering the entire central subfield and fractions of the side ones. We discarded all galaxies with poor photometric redshift fits, having $\chi^2$\,$<$\,100. To separate between point-like and extended sources, we applied the classification described above \citep[][ and Caldwell et al. in prep]{simpson13}. Moreover, we selected all galaxies with $i'$\,$\le$\,24.5, which is the band that we will use in the following analysis, and the limit at which our sample is complete, taking into account the available redshifts and photometry. The final selected sample consists on more than 68,000 galaxies. Since the photometric redshift selection is based on the K-band, we might be missing the bluest sources detected in the SXDS. We did not perform a detailed analysis with respect to this, but we estimated by cross-matching i- and K-band catalogues an upper limit of $<$\,9\% of these sources, which should not affect significantly our results.

\subsubsection[]{COSMOS} 
\label{sec_data_highz_cosmos}

The Cosmic Evolution Survey \cite[COSMOS;][]{scoville07} is a deep, 2\,deg$^2$ multiwavelength survey. In this work we use the observations from the HST ACS survey, where more than 1.2\,$\times$\,10$^6$ sources were detected down to a limiting magnitude 26.5 in the filter $F814W$. The ACS data reduction and images are described in \cite{koekemoer07}, while the photometric catalogue is presented in \cite{leauthaud07}.\\
\indent The photometric redshifts used in this work are presented in \cite{ilbert09}, and were measured using the \textit{Le Phare} code and photometric information from 30 broad, intermediate, and narrow-band filters from UV, optical, NIR and MIR bands. The obtained measurements show accuracy of $\sim$\,$\delta$z\,/\,(1\,+\,z)\,=\,0.007 for galaxies brighter than $i^{+}$\,=\,22.5. At magnitudes fainter than $i^{+}$\,$<$\,24.0 the accuracy is $\sim$\,$\delta$z\,/\,(1\,+\,z)\,=\,0.012, while for the very faint sample ($i^{+}$\,$<$\,25.0) the accuracy drops significantly with $\sim$\,20\% of outliers. All images and catalogues used in this work are available from the COSMOS webpage\footnote{http://cosmos.astro.caltech.edu}.

\subparagraph {\textit{COSMOS sample selection\\}}
The COSMOS 2\,deg$^2$ field is covered with 81 HST/ACS images \citep{koekemoer07}. Taking into account the number and size (20,480 pixels square) of the images, and the computing time that we would need to process all the data with the galSVM code (Sec~\ref{sec_method}), we are finally using an area of 0.5\,deg$^2$. Considering the depth of the COSMOS data, for the study presented in this paper, the selected area provides us with the statistical information completely comparable with the other two surveys. Moreover, we checked that the magnitude and redshift distributions of galaxies in the selected field correspond to the total ones. Our main criteria for the COSMOS sample selection is again based on the accuracy of the photometric-redshifts. We selected only sources with $i^{+}$\,$<$\,24.0, since the photo-z accuracy decreases significantly at fainter magnitudes, as described above. For the galaxy-star separation we used the classifications presented in both \cite{leauthaud07} and \cite{ilbert09}. Finally, 
the selected sample has more than 107,000 galaxies. %Among these, the morphological classification is available for $>$\,80\% of galaxies through the catalogue by \cite{tasca11}. We found that, depending on the used method, Ell/S0, spirals, and irregulars represent 14\,-\,21\%, 60\,-\,62\%, and 17\,-\,27\% of the whole population, respectively.

\section[]{Methodology}
\label{sec_method}

\indent Non-parametric methods of galaxy classification are usually applied either when dealing with high redshift and/or low resolution data, where the galaxy decomposition and profile fitting becomes impossible, or in large surveys, where due to the high number of detected sources it is necessary to use more automated and statistically based approaches. We test how much the morphological parameters described in Sec.~\ref{sec_method_param} (basically related with the distribution of light within the galaxy and its shape) are sensitive to depth and spatial resolution, and how this introduced observational bias might affect the final galaxy classification. To do this, we used a sample of local galaxies with available detailed visual classification (see Sec.~\ref{sec_data_locsample}), 
and measured their morphological parameters in two cases:\\
\indent \textbf{\textit{1) at their real redshifts (magnitudes), i.e. at z\,$\sim$\,0}} (see Sec.~\ref{sec_method_z0}), and  \\
\indent \textbf{\textit{2) moving them to higher redshifts (therefore to fainter magnitudes) and lower resolution}}, to simulate the conditions of galaxies on deep, ground- and space-based non-local surveys (see Sec.~\ref{sec_method_highz}). \\
\indent In both cases all parameters are measured in a completely consistent way, using their definitions as described in the following section. Finally, we compared the results obtained in cases 1) and 2) to quantify the impact from data depth and spatial resolution on each morphological parameter, for each of the three analysed surveys (see Sec.~\ref{sec_results_morph_simul} and ~\ref{sec_results_morphdiagram}).

\subsection[]{Description of the tested morphological parameters} 
\label{sec_method_param}

\indent \indent The six morphological parameters described here are by large, the most commonly used to distinguish between ET and LT galaxies, and to classify perturbed galaxies and interacting/merging systems. In the following definitions, when necessary, the galaxy centre is determined by minimizing the asymmetry index \citep[][see below]{abraham96}, while the total flux is defined as the one contained within 1.5 times the Petrosian radius (r$_{p}$, measured by SExtractor; see Huertas-Company et al. 2008). \\
\indent - \textbf{\textit{Abraham concentration index}} \citep[hereafter \textbf{CABR};][]{abraham96} - measured as the ratio between the flux at 30\% of the Petrosian radius (F$_{30}$) and the total flux (F$_{tot}$) : 
\begin{center}
{\begin{equation} \label{eqv_CABR}
\mathop{\mathrm{CABR}}\,=\,F_{30}\,/\,F_{tot}.  
 \end{equation}}
\end{center}
It gives values from 0 to 1, where higher values correspond to higher central light fractions.\\
\indent - \textbf{\textit{Gini coefficient}} \citep[hereafter \textbf{GINI};][]{abraham03,lotz04} - presented as the cumulative distribution function of galaxy pixel $i$ values:
\begin{center}
{\begin{equation} \label{eqv_GINI}
\mathop{\mathrm{GINI}} = 
{\frac{1}{|\bar{X}|n(n - 1)} \sum_{i}^{n} (2i - n - 1) |X_i|,}
\end{equation}}
\end{center}
where $n$ is the total number of pixels in a galaxy, X$_i$ the pixel flux value, and $\bar{X}|$ the mean over all the pixel flux values, using as an aperture 1.5\,r$_{p}$. Usually, it correlates with CABR, having also values from 0 to 1, where higher ones correspond to galaxies with higher central concentrations. However, unlike CABR, it can distinguish between galaxies with shallow light profiles and those with the light concentrated in a few pixels, but outside the galaxy centre.\\
\indent - \textbf{\textit{Conselice-Bershady concentration index}} \citep[hereafter \textbf{CCON};][]{bershady00} - measured as the logarithm of the ratio of the circular radii containing 80\% and 20\% of the total flux:
\begin{center}
{\begin{equation} \label{eqv_CCON}
\mathop{\mathrm{CCON}}\,=\,5log(r_{80}\,/\,r_{20}).
\end{equation}}
\end{center}
In general, lower CCON values correspond to lower fractions of light in the central region.\\
\indent - \textbf{\textit{M20 moment of light}} \citep[hereafter \textbf{M20};][]{lotz04} - measured as the flux (f$_i$) in each pixel (x$_i$, y$_i$) multiplied by the squared distance to the centre (x$_c$, y$_c$) of the galaxy, summed over the 20\% brightest pixels, and normalized by the total second-order moment (M$_{tot}$):
\begin{center}
{\begin{equation} \label{eqv_M20}
\mathop{\mathrm{M_{20}}} =  
log({\frac{\sum_{i} M_i}{M_{tot}}),\, \,  while \, \sum_{i} f_i < 0.2\,f_{tot}},
\end{equation}}
\end{center}
where
\begin{center}
{\begin{equation} \label{eqv_Mtot}
\mathop{\mathrm{M_{tot}}}\,=\,\sum_{i}^{n}\,f_i\,((x_i\,-\,x_c)^2\,+\,(y_i\,-\,y_c)^2). 
\end{equation}}
\end{center}
It shows the light distribution of central sources, off-centre star clusters, bars, spiral arms, etc. \\
\indent - \textbf{\textit{Asymmetry}} \citep[hereafter \textbf{ASYM};][]{abraham96} -  measured by subtracting the galaxy image rotated by 180$^{\circ}$ from the original image, taking also into account the background: 
\begin{center}
{\begin{equation} \label{eqv_ASYM}
\mathop{\mathrm{ASYM}} = 
{\frac{\displaystyle \sum (\mid I(i,j) - I_{180}(i,j) \mid/ 2)}
{\displaystyle \sum I(i,j)} - 
{\frac{\displaystyle \sum (\mid B(i,j) - B_{180}(i,j) \mid/ 2)}
{\displaystyle \sum I(i,j)} 
},}
\end{equation}}
\end{center}
where I$(i,j)$ is the flux in the $(i,j)$ pixel position on the original image, I$_{180}(i,j)$ is the flux in $(i,j)$ on the image rotated by 180$^{\circ}$, and B$(i,j)$ and B$_{180}(i,j)$ are fluxes of the background in the original and rotated image, respectively. Defined in this way, this parameter quantifies the degree to which the light of a galaxy is rotationally symmetric. It gives values from 0 to 1, where the higher ones correspond to more asymmetric galactic shapes. \\
\indent - \textbf{\textit{Smoothness}} or \textbf{\textit{clumpiness}} \citep[hereafter \textbf{SMOOTH};][]{conselice00} - quantifies the degree of small-scale structure. To measure SMOOTH, the original galaxy image is smoothed out with a boxcar of a given width and then subtracted from the original image:
\begin{center}
{\begin{equation} \label{eqv_SMOOTH}
\mathop{\mathrm{SMOOTH}} = 
{\frac{\displaystyle \sum (\mid I(i,j) - I_{S}(i,j) \mid/ 2)}
{\displaystyle \sum I(i,j)} - 
{\frac{\displaystyle \sum (\mid B(i,j) - B_{S}(i,j) \mid/ 2)}
{\displaystyle \sum I(i,j)} 
},}
\end{equation}}
\end{center}
where I$(i,j)$ (B$(i,j)$) present again the flux (background) in the pixel position $(i,j)$ on the original image, while I$_{S}(i,j)$ and B$_{S}(i,j)$ are the flux and the background, respectively, on the image smoothed by a boxcar of width 0.25\textit{r$_{p}$}. The residual image provides with information about possible regions of clumpiness, quantifying the level of small-scale structures.\\

\subsection[]{Methodology applied on the local sample} 
\label{sec_method_z0}

\indent We measured the reference morphological parameters of the local sample, at their real redshifts z\,$\sim$\,0, in three different SDSS bands: $g$, $r$, and $i$. This is needed in order to compare in a consistent way the morphologies of local and simulated galaxies in Sec.~\ref{sec_results_morph_simul} and \ref{sec_results_morphdiagram}, since the corresponding k-correction will imply a different filter for such a comparison. We used the function implemented in the galSVM\footnote{http://gepicom04.obspm.fr/galSVM/Home.html} public code of galaxy classification \citep{huertas08,huertas09} for measuring the parameters. The input information that the code needs, related with the source position, size, ellipticity parameters, and background, was obtained by running SExtractor in all three bands.

\subsection[]{Methodology applied at higher redshifts} 
\label{sec_method_highz}

\begin{figure}
\centering
\includegraphics[width=0.5\textwidth,angle=0]{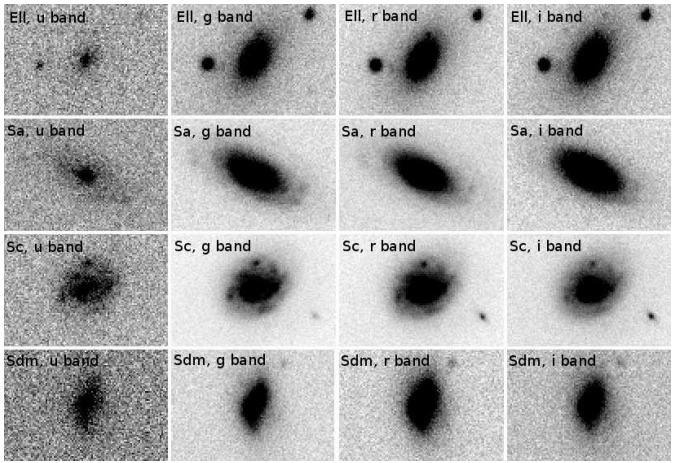}
\caption{An example of Ell, Sa, Sc, and Sdm galaxies (from top to bottom) seen in four SDSS bands, $u$, $g$, $r$, and $i$ (from left to right).
\label{fig_ugri_sample}}
\end{figure}

\indent To simulate the conditions of non-local surveys (case 2 in Sec.~\ref{sec_method}), we used again the galSVM code \citep{huertas08,huertas09}, performing the following three steps for each of the three analysed surveys:\\
\indent - First, we constructed the correspondent magnitude and redshift distributions of non-local galaxies (see Sec.~\ref{sec_results_morph_simul} and Fig.~\ref{fig_highmag_highz_dist_all}). We randomly redshifted and scaled in luminosity the selected sample of local galaxies to match the distributions of non-local ones (see Tab.~\ref{tab_surveys} fot the used photometric band in each survey). In this way, when scaling the galaxy in flux, the surface brightness dimming was directly taken into account. Moreover, we re-sampled the local galaxies with the corresponding pixel-scale for each selected non-local survey, and convolved with its PSF to match the spatial resolution. \\
\indent - Second, we dropped the simulated galaxies, obtained in the first step, into the real background of the high-redshift survey images. We made sure that the galaxies are dropped into empty regions of sky, minimising the chance of superposition with high-redshift foreground/background sources. To detect the empty sky regions, the code makes use of corresponding SExtractor segmentation images. Moreover, although the probability of overlapping is related with the survey properties, we've been strict, dropping only 10 galaxies per image. This step is then repeated until dropping all simulated galaxies, generating every time new images. With all this, we make the probability of overlapping and merger confusion practically negligible (for more information see \cite{huertas08}). Finally, by dropping the local galaxy in a real background of the high-redshift sources, we expect to reproduce the noise from the real images.\\
\indent - Third, we measured the morphological parameters described in Sec.~\ref{sec_method_param}. We took care of the k-correction effect introduced by cosmological redshift, and depending on the band selected in each survey, and also depending on the redshift to which the local galaxy was shifted, we measured the morphological parameters using the corresponding SDSS rest-frame band image. In the ALHAMBRA survey, since we are using the F613W AUTO magnitudes, the SDSS $r$ band images were used for galaxies simulated to z\_sim\,$\le$\,0.13, $g$ band for the redshift range 0.13\,-\,0.5, and $u$ band for higher redshifts. In the SXDS survey, using the $i'$ magnitudes, the $r$ band was used for redshifts z\_sim\,$\le$\,0.4, the $g$ band for 0.4\,$<$\,z\_sim\,$\le$\,0.86, and the $u$ band for z\_sim\,$>$\,0.86. Finally, in the COSMOS survey, using the F814W AUTO magnitudes, the $i$ band was used for the lowest redshifts z\_sim\,$\le$\,0.18, the $r$ band for 0.18\,$<$\,z\_sim\,$\le$\,0.5, the $g$ band for 0.5\,$<$\,z\_sim\,$\le$\,1.0, and the $u$ band for z\_sim\,$>$\,1.0. Since the filter efficiency for the SDSS $u$ band is much lower than that for $g$, $r$, and $i$ bands \citep{gunn98}, and taking into account that we are analysing three broad morphological groups (ET, LT\_et, and LT\_lt) as described in Sec.~\ref{sec_data_locsample} (instead of studying a finer morphology), we used the $g$ band images instead of $u$ when necessary. Figure~\ref{fig_ugri_sample} shows an example of $u$, $g$, $r$, and $i$ images of four galaxies from our local sample with different morphologies; note the poor information available in the $u$ image in comparison to the other bands. Finally, since the morphological classification directly depends on the source brightness, in each survey we analysed the observational bias at the three different magnitude cuts: mag1\_cut\,$\le$\,20.0, mag2\_cut\,$\le$\,21.5, and mag3\_cut\,$\le$\,23.0 in the F613W band in ALHAMBRA; $\le$\,21.0, $\le$\,23.0, and $\le$\,24.5, in the $i'$ band in SXDS; 
and $\le$\,21.0, $\le$\,23.0, and $\le$\,24.0, in the F814W band in COSMOS. In all surveys we chose the first magnitude cut such that we have a sufficient number of sources to perform our analysis. The cut at which our selected photometric/photometric redshift sample is complete, is the last one. Finally, we chose a third cut intermediate to these two.\\   

\begin{figure*}
\centering
\begin{minipage}[c]{0.7\textwidth}
\includegraphics[width=14.0cm,angle=0]{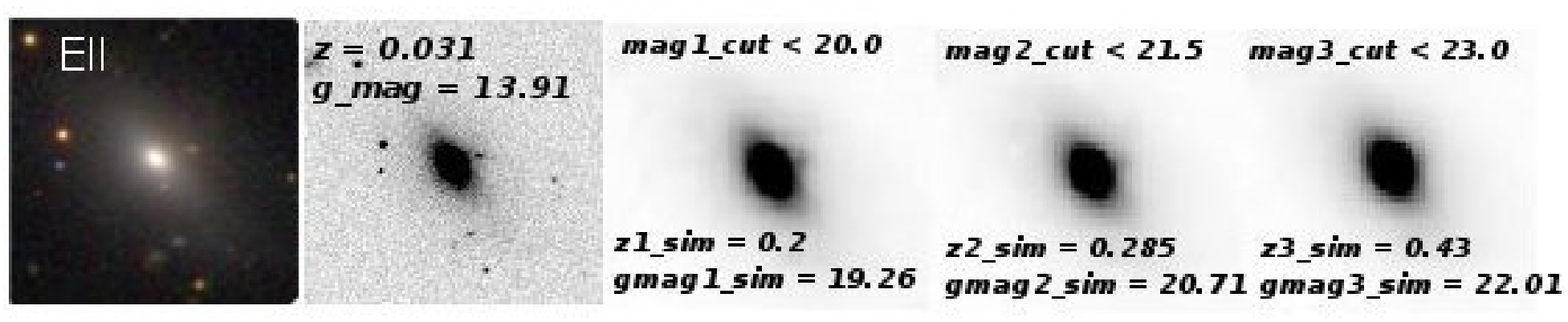}
\end{minipage}
\begin{minipage}[c]{0.71\textwidth}
\includegraphics[width=14.0cm,angle=0]{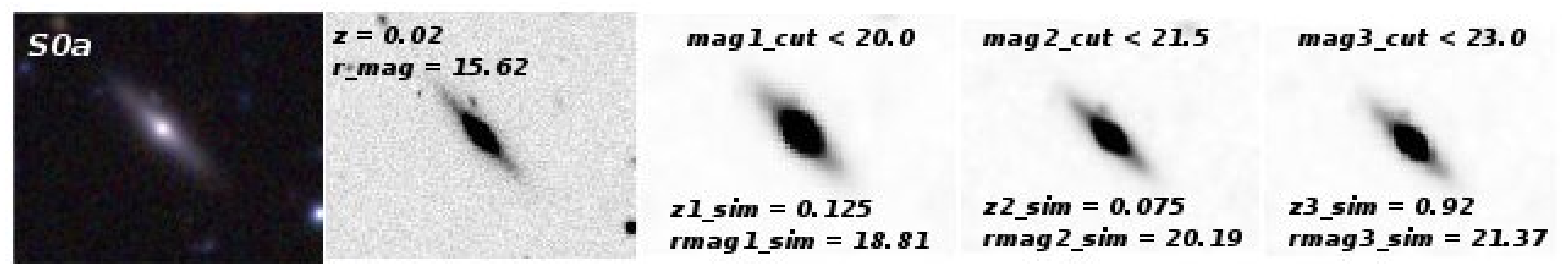}
\end{minipage}
\begin{minipage}[c]{0.7\textwidth}
\includegraphics[width=14.0cm,angle=0]{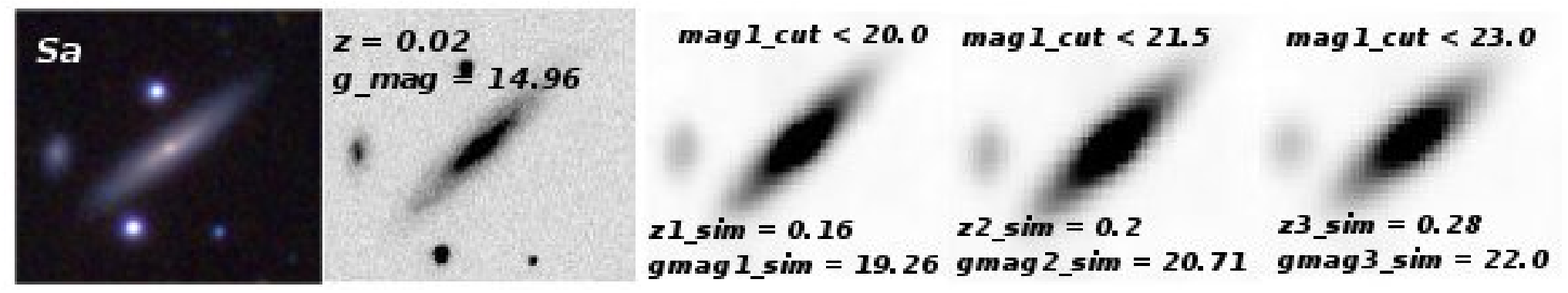}
\end{minipage}
\begin{minipage}[c]{0.7\textwidth}
\includegraphics[width=14.0cm,angle=0]{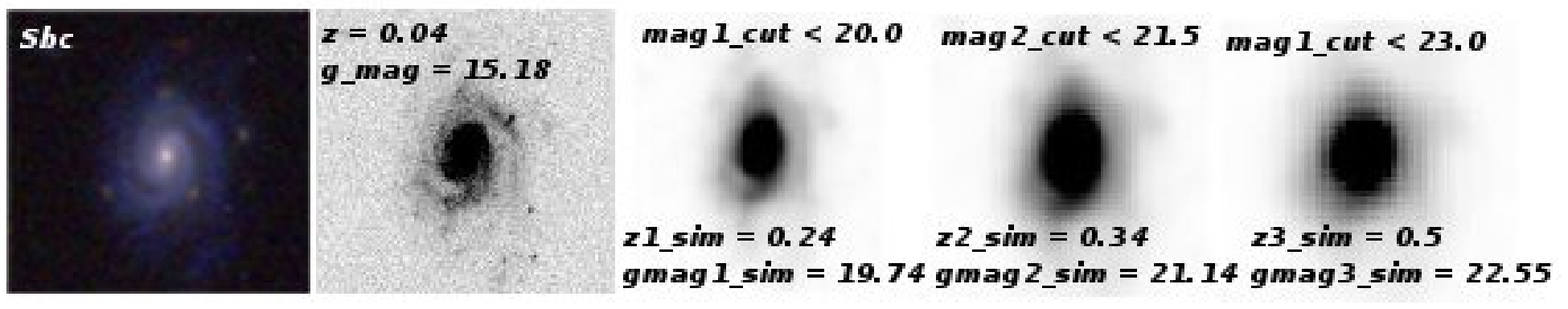}
\end{minipage}
\begin{minipage}[c]{0.7\textwidth}
\includegraphics[width=14.0cm,angle=0]{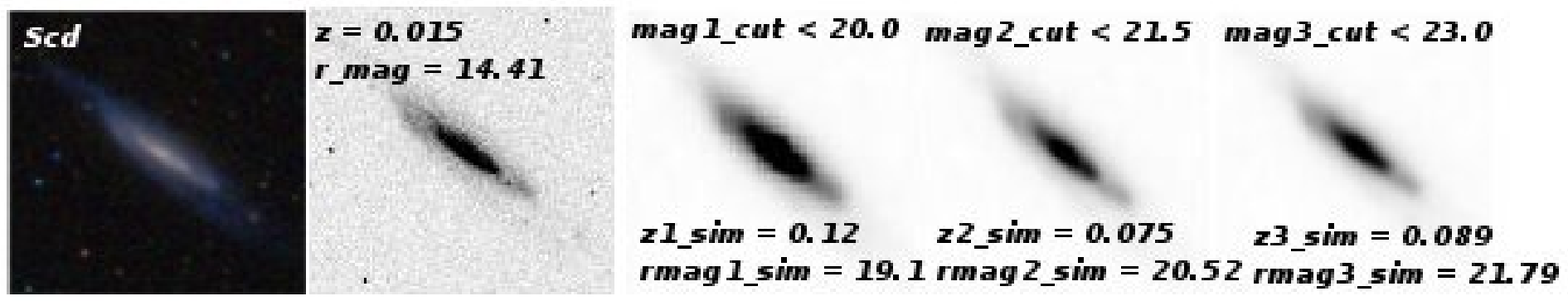}
\end{minipage}
\caption[ ]{Example of simulated images of local galaxies (with different morphologies) after being scaled to map the conditions of the ALHAMBRA survey. For each galaxy (in each row) we present the following (from left to right): colour image, used rest-frame image(s), and simulated high-redshift images at three magnitude cuts (as written on the top of each redshifted image). The morphological type of each galaxy is noted on the colour image. The real redshift and magnitude of the galaxy are noted on the corresponding rest-frame band image(s), while the simulated high-redshift and magnitude are noted on the scaled images of each magnitude cut.\\\\
(A colour version of this figure is available in the on-line journal)}
\label{fig_method_thumb_alh}
\end{figure*}

\begin{figure*}
\centering
\begin{minipage}[c]{0.68\textwidth}
\includegraphics[width=13.7cm,angle=0]{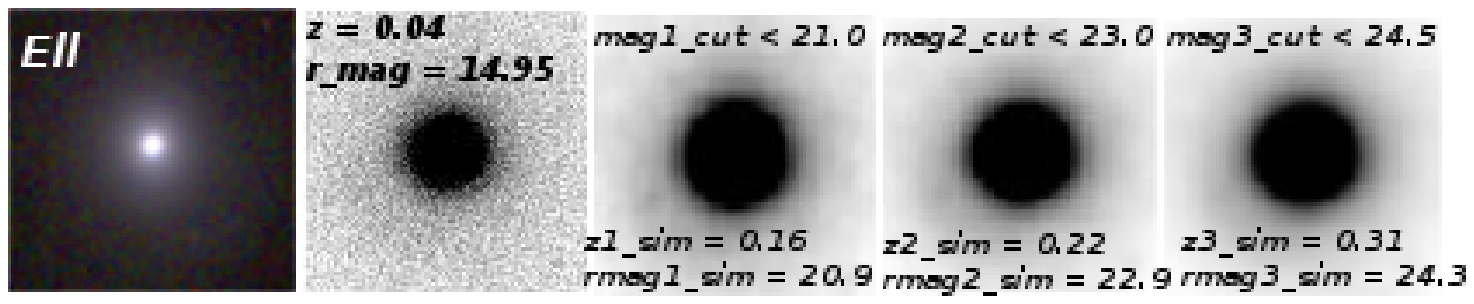}
\end{minipage}
\begin{minipage}[c]{0.7\textwidth}
\includegraphics[width=14.0cm,angle=0]{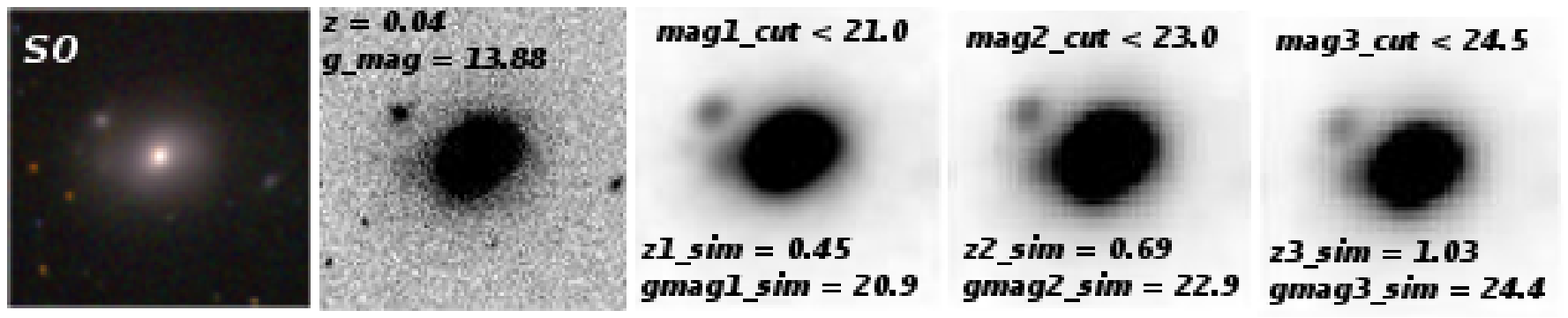}
\end{minipage}
\begin{minipage}[c]{0.7\textwidth}
\includegraphics[width=14.0cm,angle=0]{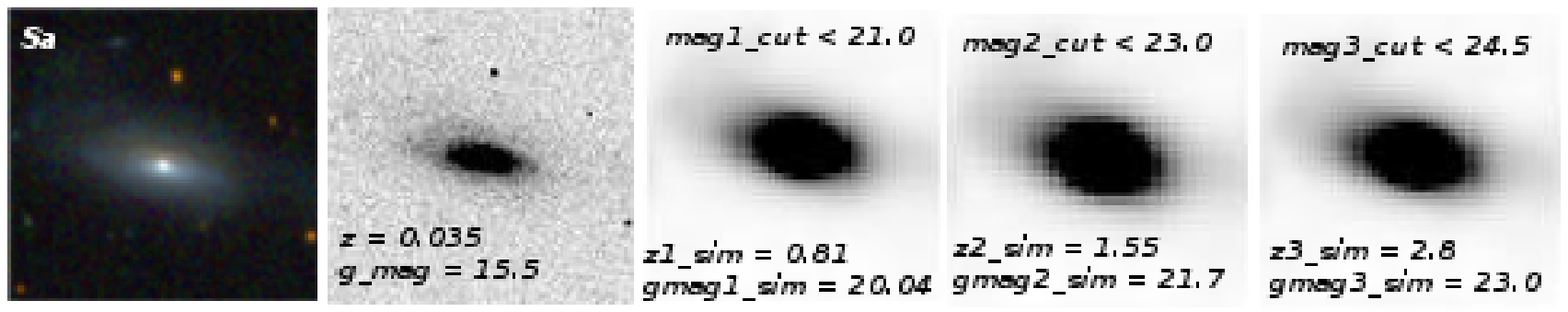}
\end{minipage}
\begin{minipage}[c]{0.7\textwidth}
\includegraphics[width=14.0cm,angle=0]{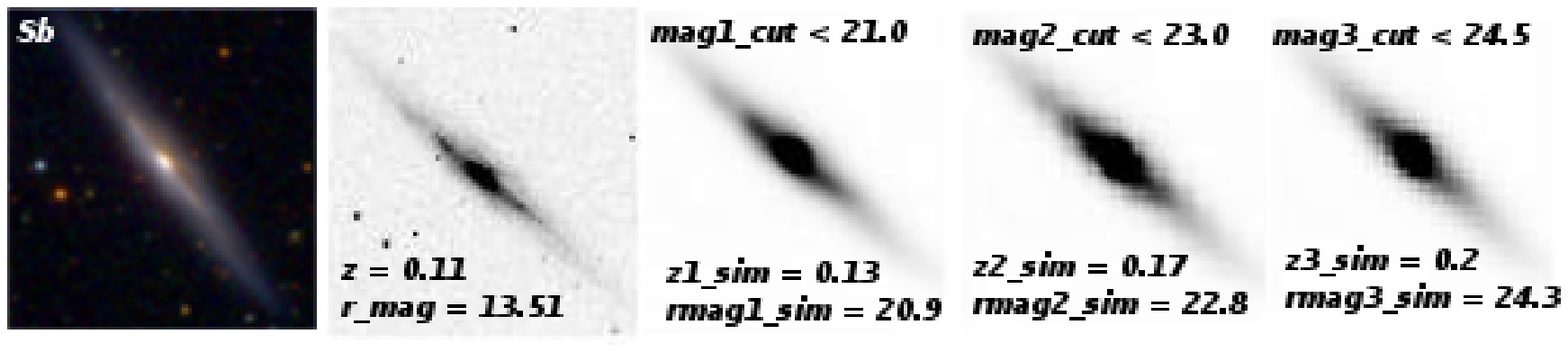}
\end{minipage}
\begin{minipage}[c]{0.72\textwidth}
\includegraphics[width=14.0cm,angle=0]{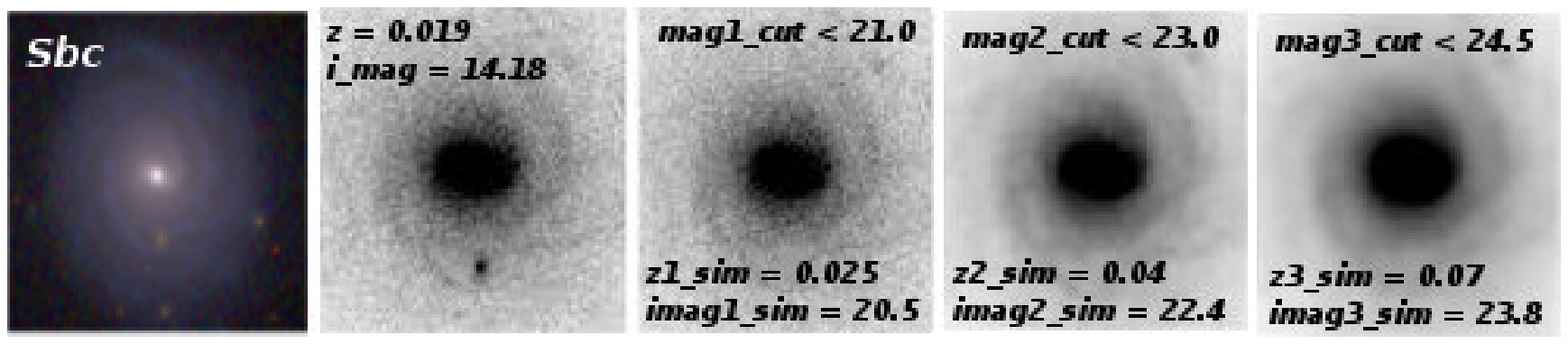}
\end{minipage}
\begin{minipage}[c]{0.69\textwidth}
\includegraphics[width=14.0cm,angle=0]{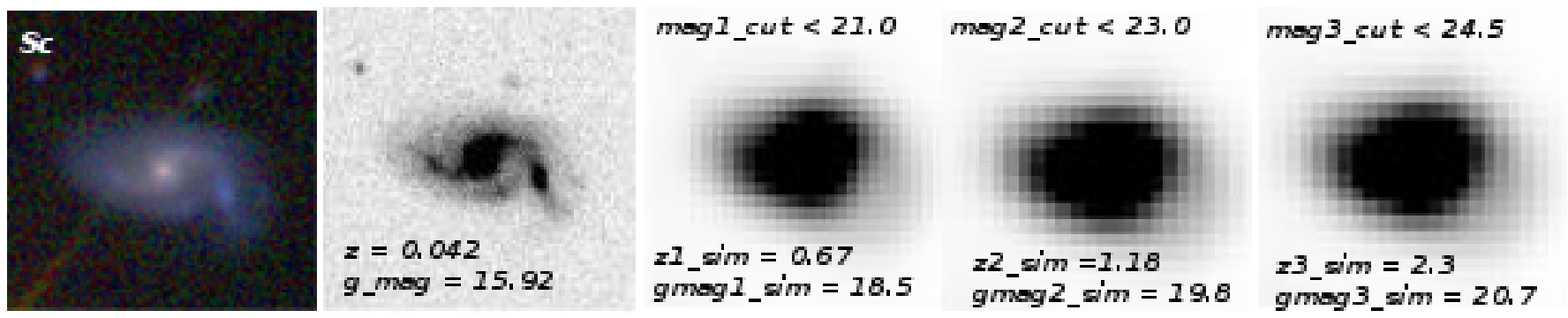}
\end{minipage}
\caption[ ]{Same as Figure~\ref{fig_method_thumb_alh}, but in the SXDS survey.\\ \\
(A colour version of this figure is available in the on-line journal)}
\label{fig_method_thumb_sxds}
\end{figure*}

\begin{figure*}
\centering
\begin{minipage}[c]{0.75\textwidth}
\includegraphics[width=15.4cm,angle=0]{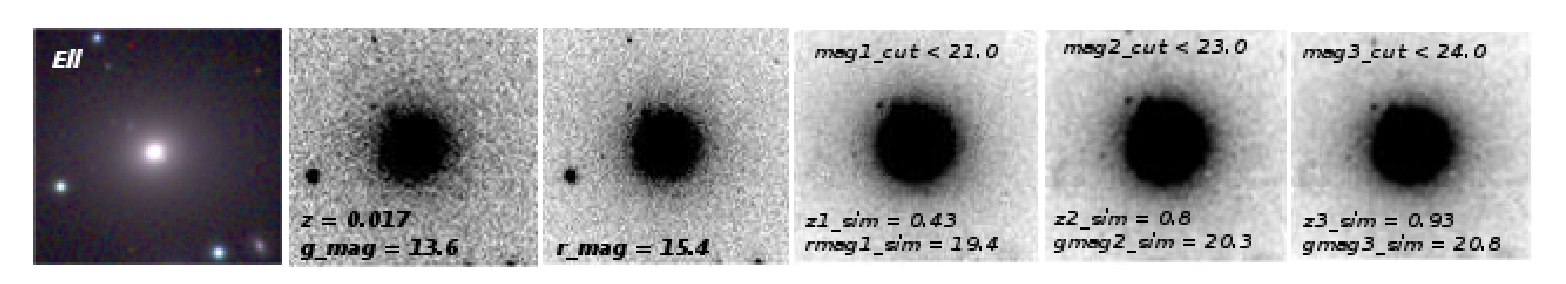}
\end{minipage}
\begin{minipage}[c]{0.7\textwidth}
\includegraphics[width=14.0cm,angle=0]{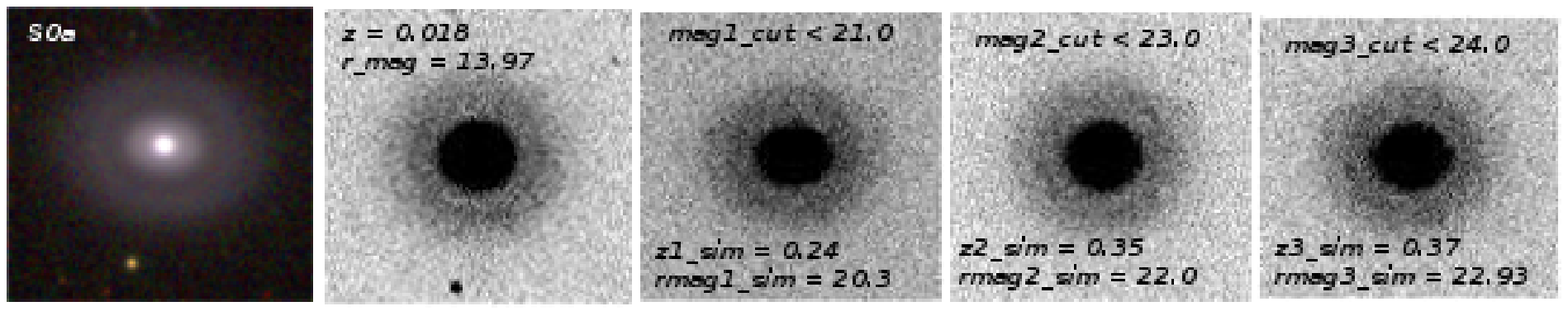}
\end{minipage}
\begin{minipage}[c]{0.75\textwidth}
\includegraphics[width=15.5cm,angle=0]{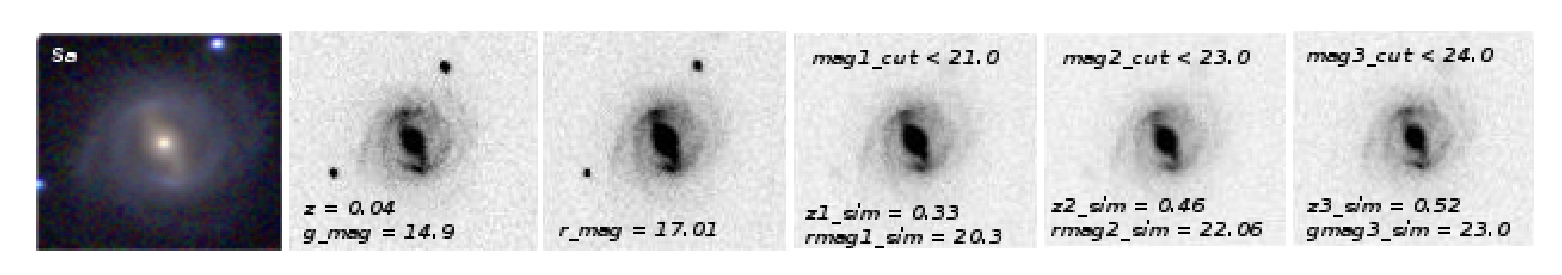}
\end{minipage}
\begin{minipage}[c]{0.7\textwidth}
\includegraphics[width=14.0cm,angle=0]{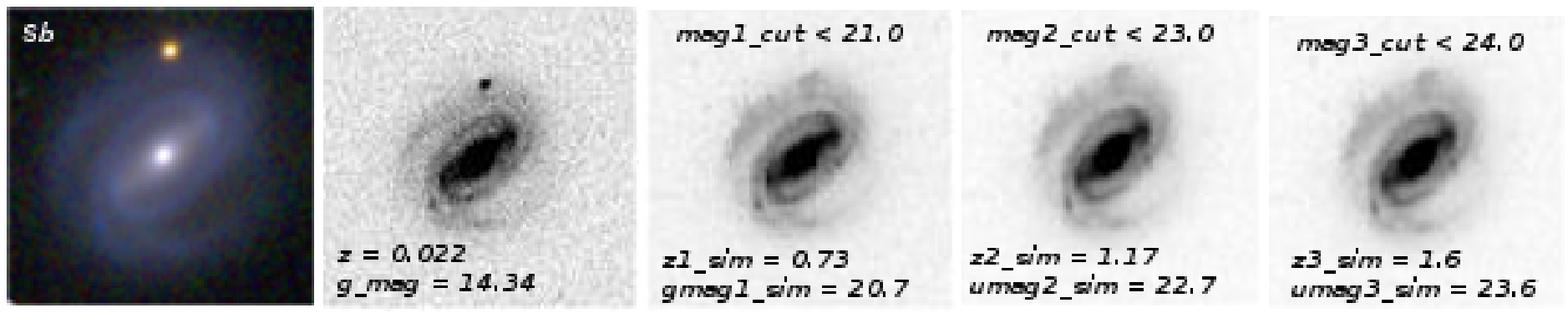}
\end{minipage}
\begin{minipage}[c]{0.68\textwidth}
\includegraphics[width=14.0cm,angle=0]{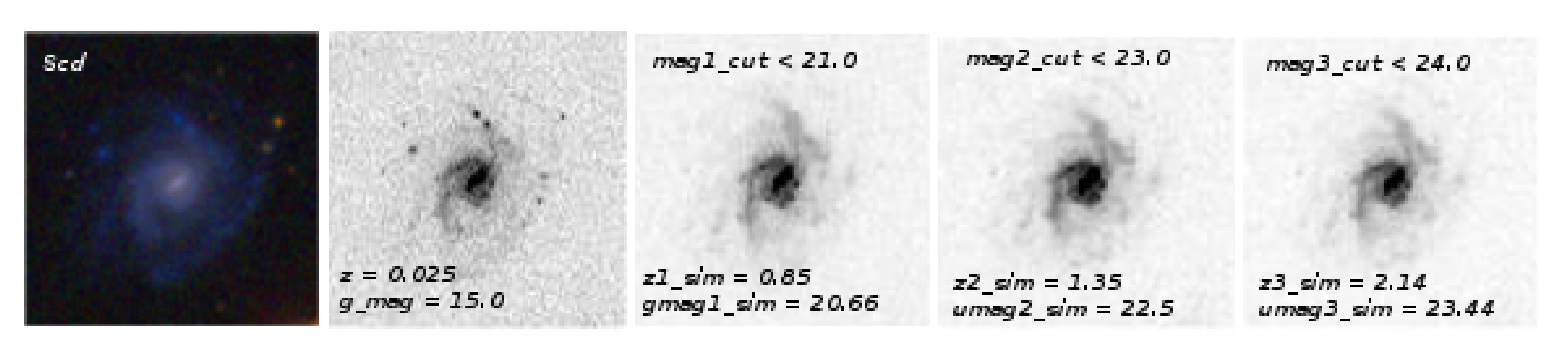}
\end{minipage}
\begin{minipage}[c]{0.7\textwidth}
\includegraphics[width=14.0cm,angle=0]{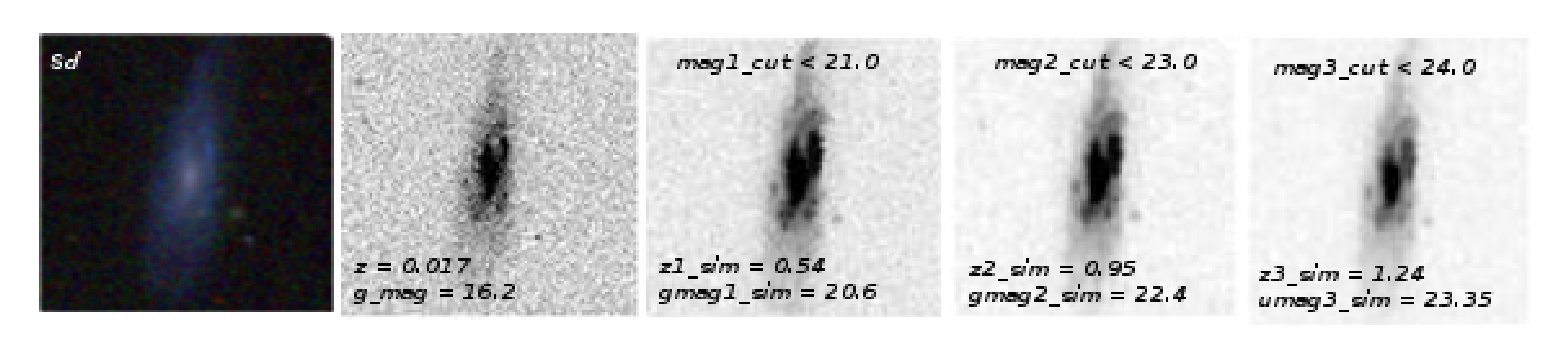}
\end{minipage}
\caption[ ]{Same as Figure~\ref{fig_method_thumb_alh}, but in the COSMOS survey.\\\\(A colour version of this figure is available in the on-line journal)}
\label{fig_method_thumb_cosmos}
\end{figure*}

\indent Figures~\ref{fig_method_thumb_alh}, \ref{fig_method_thumb_sxds}, and \ref{fig_method_thumb_cosmos} show examples of local galaxies, with different morphologies, after being scaled to the conditions of the ALHAMBRA, SXDS, and COSMOS surveys, respectively. In each survey, we show the galaxies being redshifted to the corresponding magnitude cuts, as explained above, providing for each cut the simulated values of magnitude and redshift. These images represent the output of the first step explained in this section, that we use in the second step to drop them in the real background of the non-local survey images, and to measure their morphological parameters in the third step. The colour images were gathered from the EFIGI\footnote{Extraction de Formes Idealisées de Galaxies en Imagerie; http://www.astromatic.net/projects/efigi} project, by combining the $gri$ images \citep{baillard11}, while the original images were downloaded from the SDSS DR4 database and correspond to the images used by N\&A in their 
visual classification. We can observe in each survey how the galaxy information changes when going from brighter to fainter magnitudes (in general from lower to higher redshifts), but even more, how the role that the spatial resolution plays when classifying galaxies, where in COSMOS survey the galaxy information can be conserved up to much higher redshifts in comparison with the ground-based surveys. We studied this in detail in Sec.~\ref{sec_results} and \ref{sec_discussion}. 

\subparagraph {Noise effect\\}
\indent With respect to the noise, since we are using very bright local galaxies, galSVM assumes that the noise from the galaxies themselves is negligible. Therefore, the noise coming from the galaxy itself once it has been dropped into the non-local image is not treated by the code. We tested how strong this effect could be on the measured morphological parameters. We selected 30 galaxies, 10 from each morphological group defined in Sec.~\ref{sec_data_locsample}, and we measured the morphological parameters in the case of ALHAMBRA and COSMOS in two cases: 1) without the above noise added, like in our analysis, and 2) adding the random Poisson noise, typical of each survey, to the scaled local images (before dropping them into the background of non-local surveys). We then compared the results obtained in these two cases for the whole samples, and in relation with the morphological types. We found insignificant differences of the above noise on the mean surface brightness ($<$\,1\%), and on the parameters CABR, CCON, GINI, and M20 ($<$\,6\% and $<$\,5\% in ALHAMBRA and COSMOS, respectively, at the faintest magnitude cuts), but more significant differences in the case of ASYM and SMOOTH. In the case of ALHAMBRA, the differences for these two parameters go up to $\sim$\,40\% and $\sim$\,50\%, respectively. In the case of COSMOS, differences for ASYM are 13\%, 19\%, and 57\% at F814W\,$\le$\,21.0, $\le$\,23.0, and \,$\le$\,24.0, respectively. For SMOOTH the corresponding values are 15\%, 35\%, and 50\%. From this result, we suggest that for the parameters ASYM and SMOOTH the noise effect should be taken into account when using them in morphological classification of galaxies, and especially when dealing with ground-based data. Therefore, in our following analysis we discuss all results only for the four parameters CABR, GINI, CCON, and M20, that are more stable to noise and where our galSVM measurements are reliable. The results based on ASYM and SMOOTH are provided only in the case of COSMOS, for the first two magnitude cuts, where the noise effect is lower. 

\section[]{Results}
\label{sec_results}

\subsection[]{Morphological parameters of the local sample at z\,$\sim$\,0}
\label{sec_results_morph_z0}

\begin{figure}
\centering
\protect{(a)}
\begin{minipage}[c]{0.49\textwidth}
%\protect{(a)}
\includegraphics[width=8.5cm,angle=0]{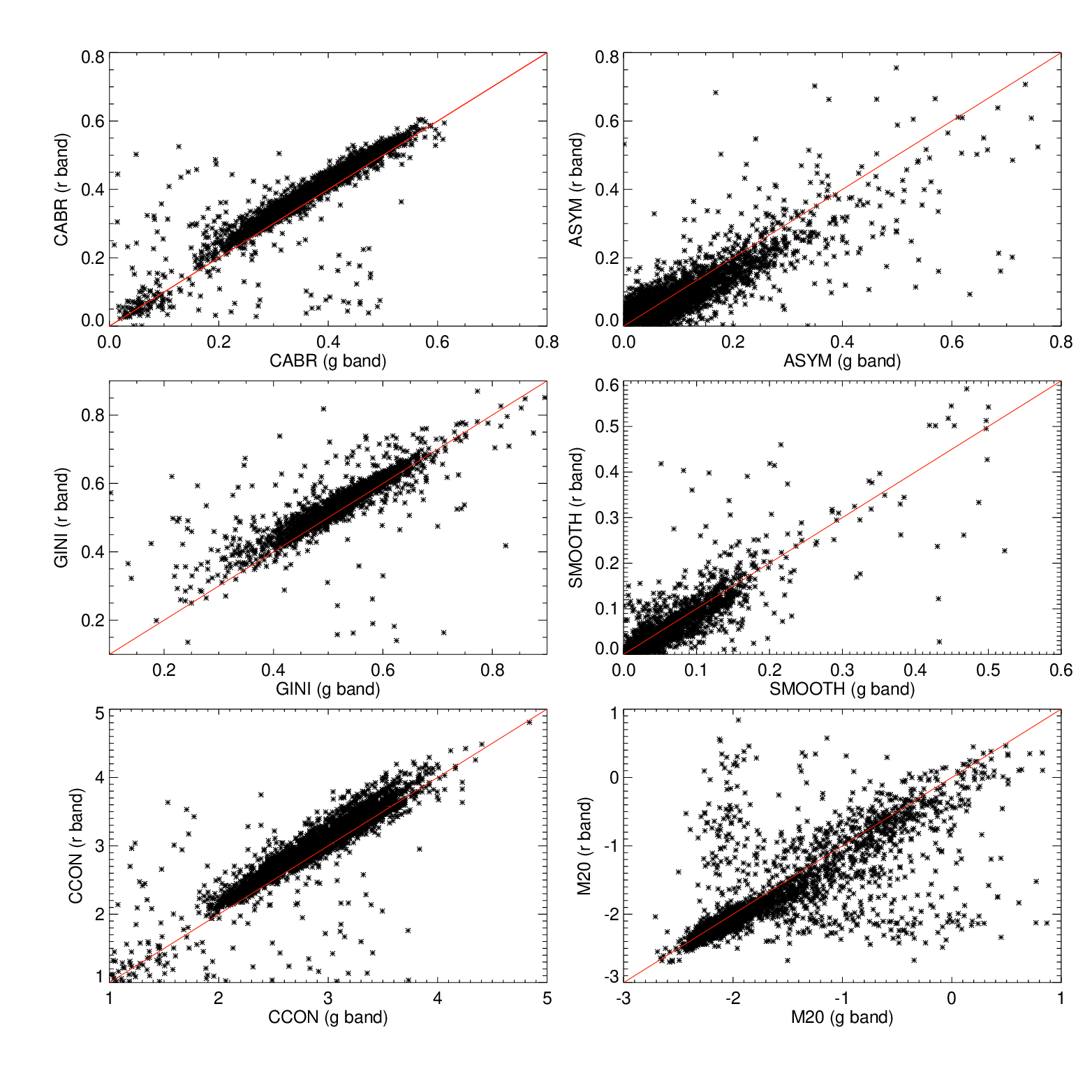}
\end{minipage}
\protect{(b)}
\begin{minipage}[c]{0.49\textwidth}
%\protect{(b)}
\includegraphics[width=8.5cm,angle=0]{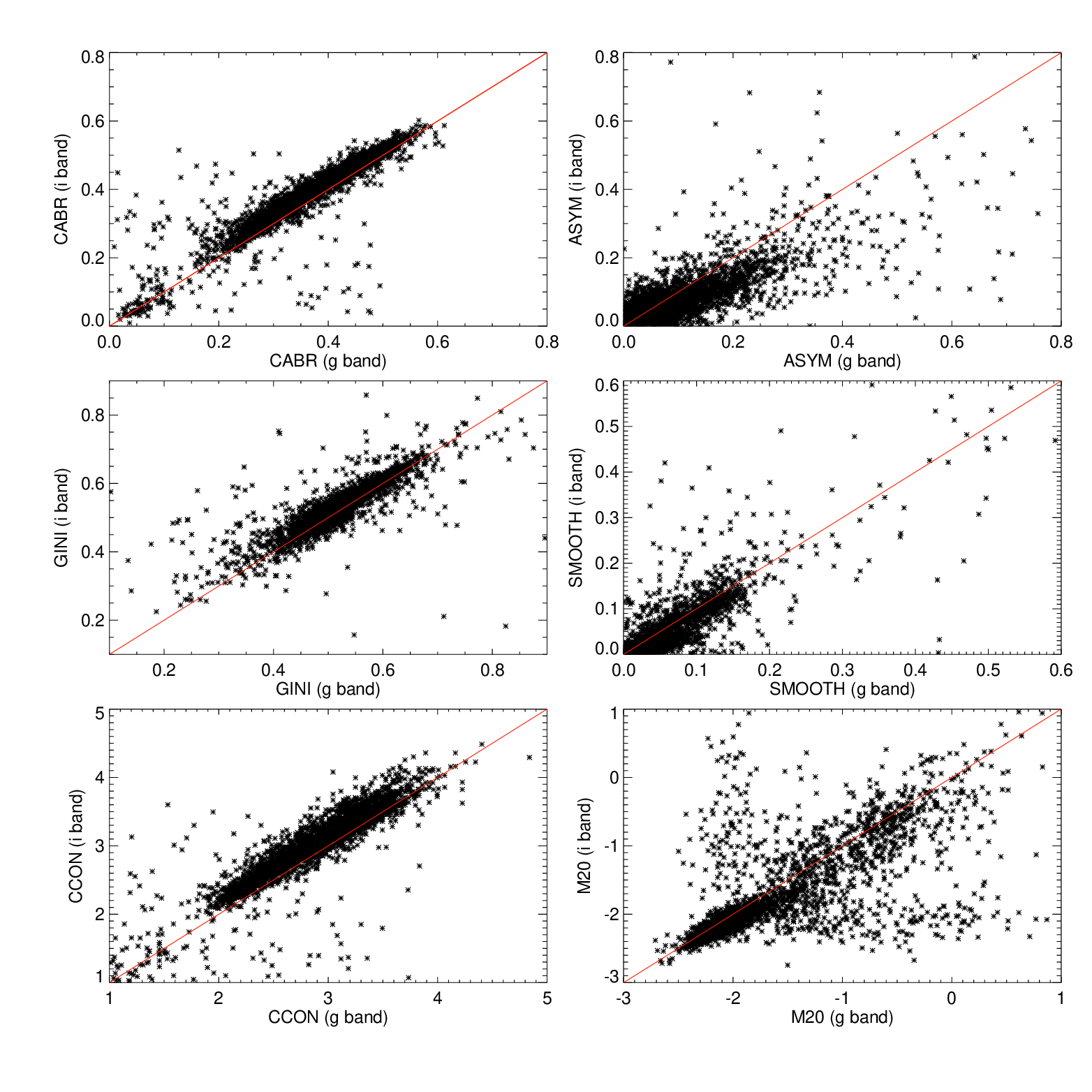}
\end{minipage}
\caption[ ]{Comparison between the six morphological parameters for the galaxies in the local sample measured in the $g$ and $r$ bands (panel a), and $g$ and $i$ bands (panel b). In each panel we compared, from top to bottom, and from left to right: Abraham concentration index, asymmetry parameter, Gini coefficient, smoothness, Conselice-Bershady concentration index, and M20 moment of light. In all plots the red solid line presents the perfect linear correlation between the two measurements.\\\\(A colour version of this figure is available in the on-line journal)}
\label{fig_analysis_gri_comparison}
\end{figure}

\indent For each galaxy, we measured the morphological parameters defined in Sec.~\ref{sec_method_param} in the $g$, $r$, and $i$ bands, as was described in Sec.~\ref{sec_method_z0}. To test how sensitive are morphological parameters to the used wavelengths and if somehow it could affect our further analysis, for each parameter, we compared its measurements obtained in the three photometric bands, finding, in general, a good linear correlation in all cases. Figure~\ref{fig_analysis_gri_comparison} shows the comparison of all six parameters between $g$ and $r$ (panel a), and $g$ and $i$ bands (panel b). In both cases, the concentration parameters, CABR, CCON, and GINI, show the best linear correlations, having similar Pearson Correlation Coefficient (PCC)\footnote{PCC is a measure of the linear correlation between two variables giving a value -1\,$\le$\,PCC\,$\le$\,+1, where +1 shows total positive correlation, 0 no correlation, and -1 total negative correlation \citep{pearson1895}.} $\sim$\,0.9. For M20, ASYM, and SMOOTH, the linear correlation is also conserved in both cases, however they are more sensitive to the selected band in comparison with the concentration indexes, and show lower PCC values (between $\sim$\,0.7 and $\sim$\,0.85). They also show a slightly higher dispersion in panel b, in comparison with panel a, with a difference in the PCC of $\sim$\,0.05. M20 is the most sensitive to the selected band, with PCC\,=\,0.733 between the $g$ and $r$ bands, and PCC\,=\,0.688 between the $g$ and $i$ measurements. The number of catastrophic outliers, placed in the 'wings' above and below the linear correlations is $\sim$\,10\% in both panels. We checked the properties of these sources (magnitude, size, and redshift), without finding any significant difference in comparison with the whole sample. We only found a small difference in relation with the morphological type, with 6\% more LTs in the wings. To have an idea about how these outliers could affect the results in our following analysis and if there are some important differences that we should study in more detail, we performed a test by excluding them from the whole sample and checking again the corresponding distributions of morphological parameters (see Sec.~\ref{sec_results_morph_simul} and Figs.~\ref{fig_alh_param_dist}, \ref{fig_sxds_param_dist}, and \ref{fig_cosmos_param_dist}), without finding any significant differences. However, this was done only to test the properties of outliers, but the all analysis and results presented in this paper were obtained using the whole sample, outliers included. For the comparison of morphological parameters between near-infrared and optical bands see \cite{huertas14}.

\subsection[]{Morphological parameters of the simulated sample}
\label{sec_results_morph_simul}

\begin{figure}
\centering
\protect{(a)}
\begin{minipage}[c]{0.48\textwidth}
\includegraphics[width=0.99\textwidth,angle=0]{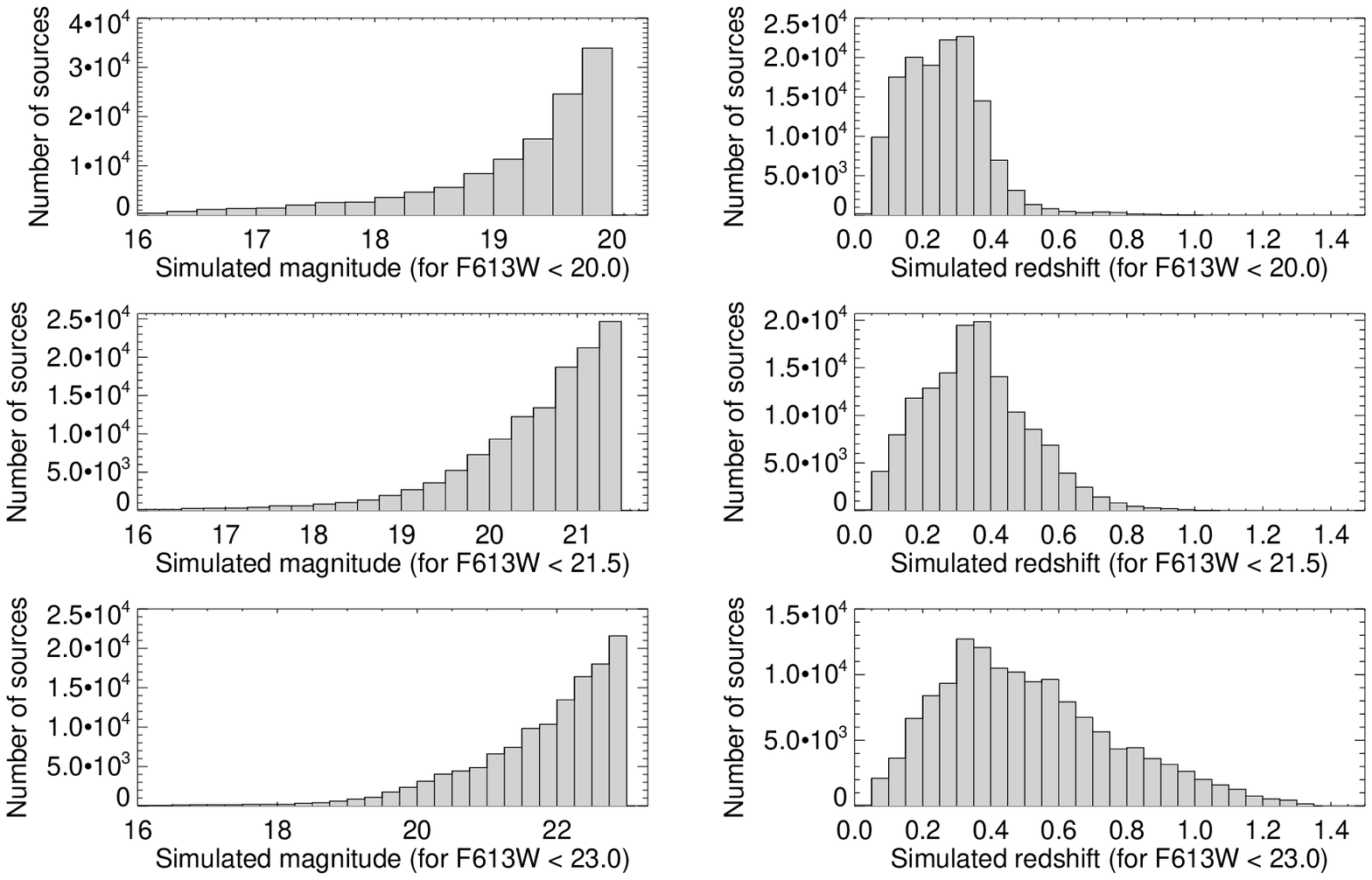}
\end{minipage}
\protect{(b)}
\begin{minipage}[c]{0.48\textwidth}
\includegraphics[width=0.99\textwidth,angle=0]{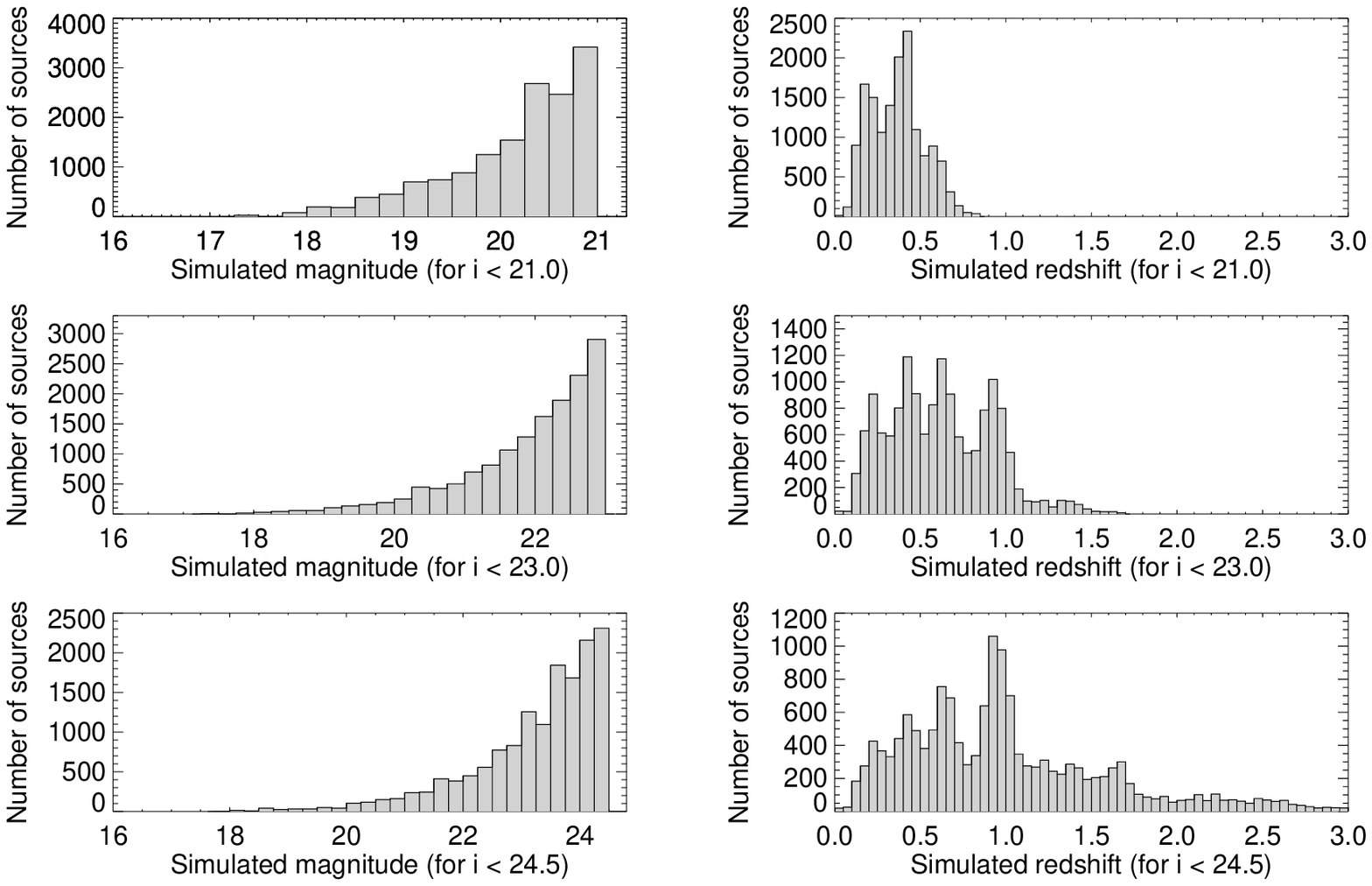}
\end{minipage}
\protect{(c)}
\begin{minipage}[c]{0.48\textwidth}
\includegraphics[width=0.99\textwidth,angle=0]{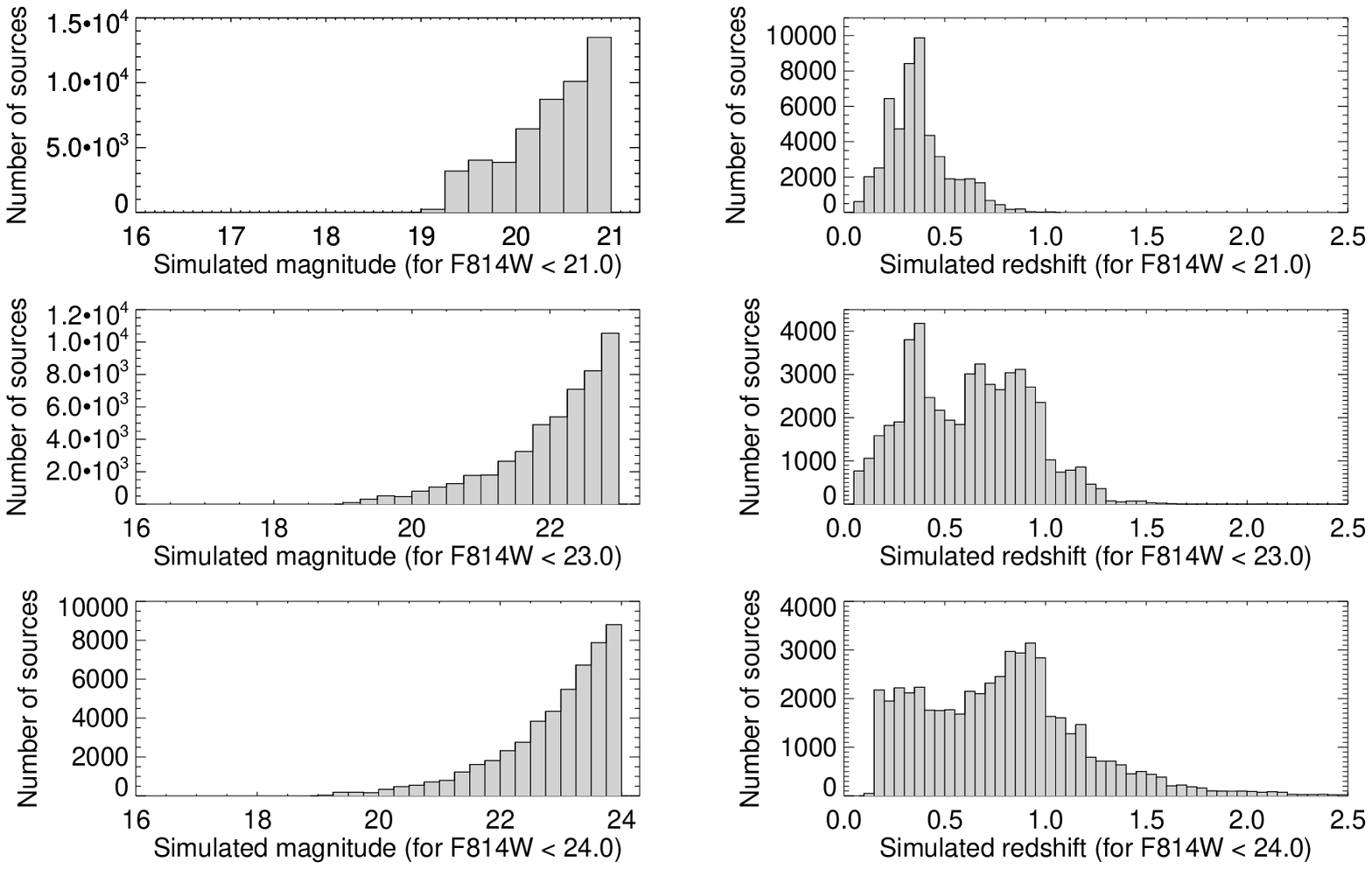}
\end{minipage}
\caption[ ]{Magnitude (left plots) and redshift (right plots) distributions of simulated galaxies for the conditions in ALHAMBRA (panel a), SXDS (panel b), and COSMOS (panel c) surveys. In each survey, they are represented for three magnitude cuts, from top to bottom: for F613W\,$\le$\,20.0, F613W\,$\le$\,21.5, and F613W\,$\le$\,23.0 in the ALHAMBRA; $i$\,$\le$\,21.0, $i$\,$\le$\,23.0, and $i$\,$\le$\,24.5 in the SXDS; and  F814\,$\le$\,21.0  (top), F814\,$\le$\,23.0 (middle), and F814\,$\le$\,24.0 in the COSMOS.}
\label{fig_highmag_highz_dist_all}
\end{figure}

\indent To measure the morphological parameters of the simulated sample, we followed the methodology described in Sec.~\ref{sec_method_highz}. We first moved randomly the local sample to map the magnitude and redshift distributions of galaxies in each high-redshift survey, at each selected magnitude cut. Figure~\ref{fig_highmag_highz_dist_all} shows the corresponding magnitude and redshift distributions of the simulated galaxies in case of ALHAMBRA (panel a), SXDS (panel b), and COSMOS (panel c). Since the magnitude cut introduces also a redshift selection, we are dealing at the same time with different redshift ranges in the three surveys (as summarized in Table~\ref{tab_surveys}). After simulating the local galaxies and dropping them into the real background of high-redshift surveys, we run again SExtractor to measure the same parameters as we did at z\,$\sim$\,0. Finally, we measured the morphological parameters of simulated galaxies, again in the same way as we did at z\,$\sim$\,0. As mentioned above, in 
each survey, depending on the used photometric band and the redshift to which the galaxy was moved, we used the corresponding rest-frame SDSS images to carry out all the measurements. \\

\indent In each non-local survey, we had to repeat all the procedure a number of times, for each separate image, in order to cover the analysed survey areas (see Table~\ref{tab_surveys}). In case of ALHAMBRA, we were dealing with 48 sub-fields \citep[4 CCD images per pointing, with 12 pointing in total (1 or 2 per each field); see][]{povic13}. In SXDS we repeated the procedure in each of five SXDS fields (see Sec.~\ref{sec_data_highz_sxds}), while to cover the selected area of COSMOS we run the code on 17 images. Moreover, in all surveys we repeated the measurements three times, since having selected three magnitude cuts (see Table~\ref{tab_surveys}). Since each time we run galSVM on one image, 3000 simulated galaxies were distributed randomly to the corresponding redshift and magnitude distributions of the non-local samples, for each galaxy we obtained a number of measurements that is equal to the number of used images. Therefore, the final catalogues contain 144000, 15000, and 51000 simulated galaxies in the case of ALHAMBRA, SXDS, and COSMOS, respectively. Each catalogue contains the columns with morphological parameters measured at higher-redshifts in the three magnitude cuts, besides the corresponding reference measurements (obtained at z\,$\sim$\,0). Once again, when assigning the reference value to the simulated one, we took care of the k-correction, assigning the value measured in the corresponding rest-frame band (that will depend on the photometric band we used for analysis in each non-local survey, and on the simulated redshift). The reader should keep this in mind for understanding easier the distributions of comparison sample of local galaxies, represented in figures from \ref{fig_morphdiag_ccon_cabr} to \ref{fig_morphdiag_cosmos_3plots_twomagbins} in Sec.~\ref{sec_results_morphdiagram}. \\  

\begin{figure*}
\centering
\begin{minipage}[c]{0.89\textwidth}
\includegraphics[width=14.5cm,angle=0]{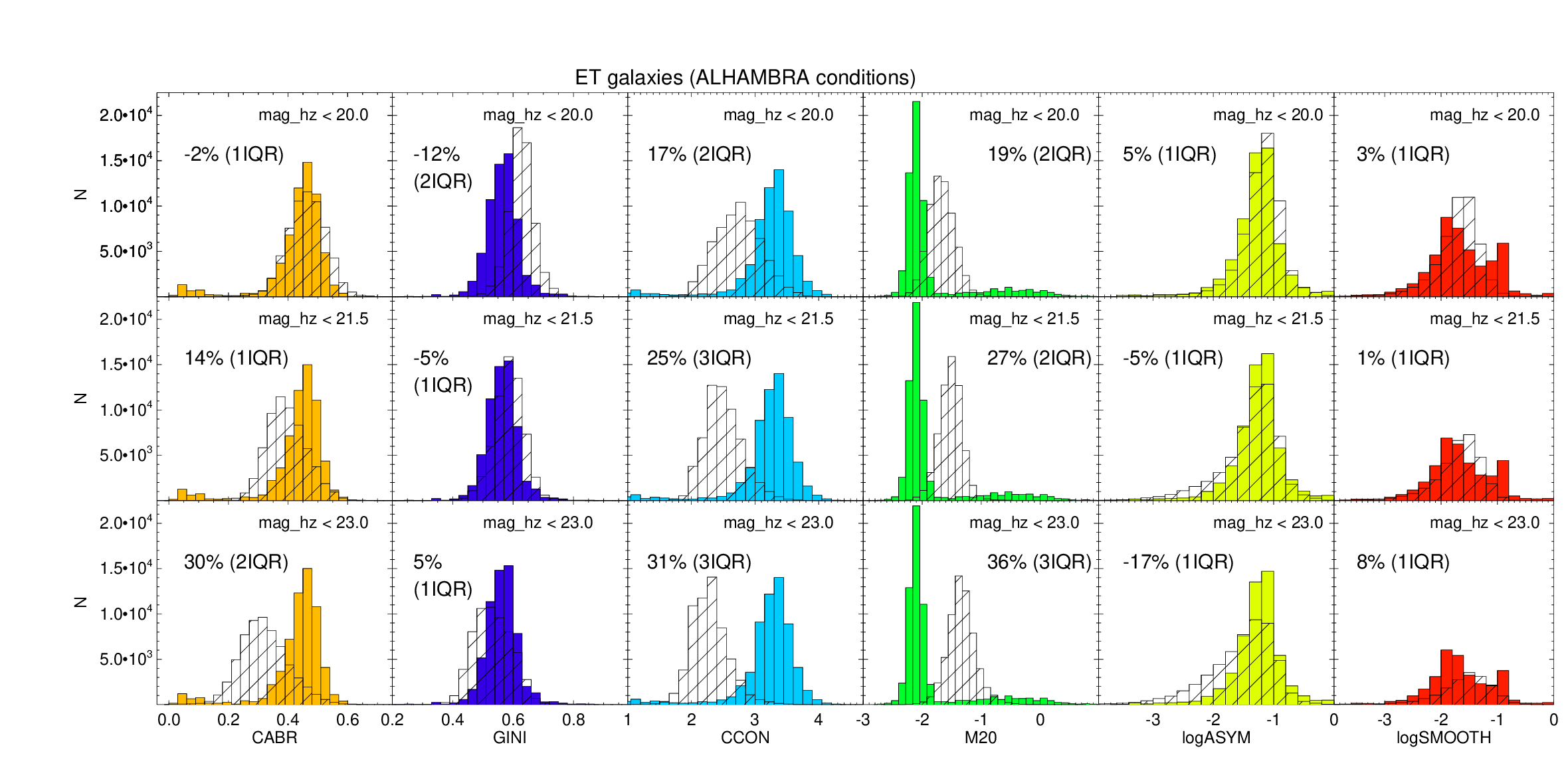}
\end{minipage}
\begin{minipage}[c]{0.89\textwidth}
\includegraphics[width=14.5cm,angle=0]{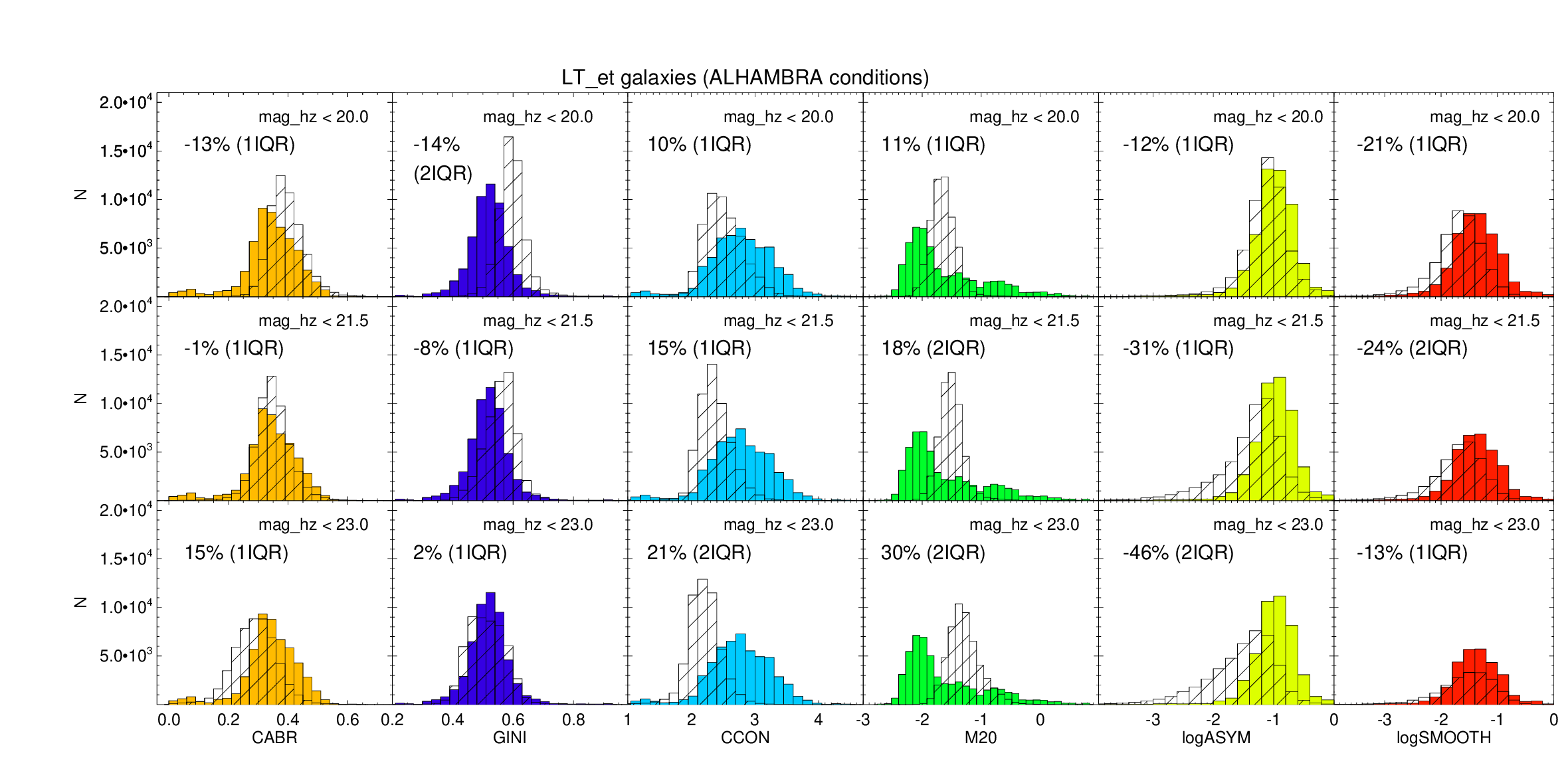}
\end{minipage}
\begin{minipage}[c]{0.89\textwidth}
\includegraphics[width=14.5cm,angle=0]{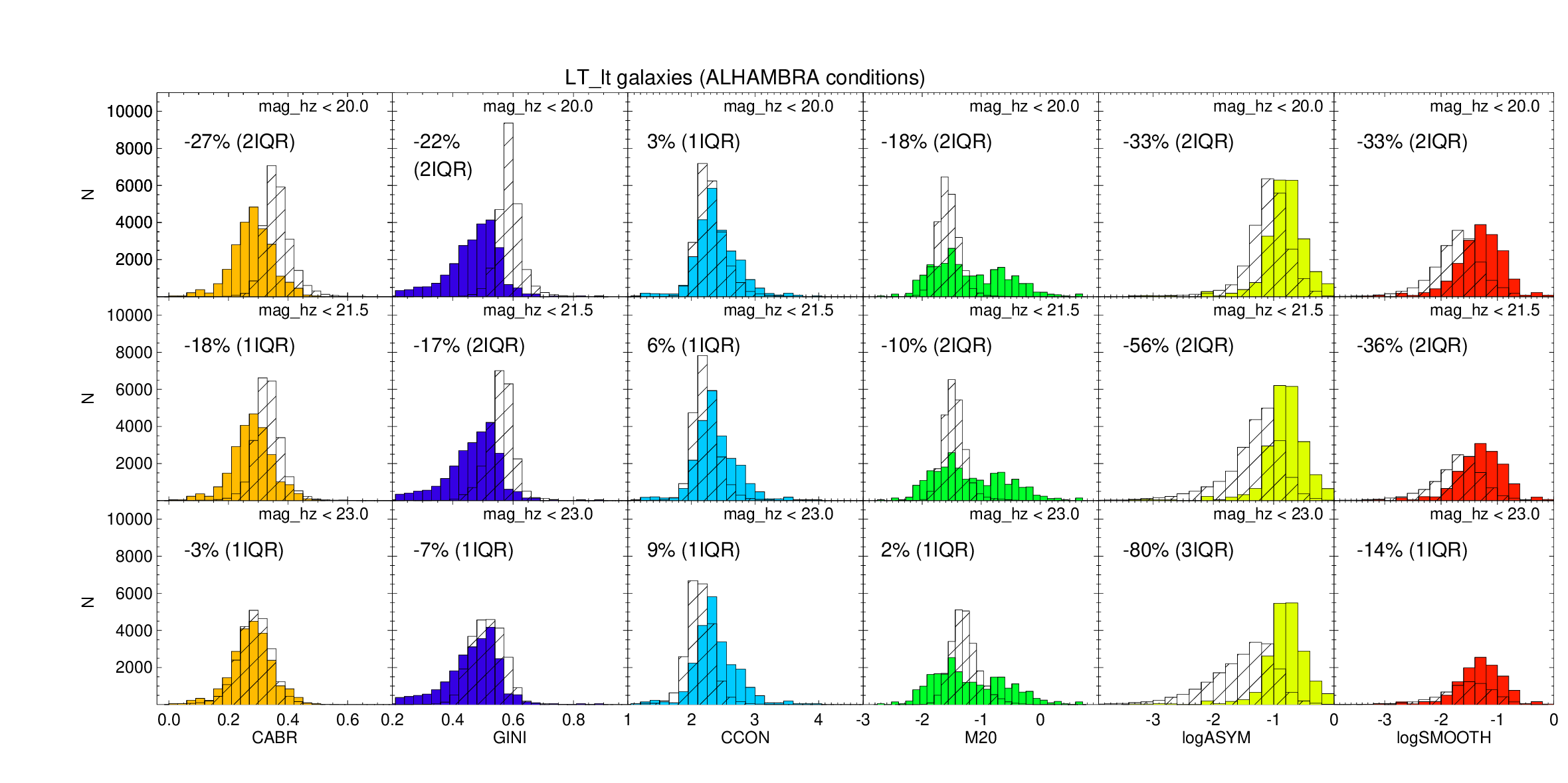}
\end{minipage}
\caption[ ]{Comparison between the morphological parameters of the sample of local galaxies measured at their real redshift (filled-colour histograms, representing the reference galaxy properties) and in conditions that correspond to the galaxies from the ALHAMBRA survey (filled-line histograms). Six parameters are compared, from left to right: Abraham concentration index, Gini coefficient, Conselice-Bershady concentration index, M20 moment of light, asymmetry, and smoothness. \textit{Top)} three magnitude cuts of the ET galaxies, \textit{middle)} LT\_et galaxies, and \textit{bottom)} LT\_lt galaxies. In all diagrams the numbers represent the difference between the median value of the reference (local) and simulated distributions, normalised with the median of the local sample (expressed in \%), and the corresponding interquartile range. (A colour version of this figure is available in the online journal)}
\label{fig_alh_param_dist}
\end{figure*}

\begin{figure*}
\centering
\begin{minipage}[c]{0.89\textwidth}
\includegraphics[width=14.5cm,angle=0]{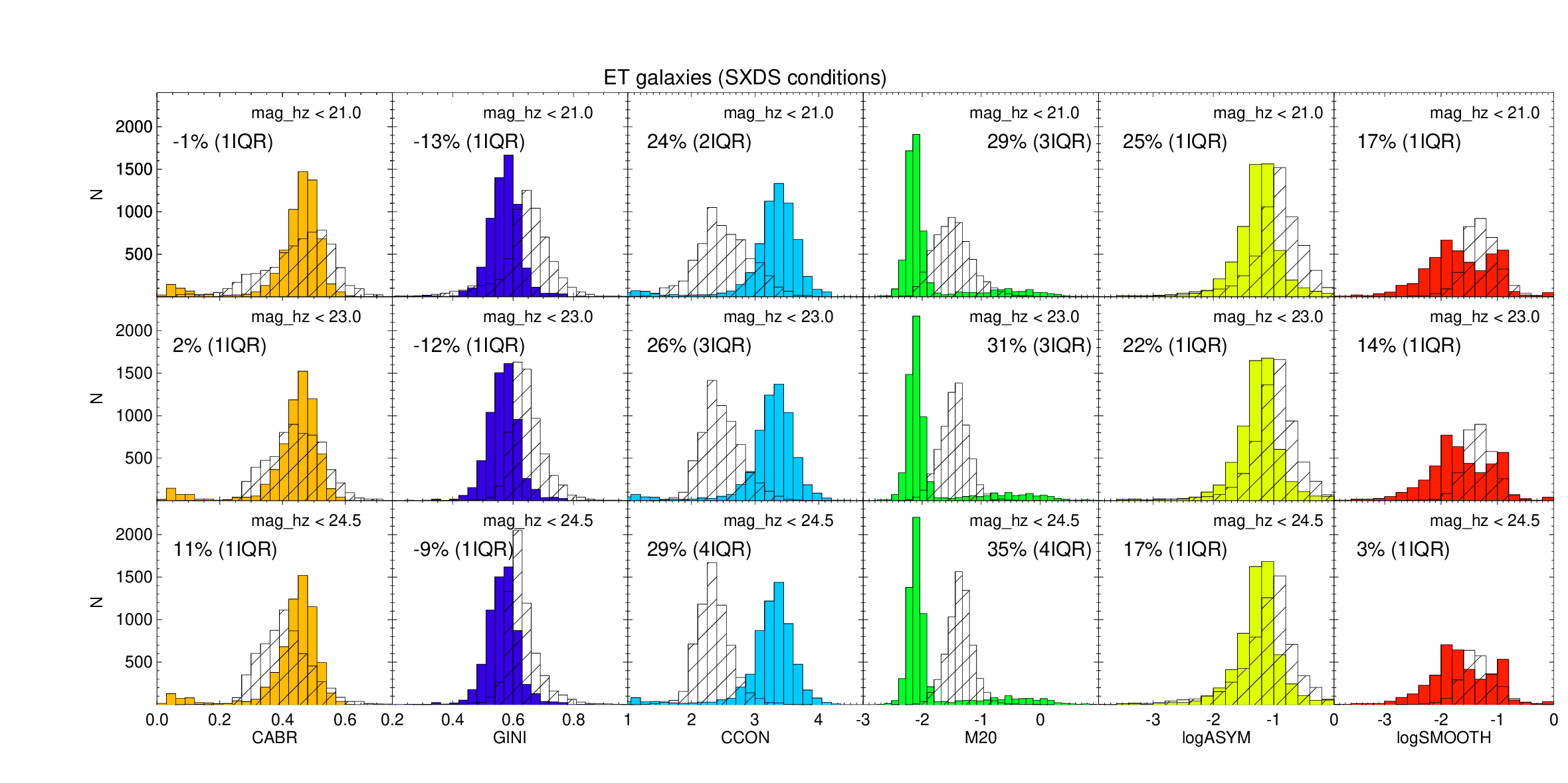}
\end{minipage}
\begin{minipage}[c]{0.89\textwidth}
\includegraphics[width=14.5cm,angle=0]{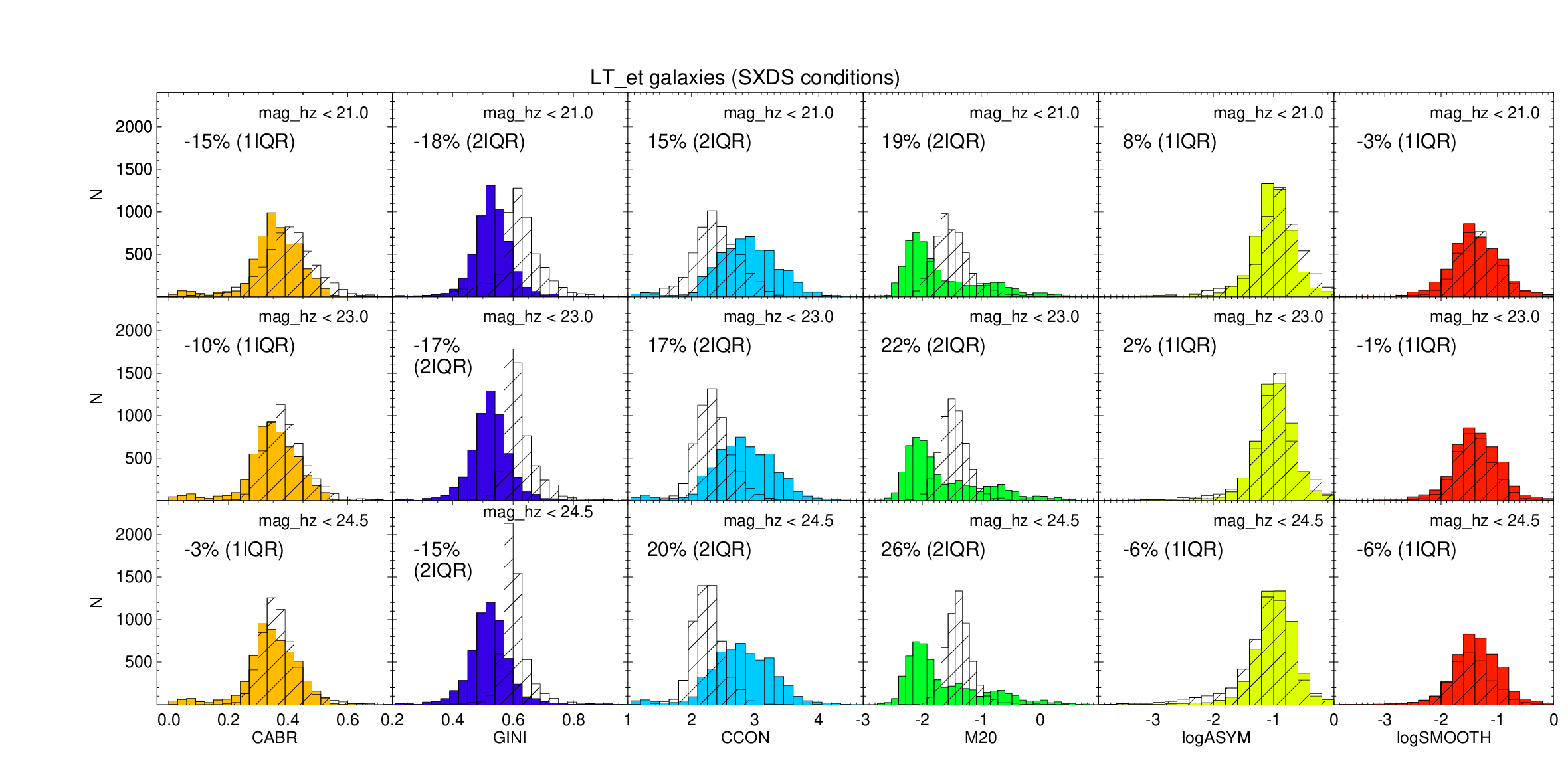}
\end{minipage}
\begin{minipage}[c]{0.89\textwidth}
\includegraphics[width=14.5cm,angle=0]{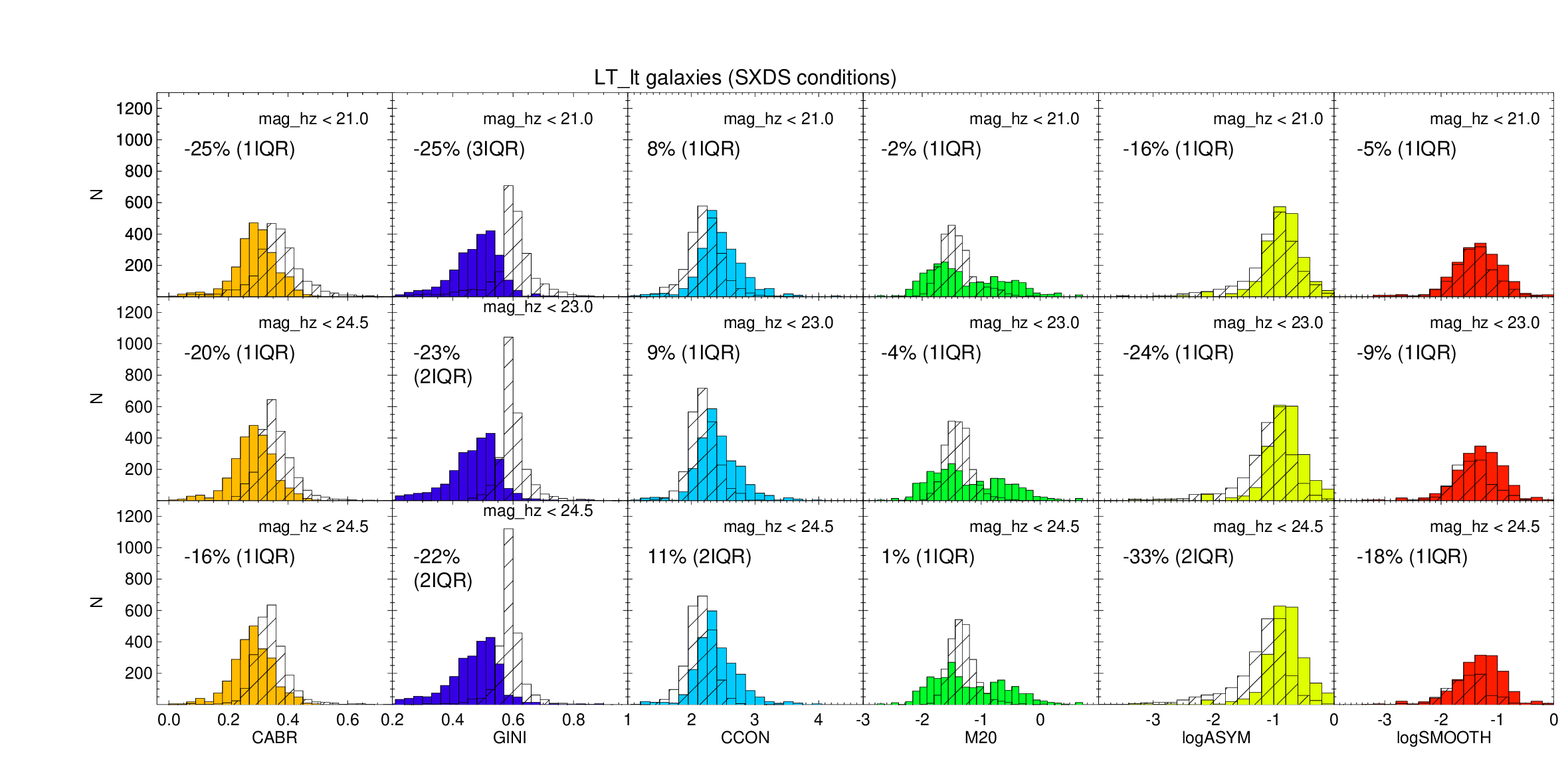}
\end{minipage}
\caption[ ]{Same as Fig.~\ref{fig_alh_param_dist}, but moving the local sample to map the conditions of the SXDS survey. \\ \\(A colour version of this figure is available in the online journal)}
\label{fig_sxds_param_dist}
\end{figure*}

\begin{figure*}
\centering
\begin{minipage}[c]{0.89\textwidth}
\includegraphics[width=14.5cm,angle=0]{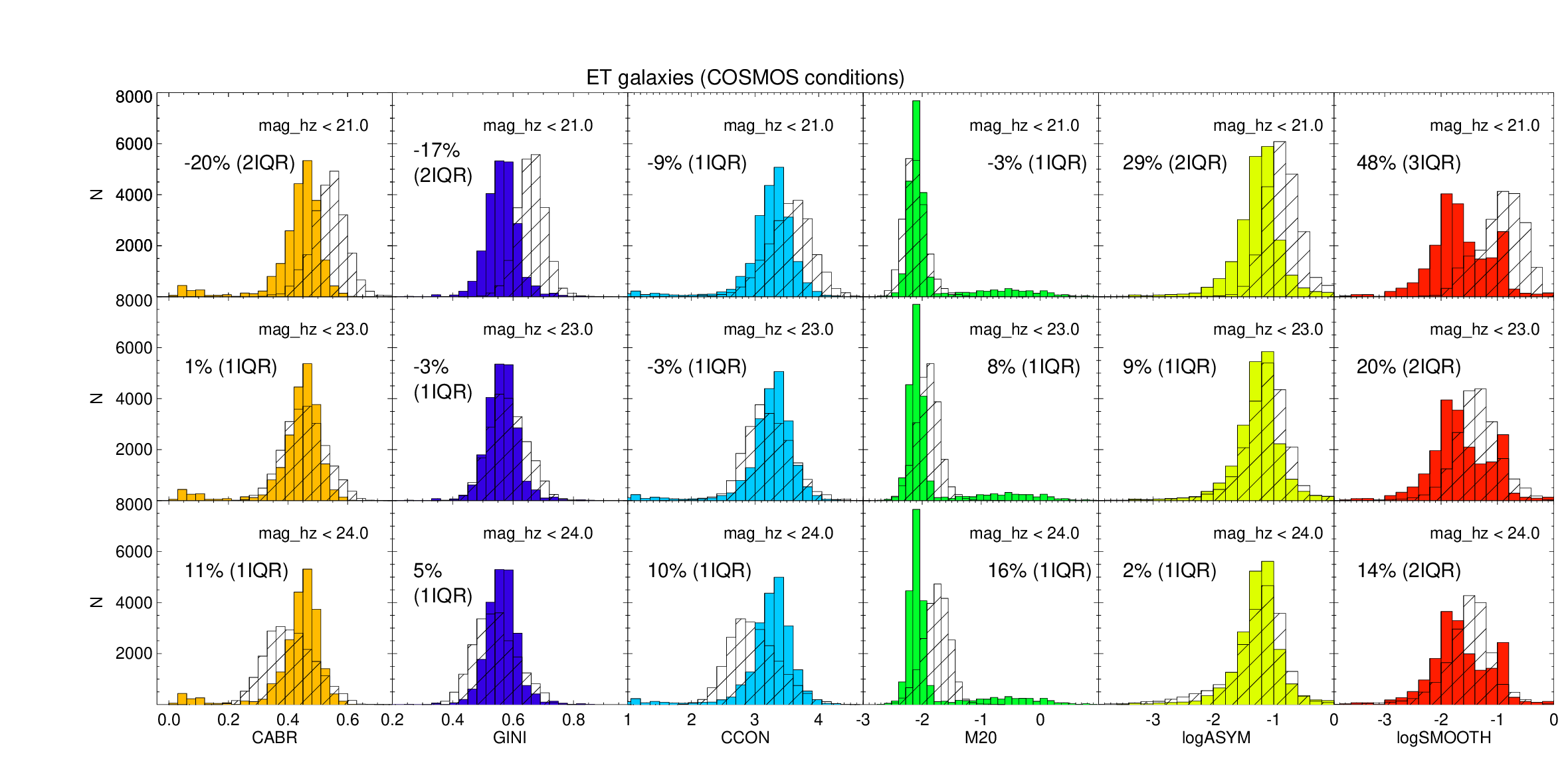}
\end{minipage}
\begin{minipage}[c]{0.89\textwidth}
\includegraphics[width=14.5cm,angle=0]{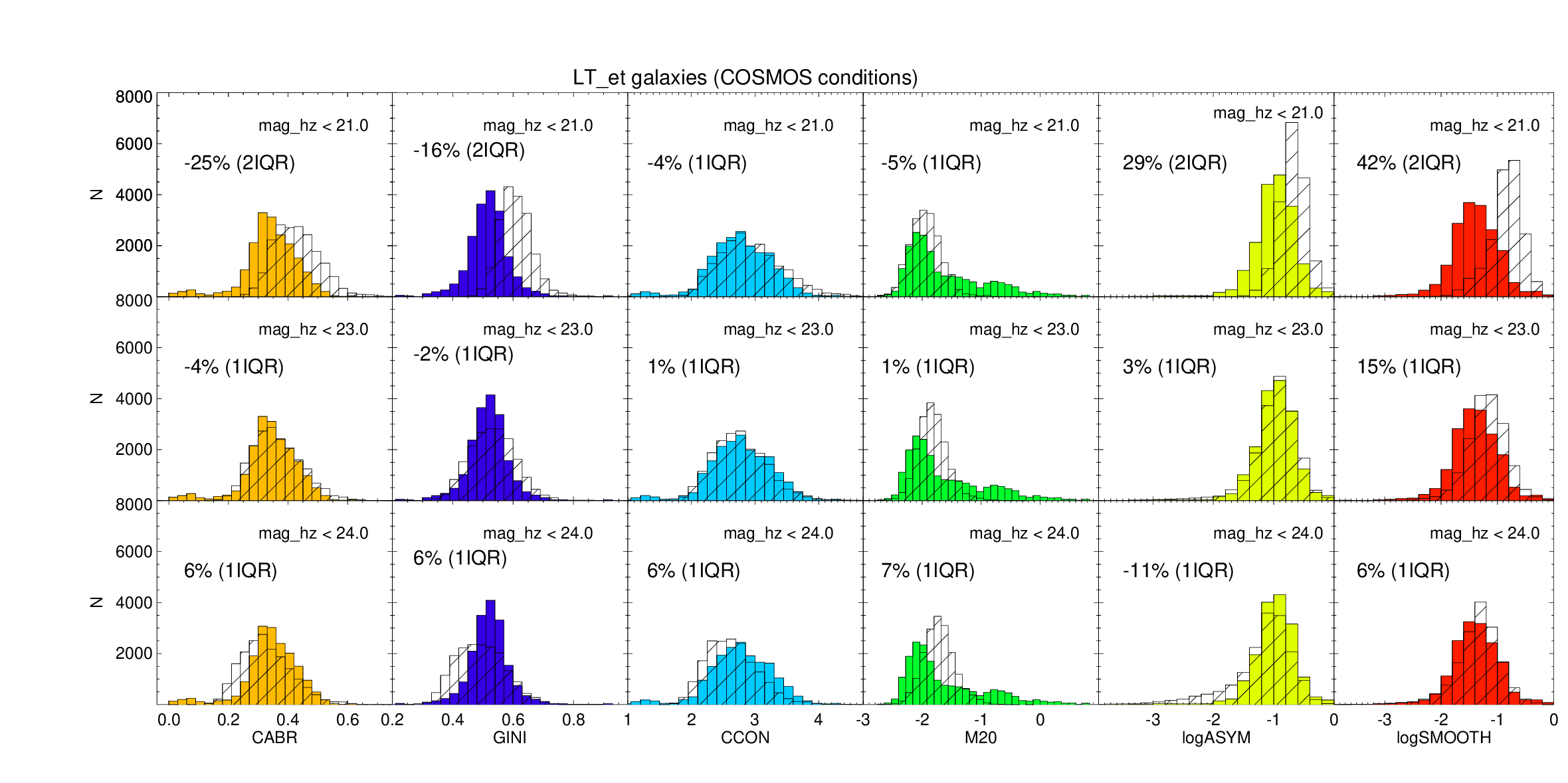}
\end{minipage}
\begin{minipage}[c]{0.89\textwidth}
\includegraphics[width=14.5cm,angle=0]{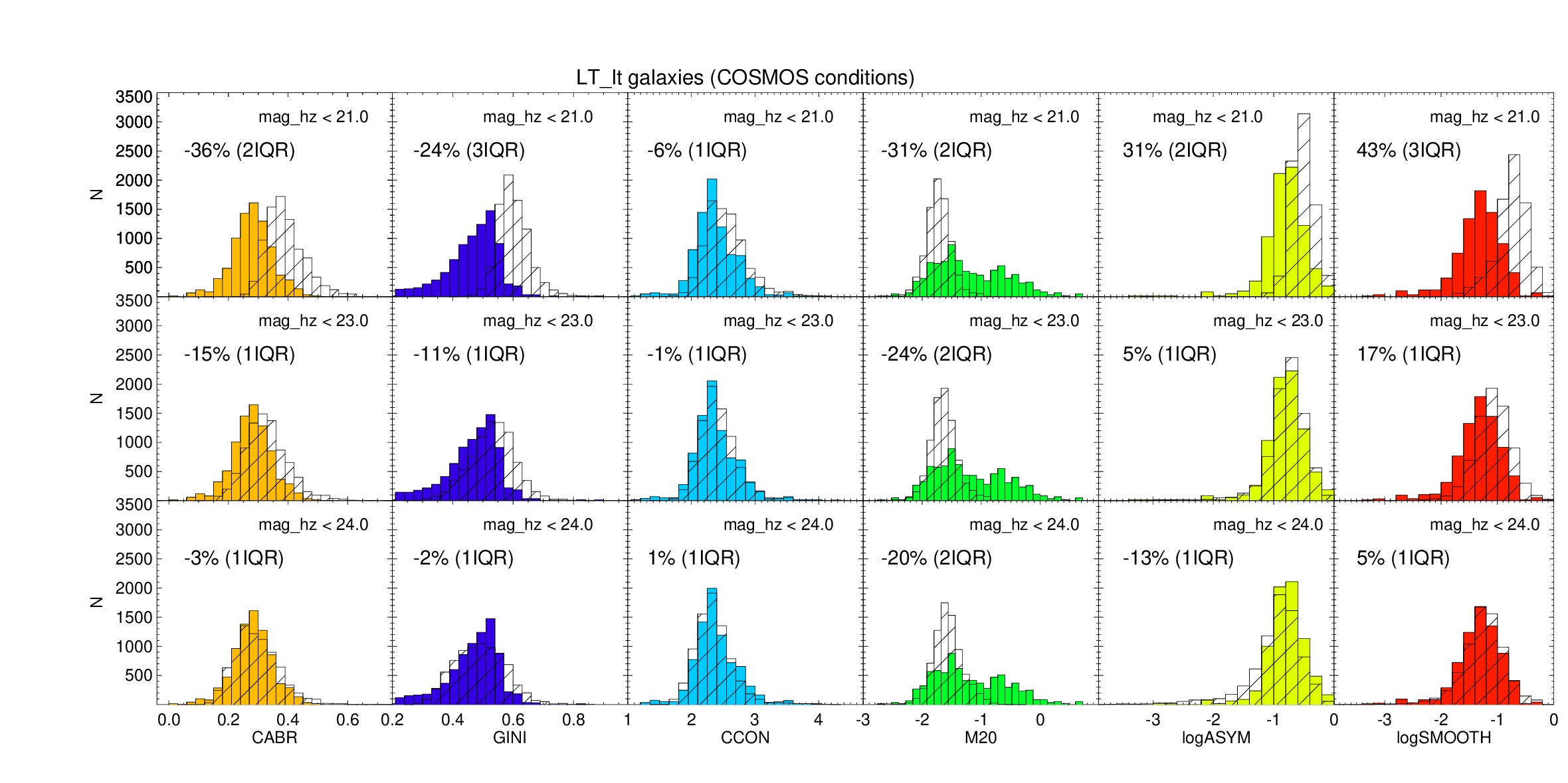}
\end{minipage}
\caption[ ]{Same as Fig.~\ref{fig_alh_param_dist}, but moving the local sample to map the conditions of the COSMOS survey. \\ \\ (A colour version of this figure is available in the online journal)}
\label{fig_cosmos_param_dist}
\end{figure*}

\indent The obtained morphological parameters of simulated galaxies are represented  with the filled-line histograms in Fig.~\ref{fig_alh_param_dist}, \ref{fig_sxds_param_dist}, and \ref{fig_cosmos_param_dist}, that reproduce the conditions of the ALHAMBRA, SXDS, and COSMOS surveys, respectively. In each figure are shown the distributions of CABR, GINI, and CCON concentration indexes, M20, asymmetry, and smoothness (from left to right panels) for three morphological groups: ET (top plots), LT\_et (middle plots), and LT\_lt (bottom plots), defined in Sec.~\ref{sec_data_locsample}. For each morphological group, we represent the distribution of parameters at the three analysed magnitude cuts. In these figures we also represent the distributions of the reference morphological parameters of local galaxies, measured in Sec.~\ref{sec_results_morph_z0}, with the filled-colour histograms.\\

\indent For all morphological parameters we can observe important differences between the reference (local) values and those obtained after moving the local galaxies to the conditions of non-local surveys, indicating the influence that the observational bias from spatial resolution and depth can have on them. To quantify this bias, we measured the difference between the median values of the local and simulated distributions, normalised with respect to the median value of the local sample:
\begin{center}
$<$\,local\,$>$\,-\,$<$\,simulated\,$>$\,/\,$<$\,local\,$>$. 
\end{center}
The obtained values are expressed in \% in each plot, together with the corresponding range of 
interquartile range (IQR; between parentheses), measured as the difference between the third and first quartiles. In the following, we describe the obtained differences, considering as significant those within $\ge$\,2\,IQR.
\subparagraph {\textit{CABR\\}}
\indent In ALHAMBRA conditions, CABR showed significant differences in the case of the faintest ETs and brightest LTs ($\sim$\,30\% within 2\,IRQ), with galaxies appearing less and more concentrated, respectively. In the conditions of SXDS we did not find any differences (all within 1\,IQR). In COSMOS, we obtained differences of $\sim$\,20\,-\,36\% (within 2\,IQR) at the brightest magnitude cuts for all three morphological types, with simulated galaxies appearing more concentrated. In ALHAMBRA, 2\,-\,3\% of sources showed invalid measurements (with values outside the typical range of this parameter, between 0 and 1), independently of the morphological type; in SXDS, this went between 3\,-\,5\% for ET and LT\_et, and 7\,-\,10\% for LT\_lt galaxies; while in COSMOS, 2\,-\,3\% for ET and LT\_et, and 8\,-\,10\% for LT\_lt galaxies.  
\subparagraph {\textit{GINI\\}}
\indent In ALHAMBRA, this parameter showed differences of 10\,-\,20\% (2\,IQR) at the first magnitude cut for all types, and at the second magnitude cut for LT\_lt galaxies. In SXDS, the difference for LT\_et galaxies was observed at all three magnitude cuts (15\,-\,18\% within 2\,IQR). The same was obtained for LT\_lt sources but with slightly higher differences. In the case of COSMOS, the only differences were observed at the first magnitude cut: $\sim$\,20\% (2\,-3\,IQR) for all three types. In all cases where the differences were detected, we obtained negative values, showing that the simulated galaxies would appear more concentrated than they really are. With respect to the measured values of simulated galaxies that are outside of the valid ranges, the populations are in agreement with those for the CABR parameter. 
\subparagraph {\textit{CCON\\}}
\indent A different behavior of CCON was detected for the conditions of the analysed ground- and space-based surveys. In the case of COSMOS, we did not detect any differences, independently of the magnitude cut and morphological type. This index showed to be especially sensitive to the spatial resolution in comparison to CABR and GINI. In the case of ALHAMBRA and SXDS, the highest differences are detected for ET galaxies at the highest magnitude cuts, $\sim$\,30\% at 3\,IQR and 4\,IQR. In all cases where differences were observed in the ground-based surveys, lower concentrations appeared for simulated galaxies in comparison with the reference (local) values, independently on the magnitude cut and morphological type. Again, the population of the obtained invalid measurements is consistent with those presented for CABR. 
\subparagraph {\textit{M20\\}}
\indent As described in Sec.~\ref{sec_method_param}, this parameter is sensitive to galaxy structures as the central light from the bulge, bars, spiral arms, strong star formation regions, etc. Therefore, for local ET galaxies it mainly maps the central light, having a much narrower distribution than in the case of LT\_et, or even more in the case of LT\_lt galaxies, where the distribution is broader (filled-colour histograms in Figs.~\ref{fig_alh_param_dist}, \ref{fig_sxds_param_dist}, and \ref{fig_cosmos_param_dist}). However, when moved to higher-redshift surveys the distribution of this parameter changed significantly (both, shape and range; filled-line histograms), and showed in all cases the information coming mainly from the centre of the galaxy, and independently on the morphological type. In ALHAMBRA, the highest differences  are of order $\ge$\,30\% within 2\,-3\,IQR, for all three types. In SXDS conditions, for ET galaxies M20 suffered similar changes as in ALHAMBRA (at higher magnitude cuts), except for LT\_LT sources where no difference were found respect to local sample. In COSMOS conditions, M20 shows higher concentrations for simulated galaxies in comparison with the local ones, affecting only LT\_lt galaxies, with differences $\le$\,-31\% (within 2\,IQR). The population of sources with invalid measurements of M20 is consistent with that for CABR. 
\subparagraph {\textit{ASYM\\}}
\indent This parameter showed the highest differences in the case of ALHAMBRA (even up to $\sim$\,80\% within 3\,IQR at the faintest magnitude cut F613W\,$\le$\,23.0), affecting LT galaxies in the sense of obtaining more symmetric galaxies in comparison with their reference (z\,$\sim$\,0) values. In COSMOS, all three morphological types suffered similar shifts ($\sim$\,30\% within 2\,IQR), being the strongest one at the lowest magnitude cut F814W\,$\le$\,21.0, but showing positive values in comparison to ALHAMBRA, and so producing more asymmetric galaxies in simulated conditions. The population with invalid measurements was higher ($<$\,20\%) than in the case of CABR, GINI, and CCON. All these measurements were obtained without taking into account the noise introduced by the galaxy itself once being dropped into the background of non-local surveys. However, as showed in Sec.~\ref{sec_method_highz}, in the case of ASYM and SMOOTH this noise could have a significant effect on these parameters, and the differences measured there should be added to those measured in this section. Therefore, we found these parameters unreliable for the use at higher redshifts in the case of ground-based surveys. In the space-based surveys similar to COSMOS, it becomes unreliable for the use in the case of galaxies fainter than F814W\,=\,24.0.  
\subparagraph {\textit{SMOOTH\\}}
\indent Without taking into account the noise studied in Sec.~\ref{sec_method_highz}, the differences in SMOOTH in the case of ALHAMBRA were detected only for LT galaxies, being lower than 36\% (within 2\,IQR); in the case of SXDS, no differences were observed between the simulated and local samples; nevertheless, in COSMOS conditions it showed 40\,-\,50\% differences (within 2\,-\,3\,IQR) at the brightest magnitude cuts F814W\,$\le$\,21.0, resulting in galaxies to appear more clumpy. Again, to these differences those related to the inclusion of noise (see Sec.~\ref{sec_method_highz}) should be added. Moreover, this parameter showed to be significantly more affected with both spatial resolution and depth, when considering the number of invalid measurements of the simulated sample (up to 50\% and 20\% at the faintest magnitude cuts in the case of ALHAMBRA and COSMOS, respectively). Taking all this into account, we found this parameter especially sensitive to all, noise, spatial resolution, and data depth, and therefore unreliable for the use in the morphological classification of galaxies in the case of ground-based surveys, and for faintest sources in space-based surveys.\\

\subsection[]{Behaviour of morphological diagnostic diagrams}
\label{sec_results_morphdiagram}

\begin{figure}
\centering
\begin{minipage}[c]{0.49\textwidth}
\includegraphics[width=8.2cm,angle=0]{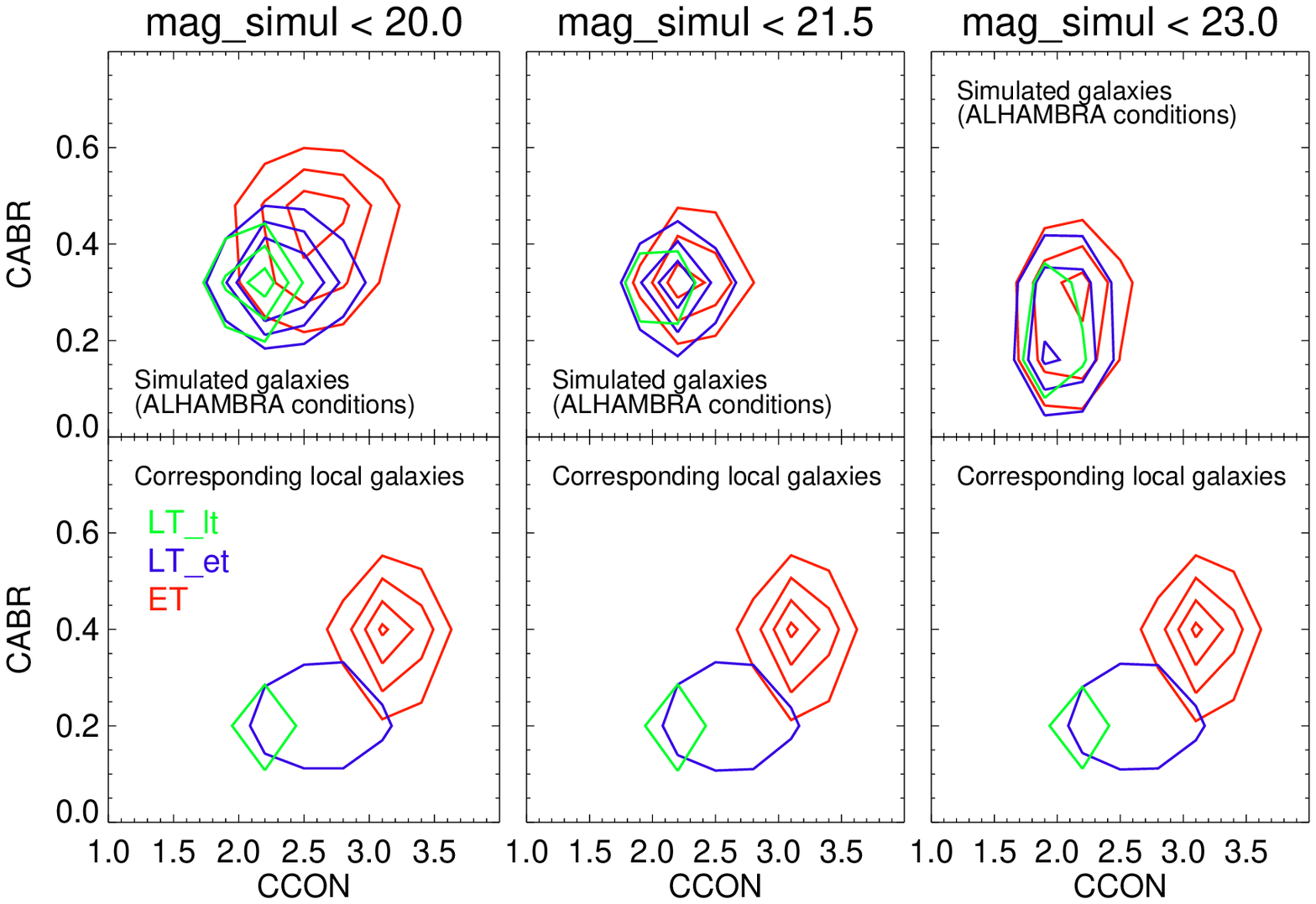}
\end{minipage}
\begin{minipage}[c]{0.49\textwidth}
\includegraphics[width=8.2cm,angle=0]{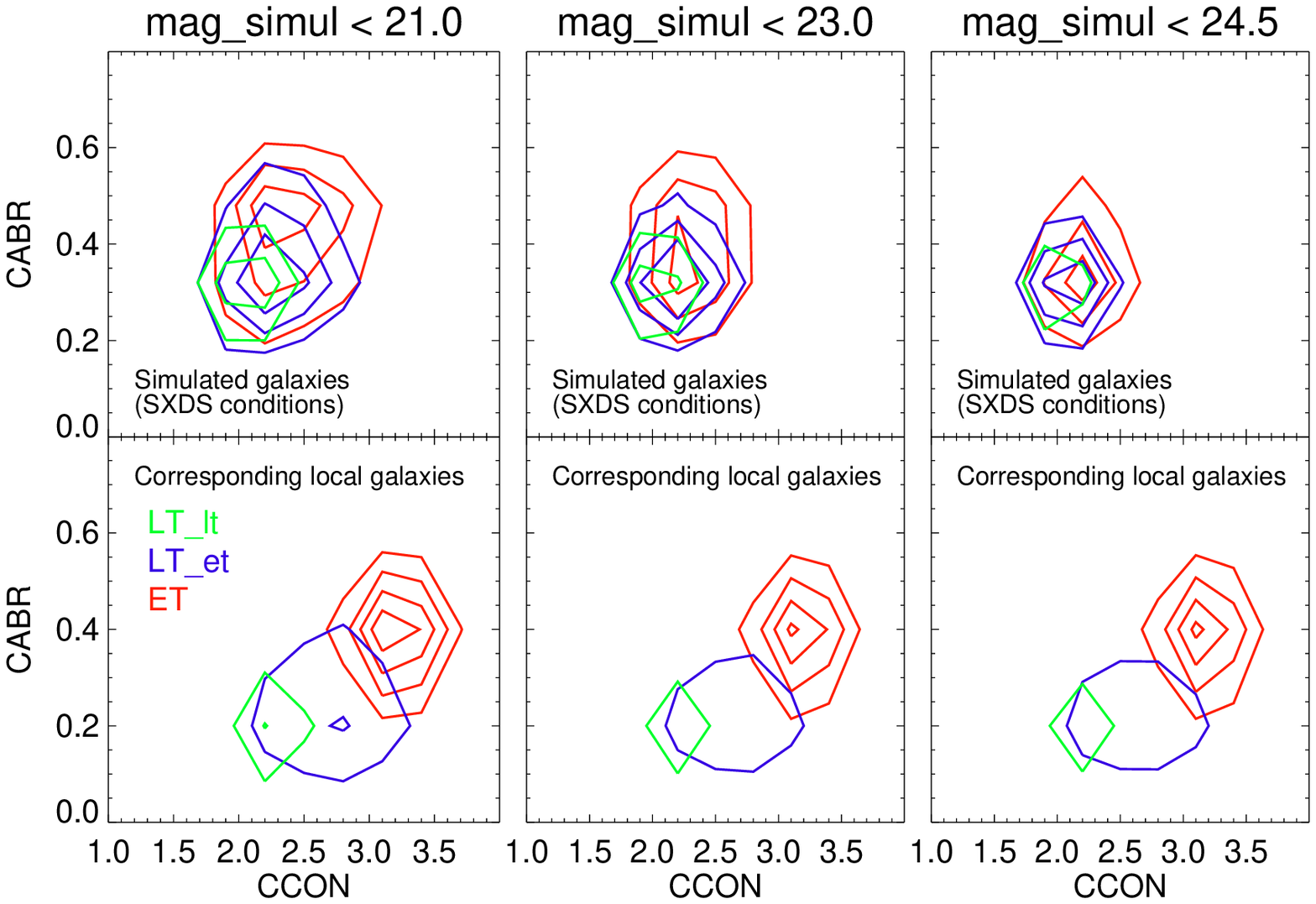}
\end{minipage}
\begin{minipage}[c]{0.49\textwidth}
\includegraphics[width=8.2cm,angle=0]{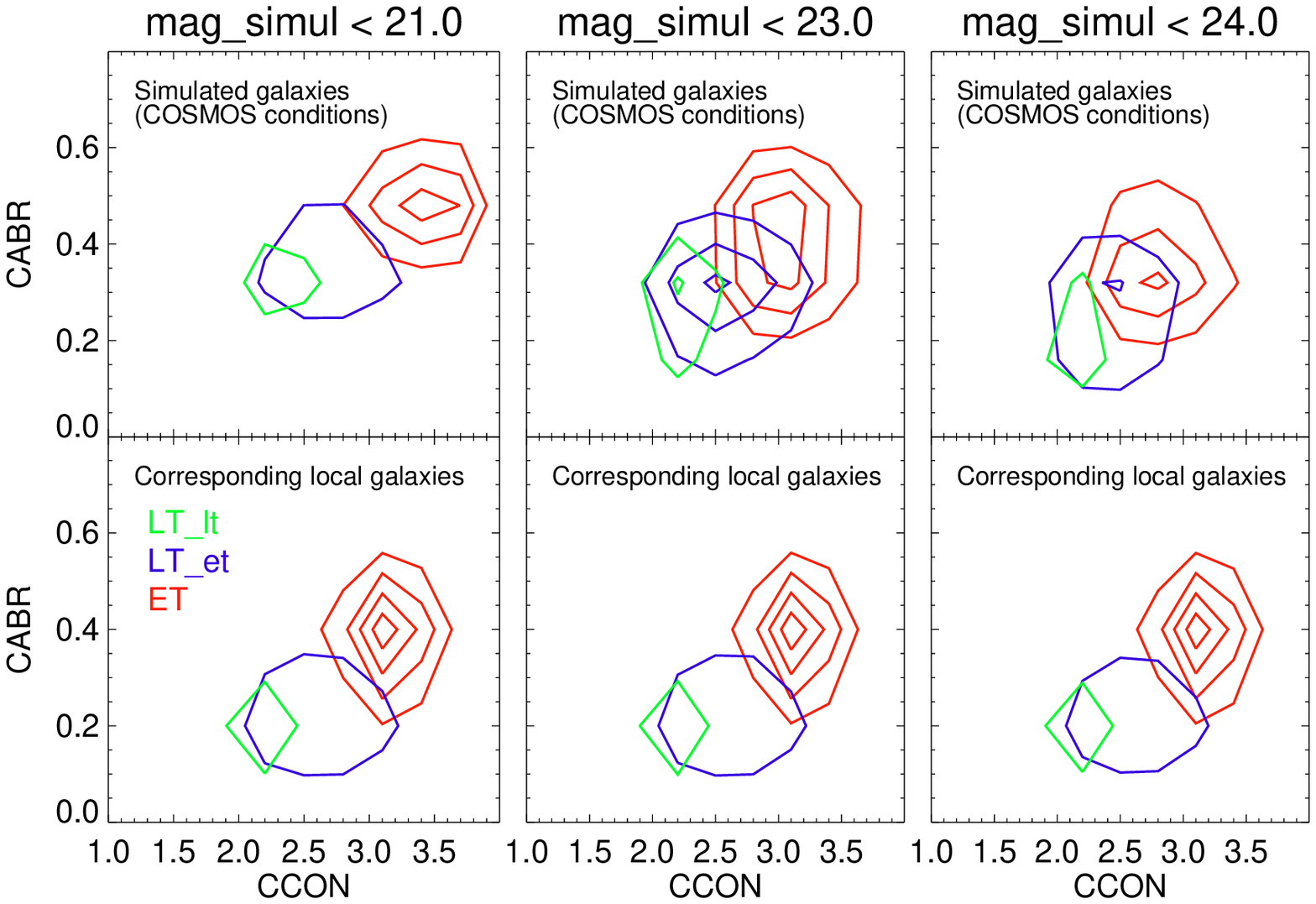}
\end{minipage}
\caption[ ]{Relation between the Conselice-Bershady and Abraham concentration indexes, in the three analysed non-local surveys: ALHAMBRA \textit{(top plots)}, SXDS \textit{(middle plots)}, and COSMOS \textit{(bottom plots)}. In each survey, \textit{top rows} represent the morphological parameters of simulated sample obtained after moving the local galaxies to higher redshifts, considering three magnitude cuts (the first and the last column showing the lowest and the highest analysed magnitude cut, respectively), while \textit{bottom rows} show the corresponding reference (local) values. The red, blue, and green contours represent ET, LT\_et, and LT\_lt galaxies, respectively. \\\\(A colour version of this figure is available in the online journal)}
\label{fig_morphdiag_ccon_cabr}
\end{figure}

\begin{figure}
\centering
\begin{minipage}[c]{0.49\textwidth}
\includegraphics[width=8.2cm,angle=0]{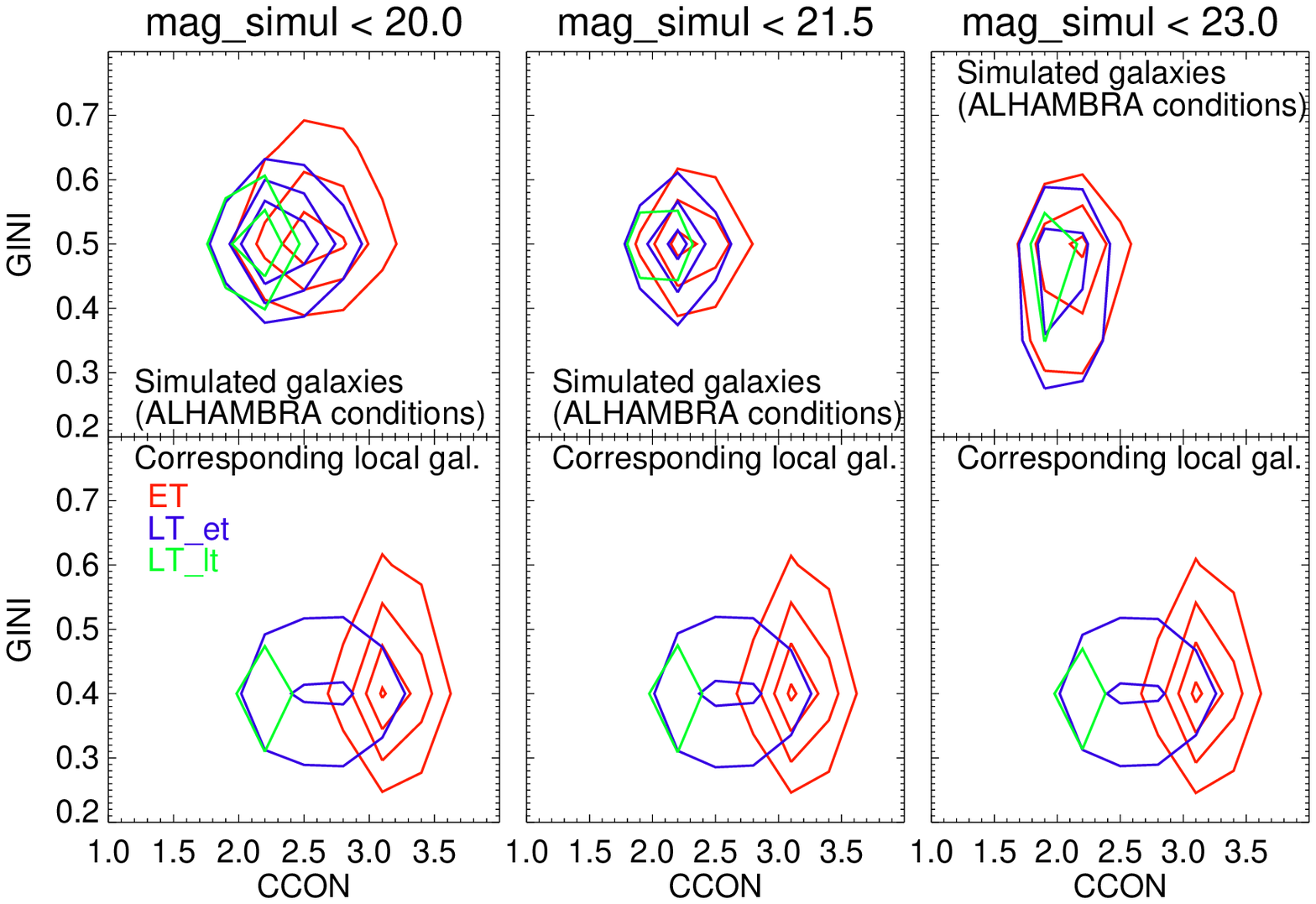}
\end{minipage}
\begin{minipage}[c]{0.49\textwidth}
\includegraphics[width=8.2cm,angle=0]{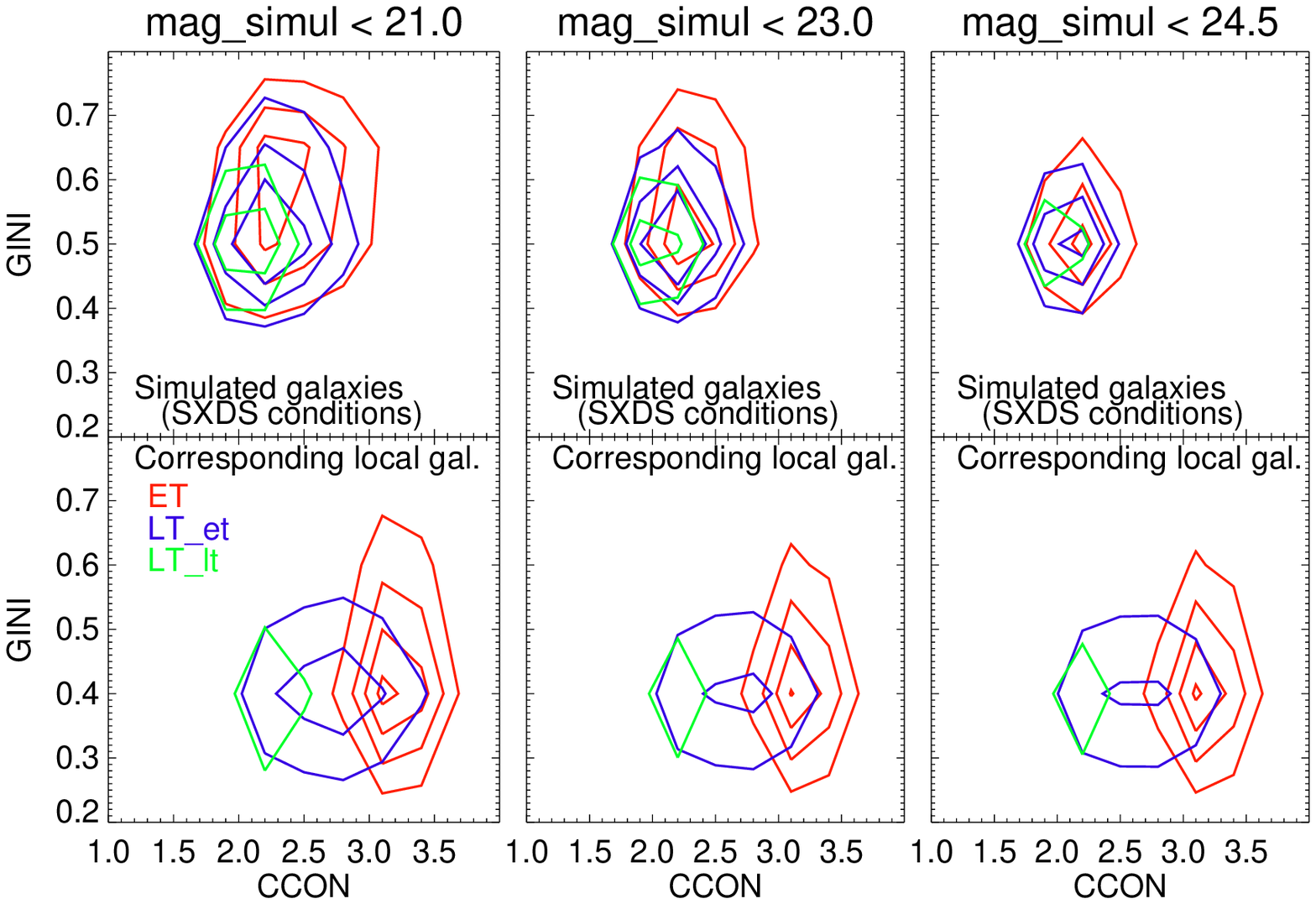}
\end{minipage}
\begin{minipage}[c]{0.49\textwidth}
\includegraphics[width=8.2cm,angle=0]{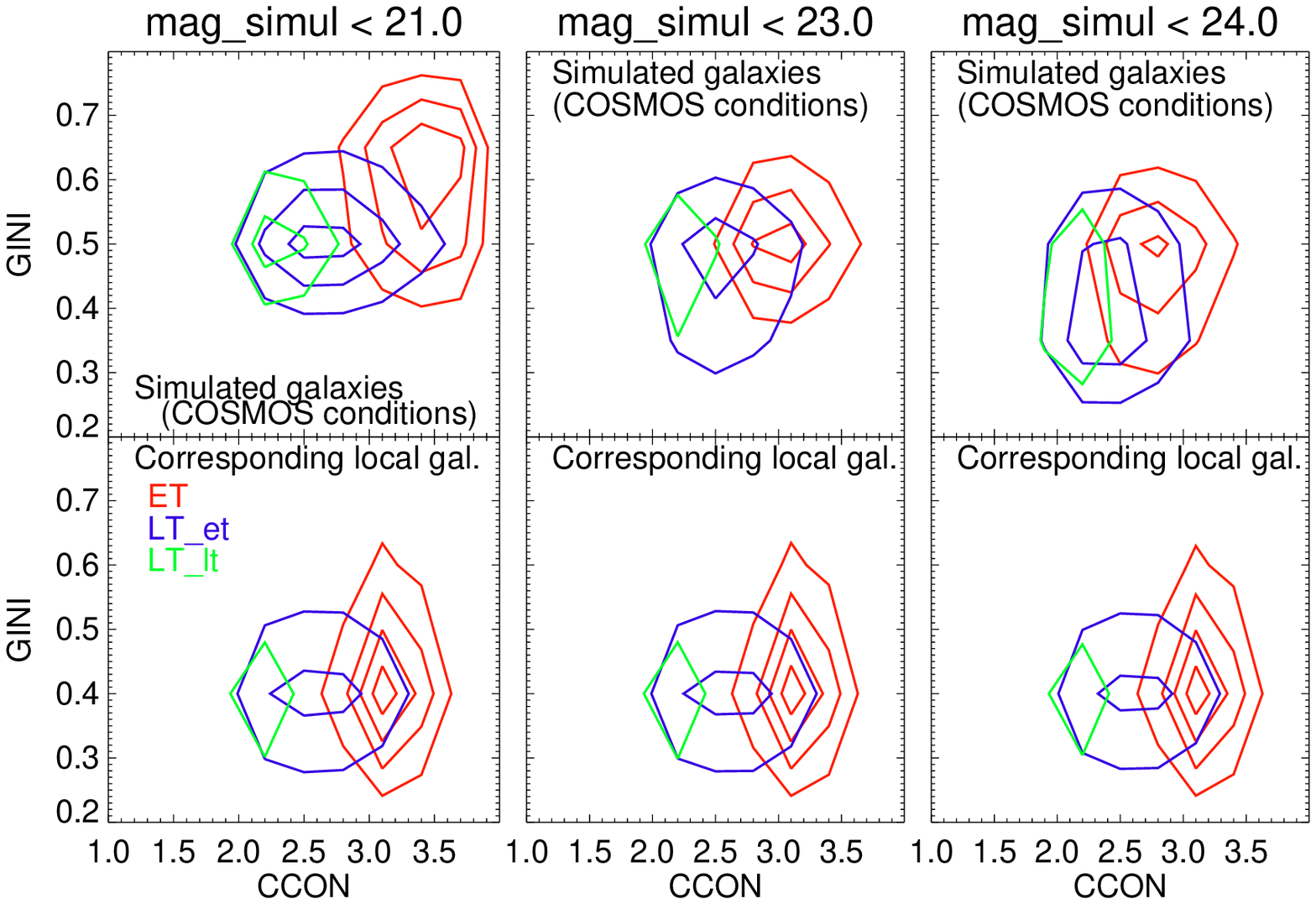}
\end{minipage}
\caption[ ]{Same as Fig.~\ref{fig_morphdiag_ccon_cabr}, but representing the relation between the Conselice-Bershady concentration index and Gini coefficient.\\\\(A colour version of this figure is available in the online journal)}
\label{fig_morphdiag_ccon_gini}
\end{figure}

\begin{figure}
\centering
\begin{minipage}[c]{0.49\textwidth}
\includegraphics[width=8.2cm,angle=0]{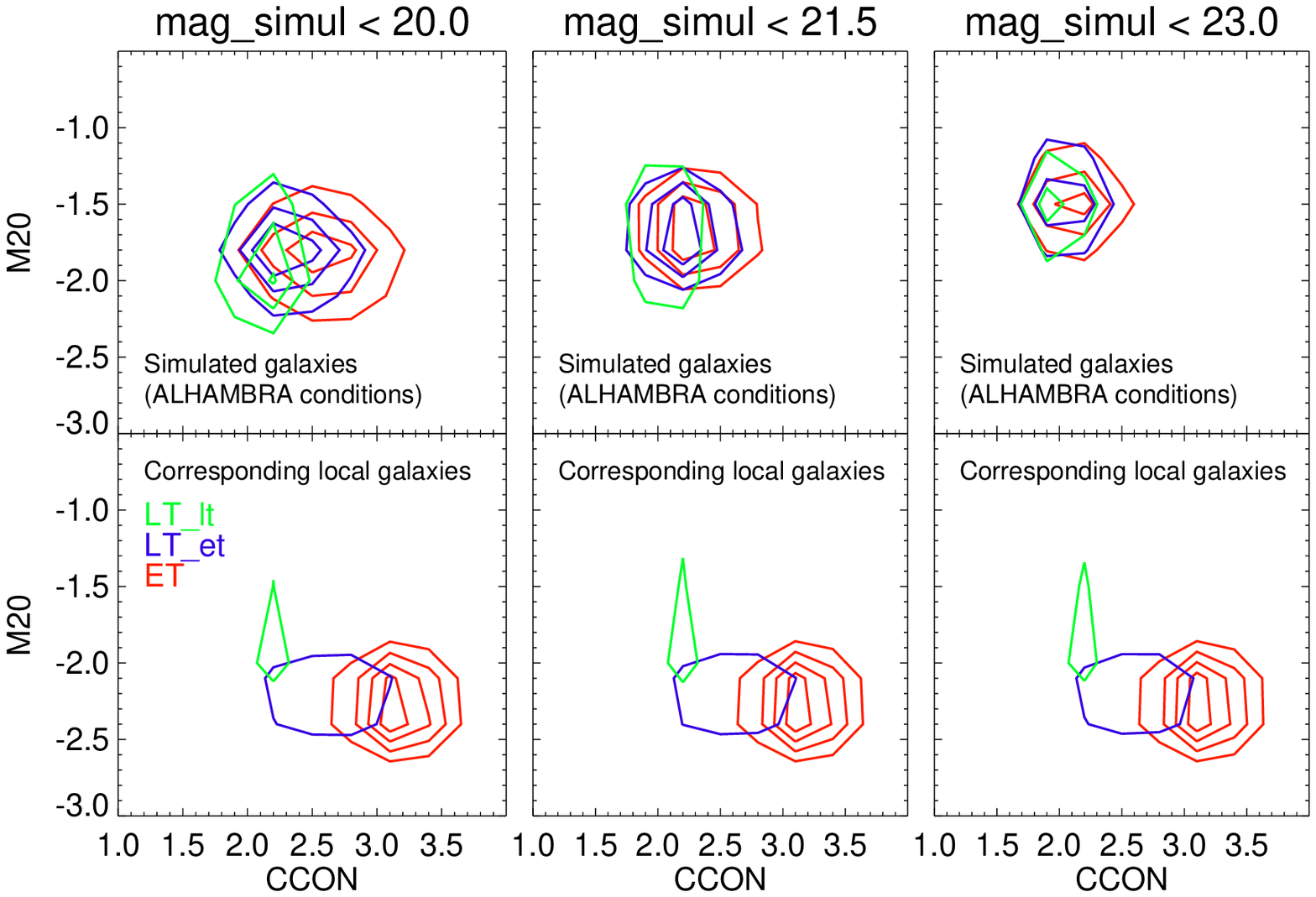}
\end{minipage}
\begin{minipage}[c]{0.49\textwidth}
\includegraphics[width=8.2cm,angle=0]{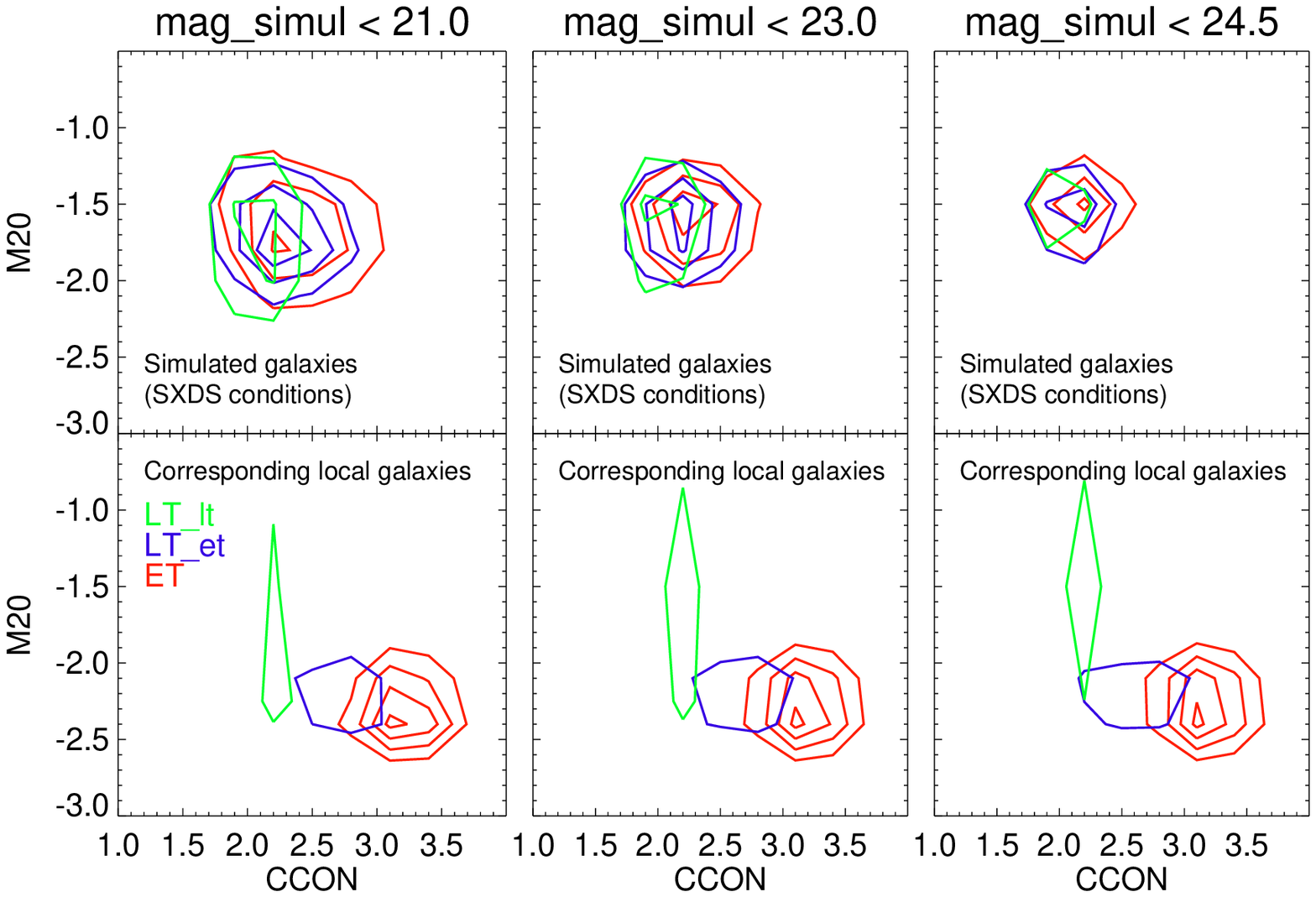}
\end{minipage}
\begin{minipage}[c]{0.49\textwidth}
\includegraphics[width=8.2cm,angle=0]{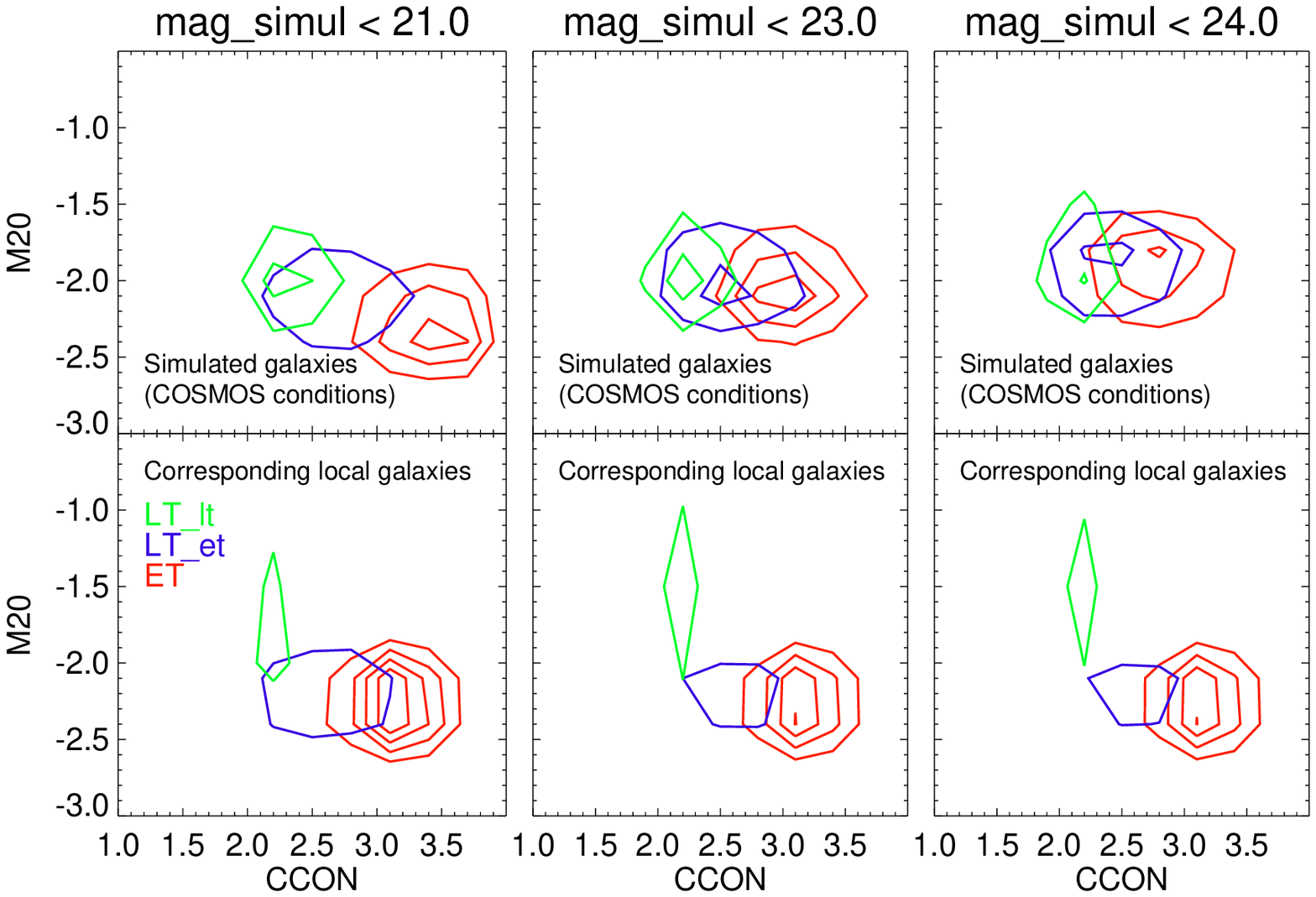}
\end{minipage}
\caption[ ]{Same as Fig.~\ref{fig_morphdiag_ccon_cabr}, but representing the relation between the Conselice-Bershady concentration index and M20 moment of light.\\\\(A colour version of this figure is available in the online journal)}
\label{fig_morphdiag_ccon_m20}
\end{figure}

\begin{figure}
\centering
\begin{minipage}[c]{0.49\textwidth}
\includegraphics[width=8.1cm,angle=0]{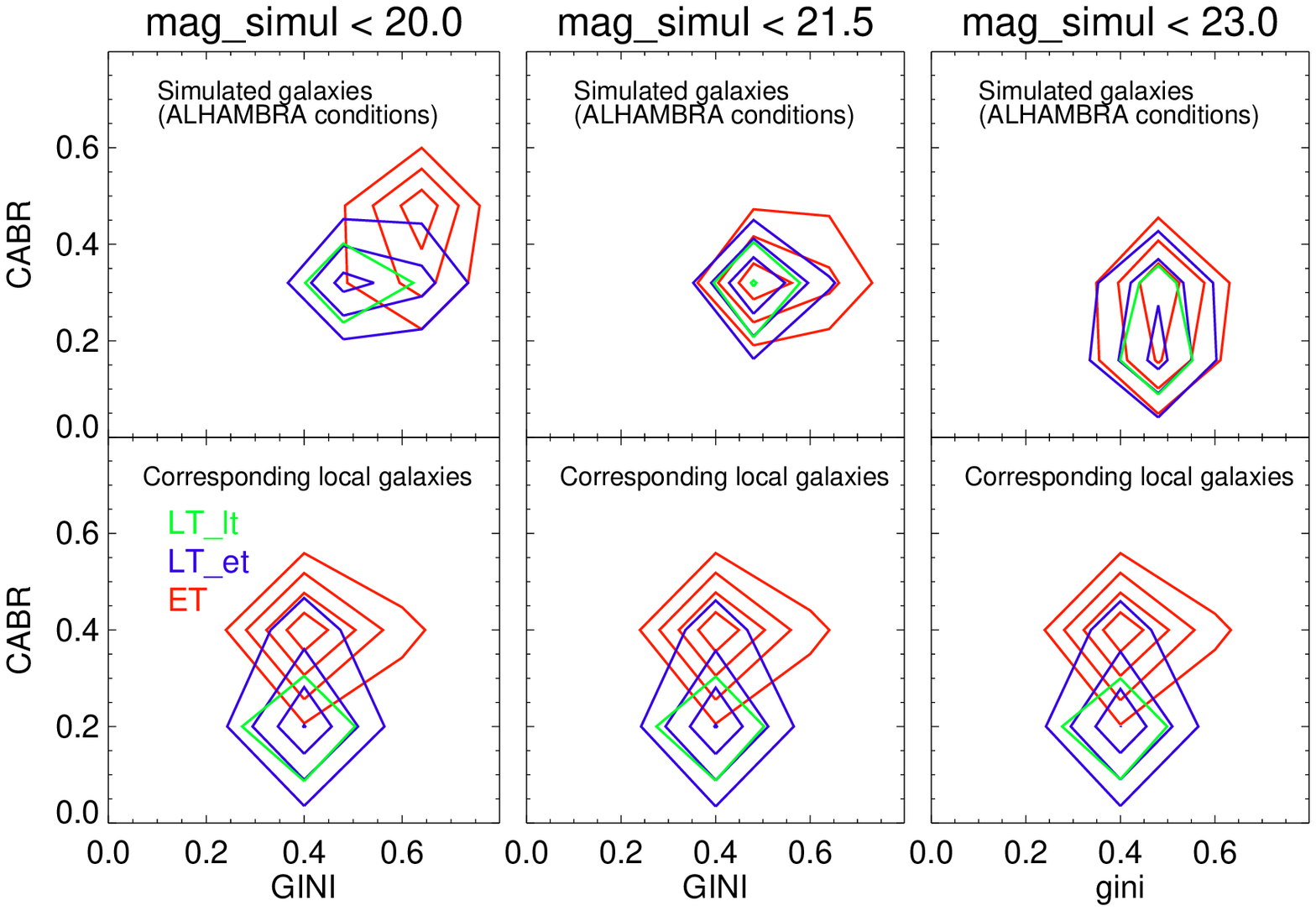}
\end{minipage}
\begin{minipage}[c]{0.49\textwidth}
\includegraphics[width=8.1cm,angle=0]{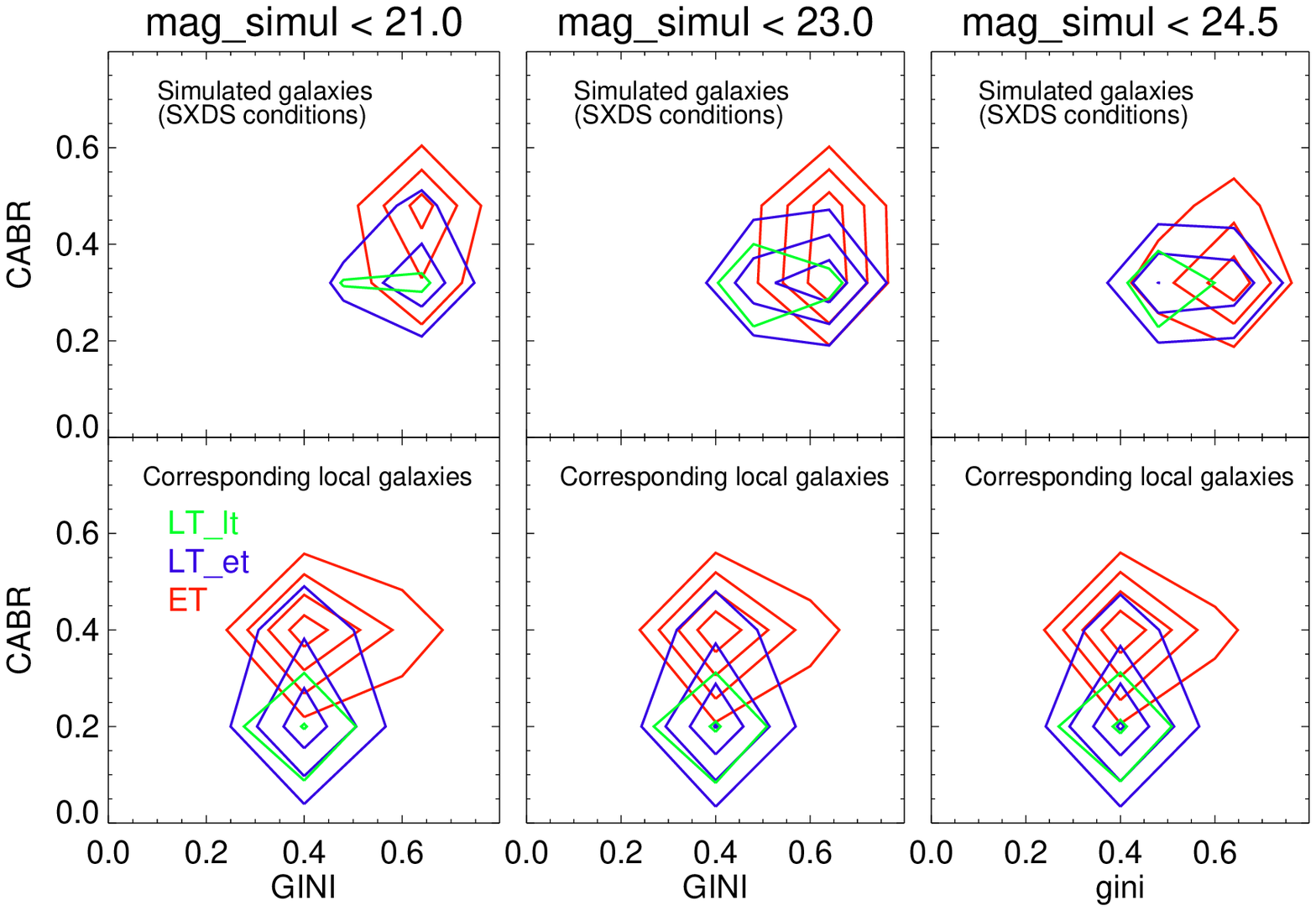}
\end{minipage}
\begin{minipage}[c]{0.49\textwidth}
\includegraphics[width=8.1cm,angle=0]{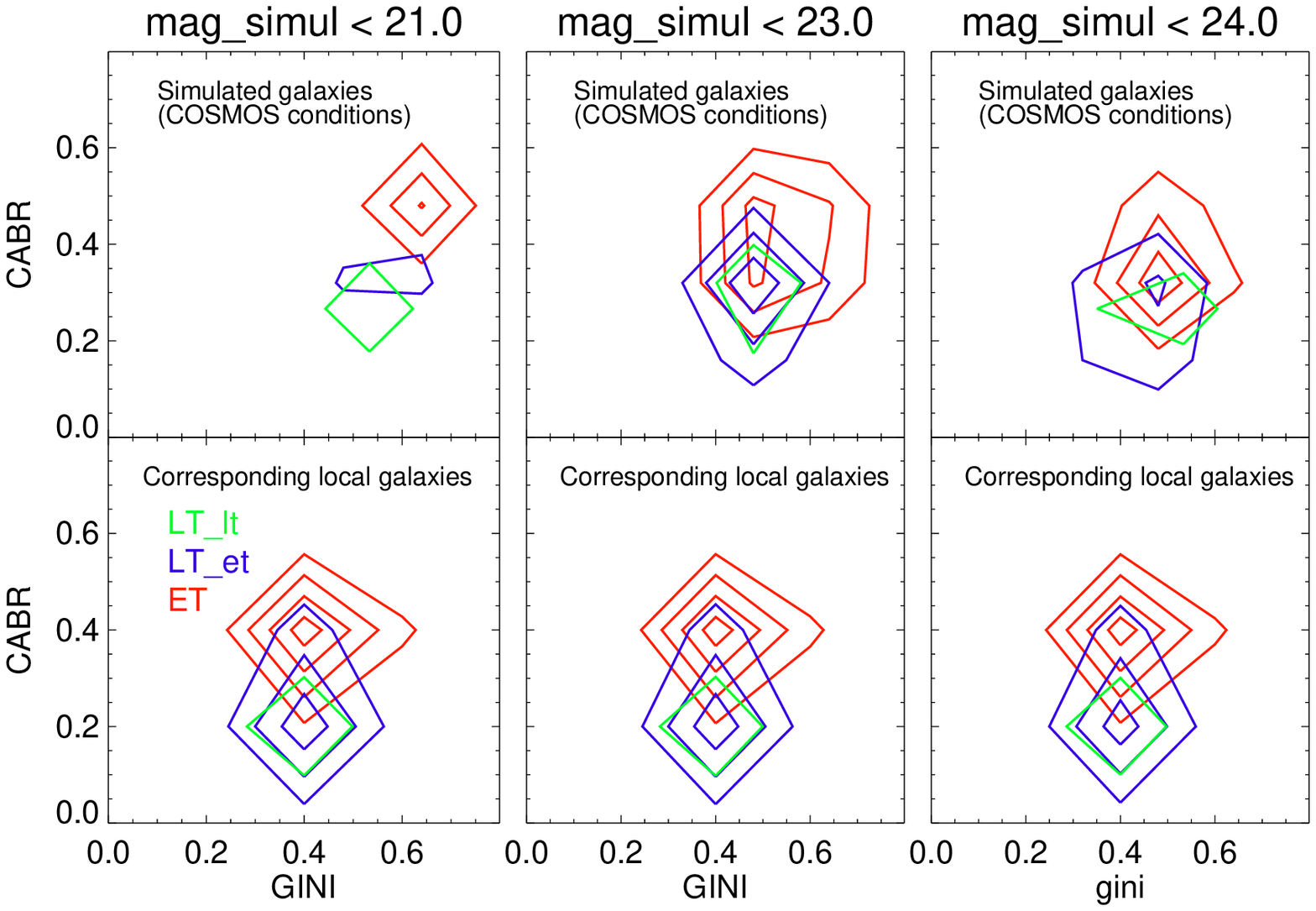}
\end{minipage}
\caption[ ]{Same as Fig.~\ref{fig_morphdiag_ccon_cabr}, but representing the relation between the Gini coefficient and Abraham concentration index. \\\\(A colour version of this figure is available in the online journal)}
\label{fig_morphdiag_gini_cabr}
\end{figure}

\begin{figure}
\centering
\begin{minipage}[c]{0.49\textwidth}
\includegraphics[width=8.1cm,angle=0]{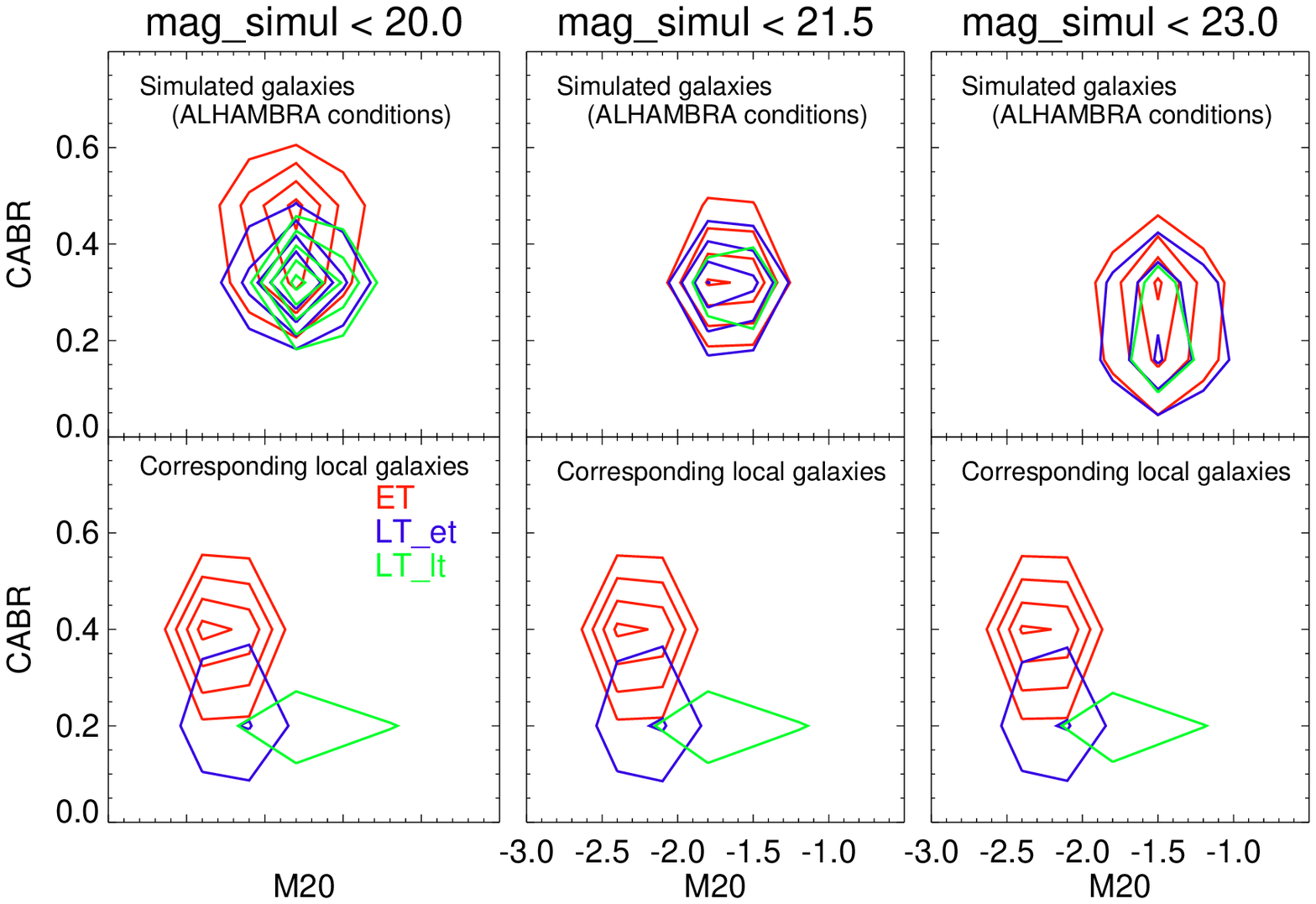}
\end{minipage}
\begin{minipage}[c]{0.49\textwidth}
\includegraphics[width=8.1cm,angle=0]{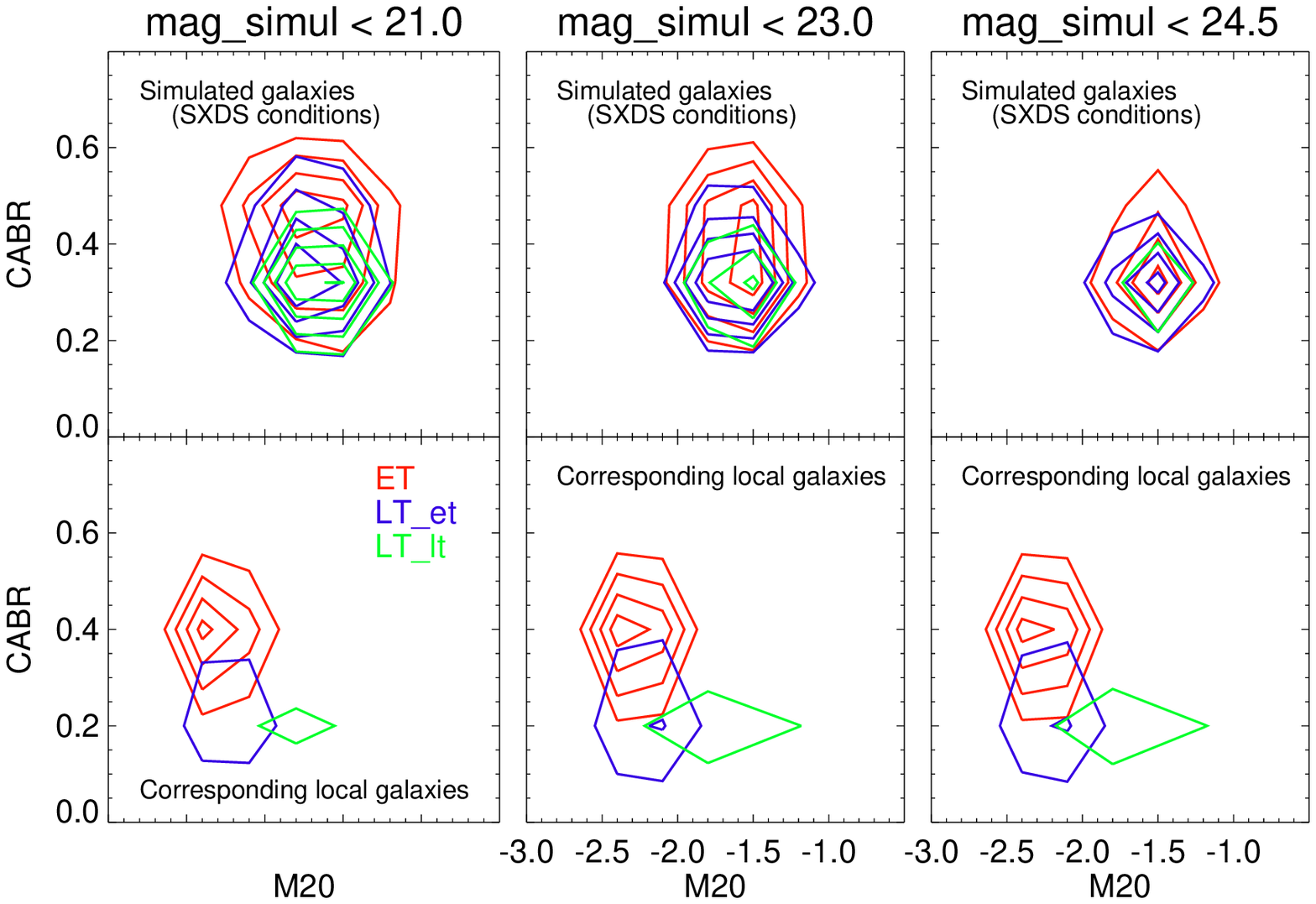}
\end{minipage}
\begin{minipage}[c]{0.49\textwidth}
\includegraphics[width=8.1cm,angle=0]{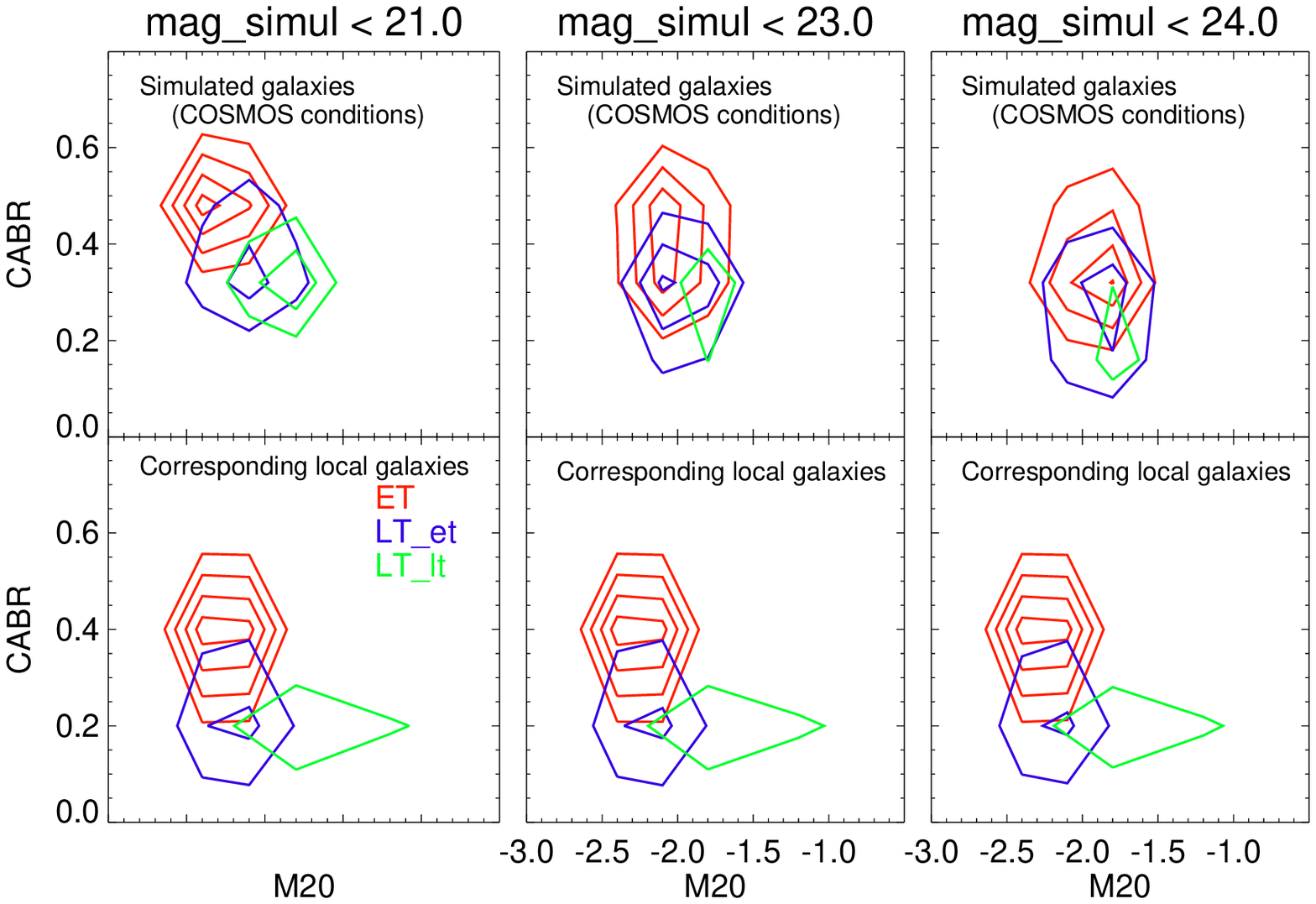}
\end{minipage}
\caption[ ]{Same as Fig.~\ref{fig_morphdiag_ccon_cabr}, but representing the relation between the M20 moment of light and Abraham concentration index. \\\\(A colour version of this figure is available in the online journal)}
\label{fig_morphdiag_m20_cabr}
\end{figure}

\begin{figure}
\centering
\begin{minipage}[c]{0.49\textwidth}
\includegraphics[width=8.1cm,angle=0]{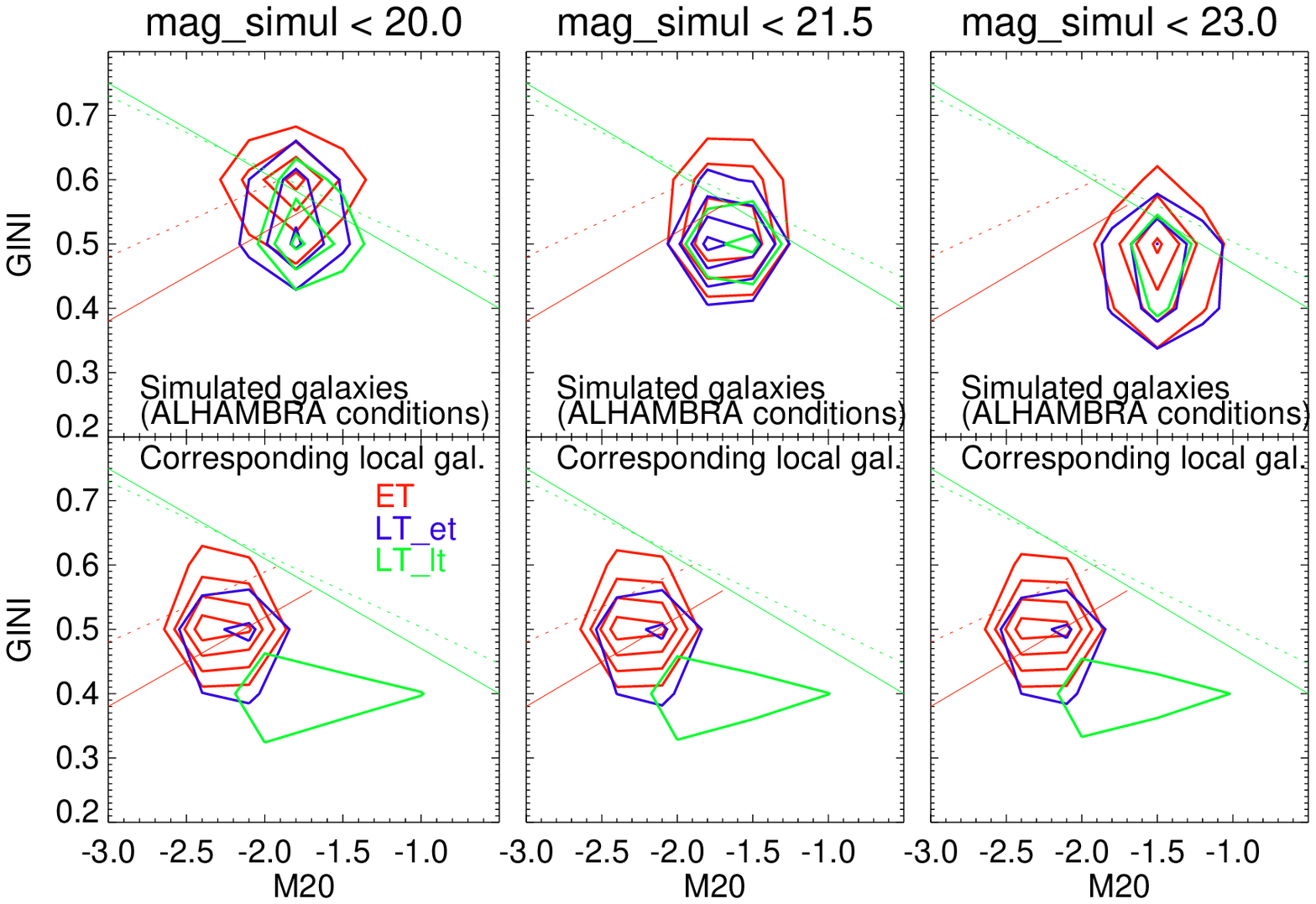}
\end{minipage}
\begin{minipage}[c]{0.49\textwidth}
\includegraphics[width=8.1cm,angle=0]{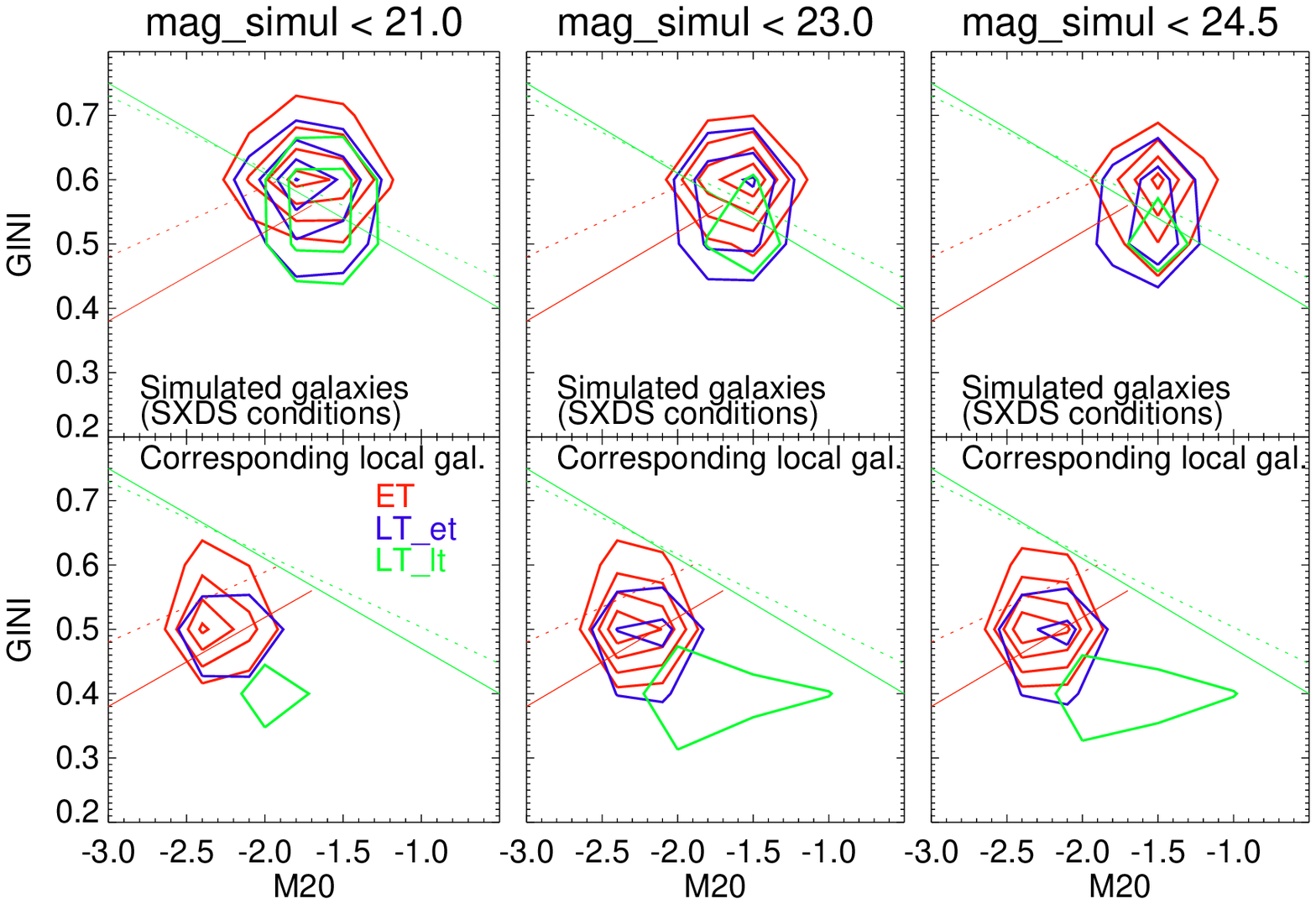}
\end{minipage}
\begin{minipage}[c]{0.49\textwidth}
\includegraphics[width=8.1cm,angle=0]{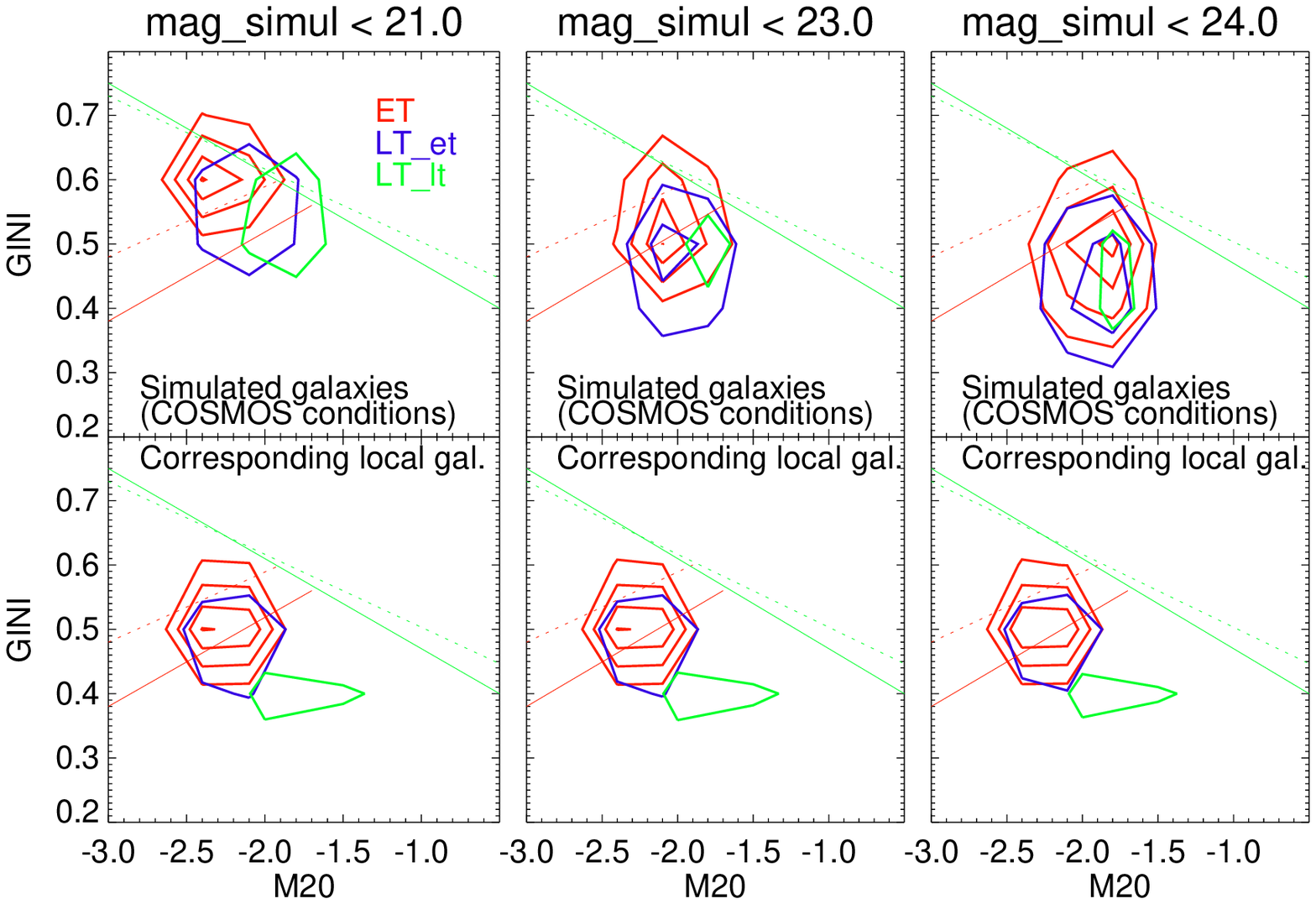}
\end{minipage}
\caption[ ]{Same as Fig.~\ref{fig_morphdiag_ccon_cabr}, but representing the relation between the M20 moment of light and Gini coefficient. The green and red lines represent the limits of \cite{lotz08} to distinguish between the normal galaxies and mergers, and between 
ET and LT galaxies, respectively. In their classifications, mergers occupy the regions above the green lines, ETs (LTs) below the green lines and above (below) the red ones. Dotted lines (both green and red) provided this separation for a sample of local galaxies, while solid lines were used to classify sources at 0.2\,$<$\,z\,$<$\,0.4. \\(A colour version of this figure is available in the online journal)}
\label{fig_morphdiag_m20_gini}
\end{figure}

\begin{figure}
\centering
\begin{minipage}[c]{0.49\textwidth}
\includegraphics[width=7.4cm,angle=0]{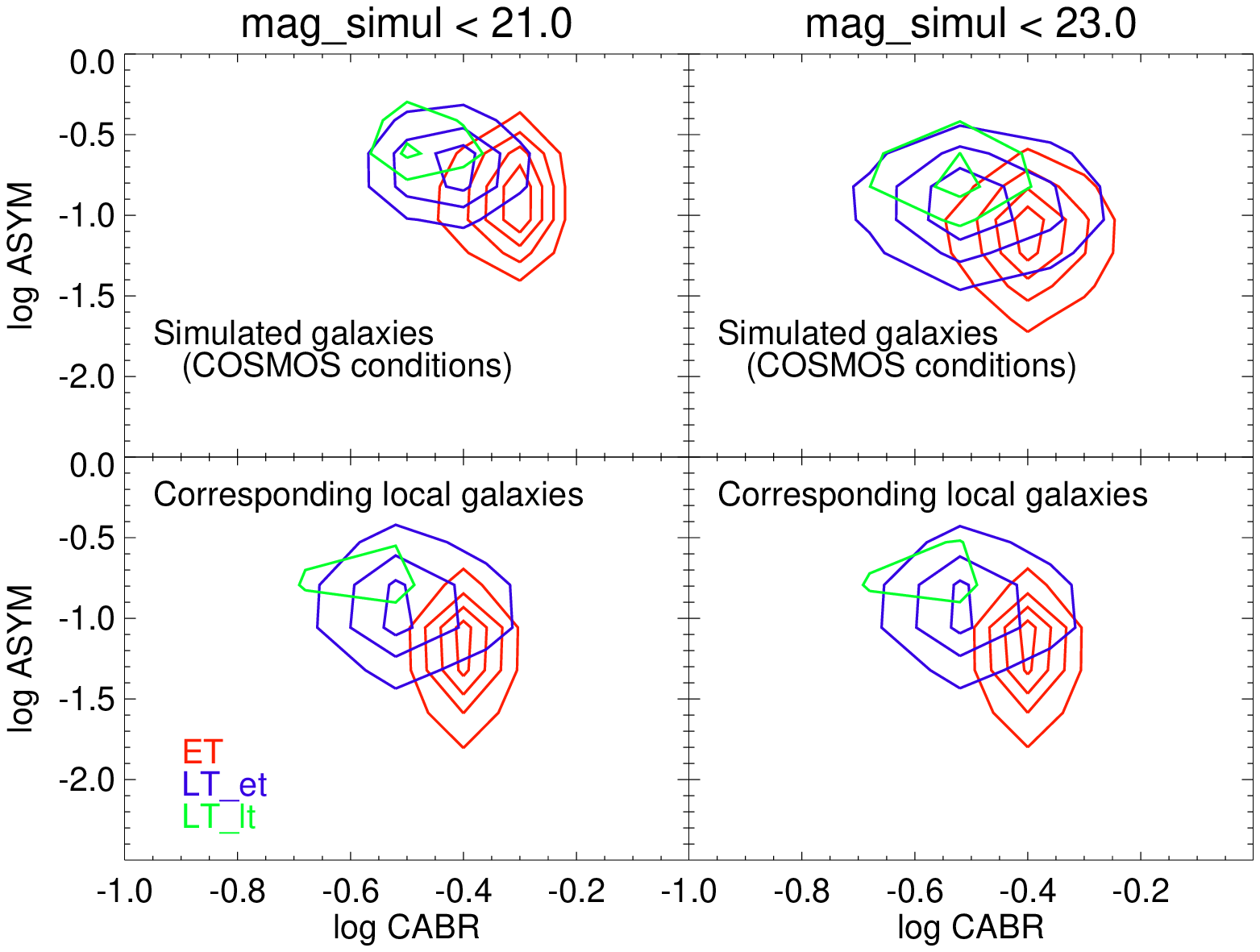}
\end{minipage}
\begin{minipage}[c]{0.49\textwidth}
\includegraphics[width=7.4cm,angle=0]{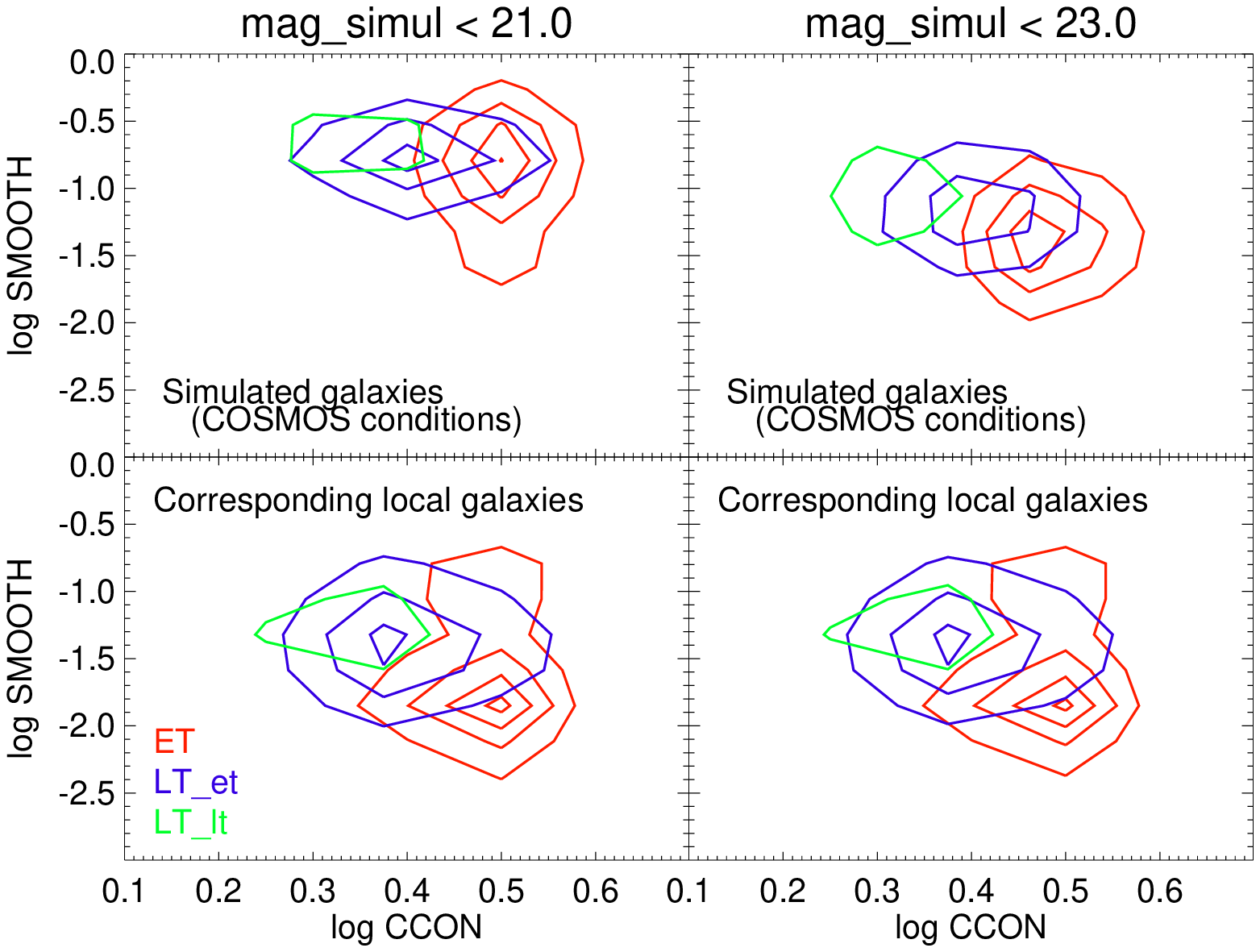}
\end{minipage}
\begin{minipage}[c]{0.49\textwidth}
\includegraphics[width=7.4cm,angle=0]{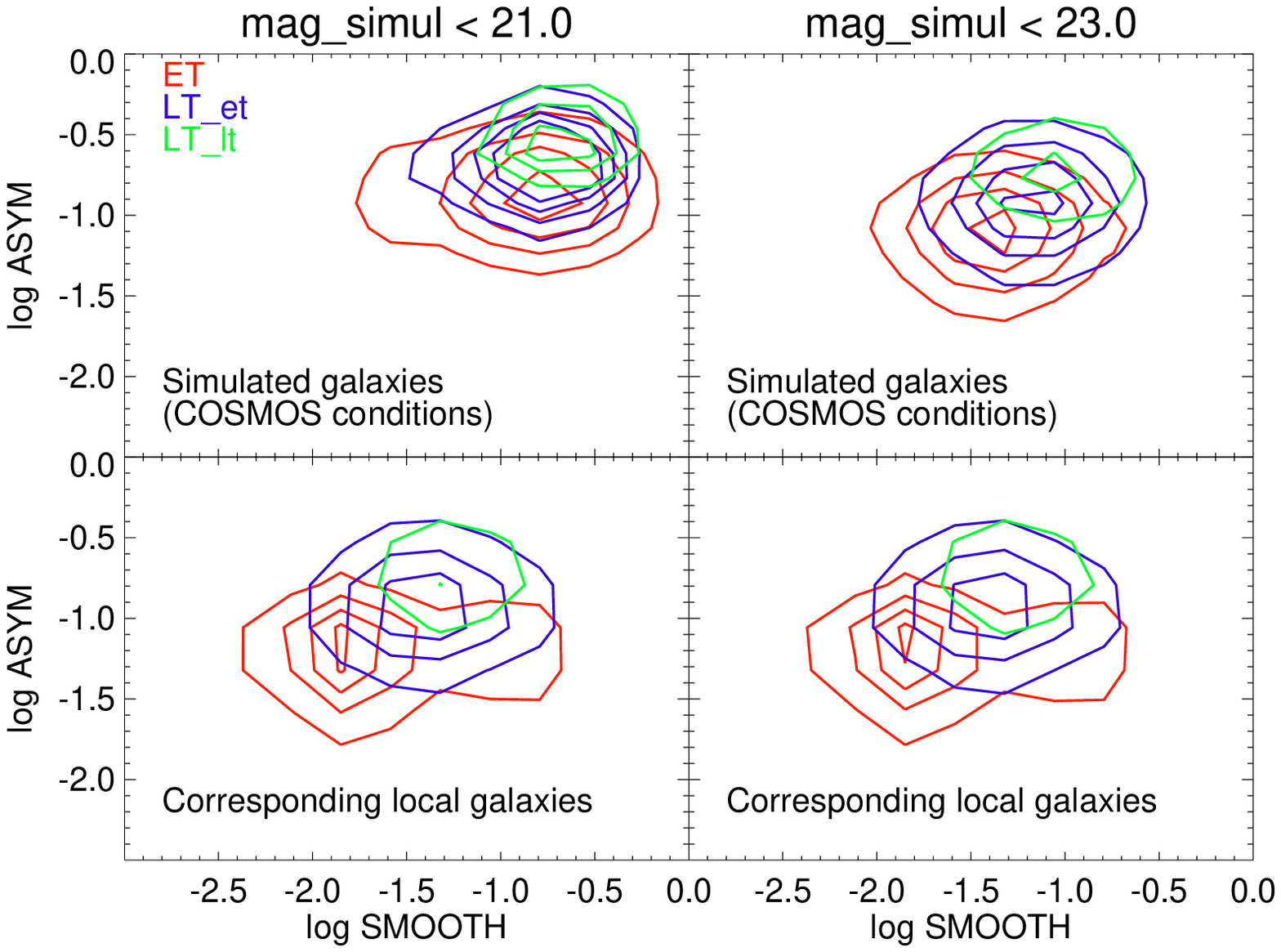}
\end{minipage}
\begin{minipage}[c]{0.49\textwidth}
\includegraphics[width=7.4cm,angle=0]{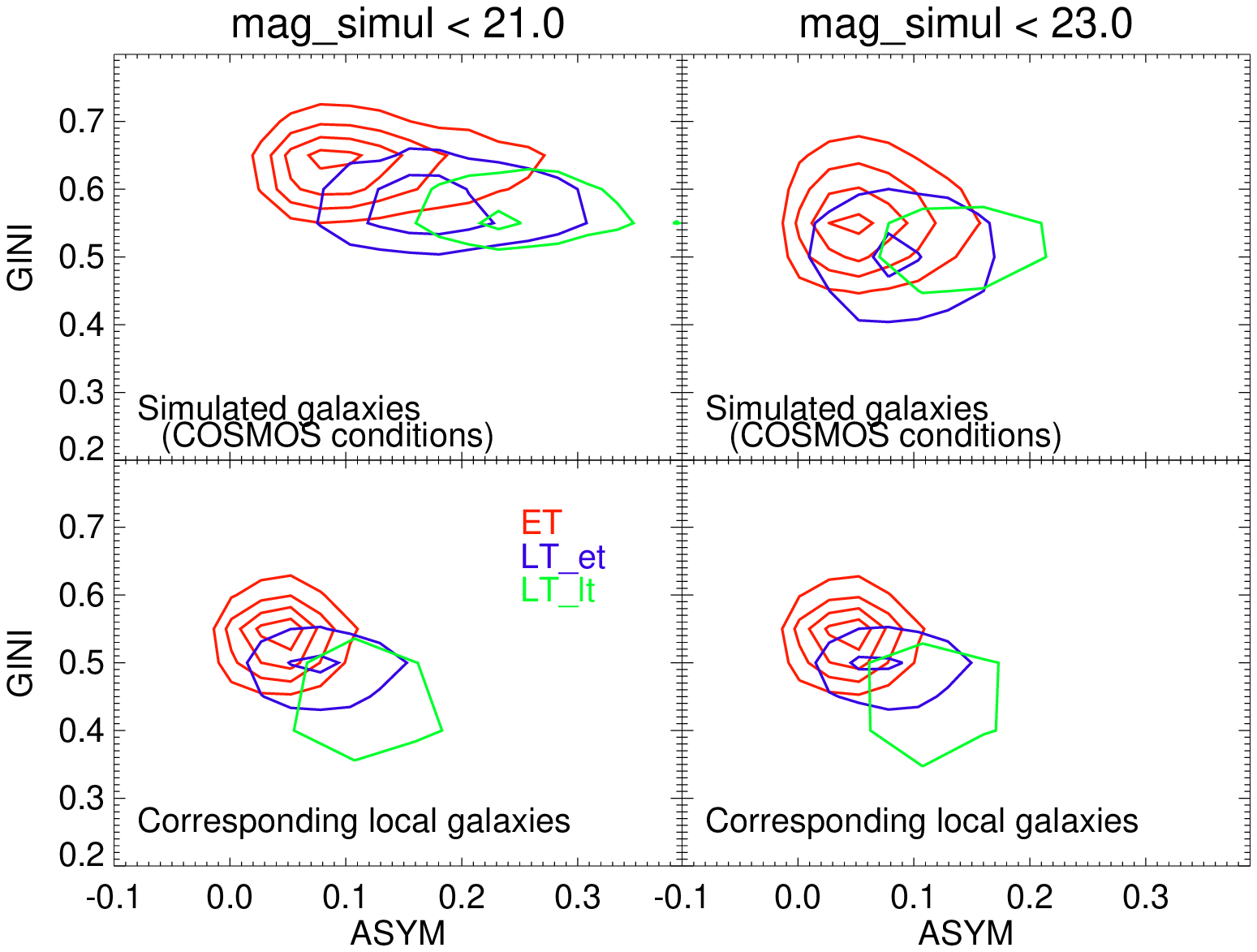}
\end{minipage}
\caption[ ]{Relation between the Abraham concentration index and asymmetry \textit{(top)}, Conselice-Bershady concentration index and smoothness \textit{(middle, top)}, asymmetry and smoothness \textit{(middle, bottom)}, and asymmetry and Gini coefficient \textit{(bottom)} in COSMOS, in the two analysed magnitude cuts, where the noise effect on ASYM and SMOOTH parameters is still insignificant (see Sec.~\ref{sec_method_highz}). For more explications see the caption of Fig.~\ref{fig_morphdiag_ccon_cabr}. (A colour version of this figure is available in the online journal)}
\label{fig_morphdiag_cosmos_3plots_twomagbins}
\end{figure}

\indent In this section we analyse how the observational bias, discussed in the previous section, affects some of the commonly used diagnostic diagrams, that allow to separate between the ET and LT galaxies, to select disturbed galaxies, and/or interacting (merging) systems. To do this, we analysed the simulated sample in the conditions of the three non-local surveys at different magnitude cuts, and compared to the corresponding local sample. We can directly use the obtained diagrams to quantify how strong is the bias for each of these diagrams when taking into account the survey observational properties: basically the spatial resolution (comparing the results between the ground-based and space-based surveys) and survey depth (observing the changes at different magnitude cuts). \\
\indent As already mentioned in Sec.~\ref{sec_method_highz}, we analysed diagrams based on CABR, GINI, CCON, and M20 in all surveys, since these parameters are more stable to the noise effects. Figures \ref{fig_morphdiag_ccon_cabr}, ~\ref{fig_morphdiag_ccon_gini}, ~\ref{fig_morphdiag_ccon_m20}, ~\ref{fig_morphdiag_gini_cabr}, ~\ref{fig_morphdiag_m20_cabr}, and ~\ref{fig_morphdiag_m20_gini} show the relation between CCON and CABR, CCON and GINI, CCON and M20, GINI and CABR, M20 and CABR, and finally M20 and GINI, respectively. In each figure are shown the results obtained for the conditions in the three surveys: ALHAMBRA (top), SXDS (middle), and COSMOS (bottom). For each survey we plot the locus of the three analysed morphological groups, ET (red contours), LT\_et (blue contours), and LT\_lt (green contours), obtained once local galaxies (bottom plots) were moved to higher-redshift conditions (top plots). Although initially we simulated the same sample of 3000 local galaxies to the conditions of the three non-local surveys, in some plots of the reference (local) values small differences can be observed, which is the reason why we represent them in each survey and for each magnitude cut. To understand these differences it has to be considered that, first, the local measurements correspond to the rest-frame band of the simulated sample, and hence slightly changes depending on the magnitude cut and survey (see Sec.~\ref{sec_results_morph_simul}). Secondly, the simulated galaxies with invalid measurements of morphological parameters (discussed in the previous section) are excluded in all plots, and therefore the corresponding local galaxies are also excluded from the bottom diagrams. ASYM and SMOOTH were only used in the case of COSMOS and its first two magnitude cuts. Figure \ref{fig_morphdiag_cosmos_3plots_twomagbins} shows some of the commonly used relations, representing the relation between CABR and ASYM (top plots), CCON and SMOOTH (middle plots), and finally ASYM and GINI (bottom plots). In all three plots, we represent the same comparisons as in Fig.~\ref{fig_morphdiag_ccon_cabr} to \ref{fig_morphdiag_m20_gini}.\\
\indent In all figures, when observing the reference parameters for local galaxies, we can clearly separate the regions typically occupied by ET, LT\_et, and  LT\_lt galaxies, as expected. On the other side, we can observe how the position and the shape of the same regions change once we go to fainter magnitudes (higher redshifts). We measured the level of contamination for the highest density population (50\%) of each analysed morphological group (ET, LT\_et, and LT\_lt) with the other two types, for the morphological diagrams presented in Fig.~\ref{fig_morphdiag_ccon_cabr} to \ref{fig_morphdiag_m20_gini}. We applied the Tukey's Five Number Summary statistic \citep{tukey77}, and measured for each parameter, and each morphological type: the sample minimum, first quartile, median value, third quartile, and sample maximum. The first and third quartiles, present the median values of the lower half and the upper half of the sample, respectively, with respect to the median value of the total sample. Therefore, 
measuring the first and the third quartile of each parameter on the particular diagnostic diagram, we are able to determine the regions that correspond to the highest population (50\% of sources) for ET, LT\_et, and LT\_lt around their median values. From this, for each morphological type, we can estimate the level of contamination by the other two types, within their densest regions: e.g., contamination of the densest ET region with the densest region of the LT\_et galaxies (as a number of LT\_et galaxies in the overlapping area, normalized with the number of ET galaxies), and/or with the densest region of the LT\_lt galaxies, and vice-versa. Table~\ref{tab_cont_alh_sxds_cosmos} shows these measurements for simulated galaxies in ALHAMBRA (top table), SXDS (middle table), and COSMOS (bottom table), for their corresponding magnitude cuts. These measurements represent a quantification of the bias represented in the diagrams \ref{fig_morphdiag_ccon_cabr} to \ref{fig_morphdiag_m20_gini} in all the three surveys, and in Fig.~\ref{fig_morphdiag_cosmos_3plots_twomagbins} in the case of COSMOS. We can again observe that the contamination levels in the morphological types increase with the data depth in all three surveys, being however significantly lower in the case of COSMOS.  \\

\section[]{Discussion}
\label{sec_discussion}

\subsection{Individual parameters}
\label{sec_discussion_indivparam}

\indent As seen in Sec.~\ref{sec_results_morph_simul}, the morphological parameters show complex behaviors when going from one survey to another, from one magnitude cut to another, and when dealing with different morphological types. In principle, they suffer changes due to the impact of spatial resolution and survey depth, and depending on how strong these two effects are, they might cause:\\
\indent - only a re-distribution of the light within the galaxy, without changing much the total flux, resulting in a weak impact, and/or\\ 
\indent - the loss of the galaxy low surface brightness structures and the change of the total flux, hence producing a more significant impact. \\
\indent However, how the distributions of morphological parameters behave in each survey when going from brighter to fainter magnitude cuts (from lower to higher redshifts), depends on their definitions. In general, for weak impacts of spatial resolution and data depth, the galaxies might appear as more concentrated, asymmetric and clumpy, while for stronger impacts, when the galaxy total flux can be affected significantly, the galaxies might appear as less concentrated, more symmetric, and less clumpy. To trace the real impact from spatial resolution and depth on the parameters in Fig.~\ref{fig_alh_param_dist}, \ref{fig_sxds_param_dist}, and \ref{fig_cosmos_param_dist}, it is not only important to take into account the difference between the simulated and real values (and its standard deviation), but also the sign of that difference. In general, for the concentration parameters (CABR, GINI, CCON), the distributions will be the most affected when the final differences have positive values (showing that the simulated galaxies are less concentrated in comparison with the real ones); negative in case of M20, ASYM, and SMOOTH due to the use of the logarithm scale in our analysis.

\subsection{Diagnostic Diagrams: impact from spatial resolution and data depth}
\label{sec_discussion_diagrams}

\indent To evaluate the separate effect of resolution and data depth, we compare two cases: first, the results obtained for the two ground-based surveys, where both of them are affected with the spatial resolution, and we can study how important is the survey depth, and second, the results obtained for space-based data in comparison to the ground-based surveys, where we can see the role that play the spatial resolution in morphological classification.  

\subsubsection{The effect of survey depth in deep ground-based surveys: ALHAMBRA and SXDS}
\label{sec_discussion_morphdiagram_Icase}

\indent \indent From the morphological diagrams and from the Table~\ref{tab_cont_alh_sxds_cosmos}, for the deep ground-based surveys, like ALHAMBRA and SXDS, diagrams combined with M20 show to be less effective in separating galaxies. On the other side, as seen in all figures, survey depth is crucial when classifying galaxies. In the case of ALHAMBRA, at its lowest magnitude cut (F613W\,$\le$\,20.0), although in all diagrams the position/shape of locus of the three morphological types change significantly in comparison with the same sample at z\,$\sim$\,0, in most of diagrams we are still able to distinguish between the ET and LT galaxies, or at least the regions of the highest source densities do not overlap. Observing for each type their contours enclosing 50\% of galaxies, we found a $\sim$\,5\,-\,25\% contamination of ET with LT\_et sources, and a $\sim$\,3\,-\,35\%, of LT\_et with ET sources, where the contamination level depends on the used diagram, with GINI vs. CABR and CCON vs. CABR diagrams showing the lowest contamination ($<$\,5\%), and CCON vs. M20, and M20 vs. GINI the highest one. Similar conclusions are obtained for classifying LTs, but with higher contamination by ET sources. When including fainter galaxies (second magnitude cut F613W\,$\le$\,21.5), the mixing between the three regions increases significantly, but still for diagrams that do not use M20 parameter the contamination levels of ET and LT\_et are between 20\,-\,30\%, depending on the diagram. In the case of the last magnitude cut (F613W\,$\le$\,23.0) the mixing is much higher, $\sim$\,60\,-\,100\%, making each of the diagrams useless if used separately. On the other side, in the case of SXDS we observe a similar situation at the different magnitude cuts as in the case of ALHAMBRA, but when including fainter galaxies at higher redshifts. 
The contamination levels obtained at the lowest magnitude cut ($i'$\,$\le$\,21.0) in SXDS (taking into account the central wavelength of the $i'$ band, of $\sim$\,760\,nm, this would corresponds to even higher magnitudes in ALHAMBRA), are higher than those obtained at the lowest analysed magnitude cut in ALHAMBRA (F613W\,$\le$\,20.0), but lower than the second magnitude cut (F613W\,$\le$\,21.5). Moreover, at the highest magnitude cut in the SXDS ($i'$\,$\le$\,24.5), we obtained $\sim$\,10\,-\,80\% lower contaminations, depending on the type, than in the case of the last magnitude cut in ALHAMBRA (F613W\,$\le$\,23.0). The combination of other parameters with M20 (Fig.~\ref{fig_morphdiag_ccon_m20}, ~\ref{fig_morphdiag_m20_cabr}, and ~\ref{fig_morphdiag_m20_gini}), which seems to be the most affected with the resolution, again shows to be the most ineffective. Finally, we should notice that either the effectiveness in separating between ET and LT\_et changes from one diagram to another and from one magnitude cut to another, LT\_lt galaxies in all cases, in both surveys, suffer very high levels of contamination ($>$\,80\% in most of cases) from other two morphological types.

\subsubsection{The effect of resolution. COSMOS as a reference}
\label{sec_discussion_morphdiagram_IIcase}

\indent \indent Since the spatial resolution affects little the measurements of the morphological parameters in the case of COSMOS data, in comparison with the results obtained in ALHAMBRA and SXDS surveys, we can evaluate how much impact does it have when classifying galaxies. In all diagnostic diagrams (Fig.~\ref{fig_morphdiag_ccon_cabr} to \ref{fig_morphdiag_m20_gini}) the loci of galaxies simulated to the conditions of COSMOS, irrespective of their morphological type, change significantly less at all magnitude cuts than in case of ALH and SXDS. Moreover, the diagrams that include ASYM and SMOOTH parameters provide the additional information, and can be used up to F814W\,=\,24.0 magnitudes. However, the data are still affected by the survey depth, as can be seen in all figures when comparing the diagrams at different magnitude cuts. When including only the brightest simulated galaxies in the sample (F814W\,$\le$\,21.0) the locus of the three morphological groups basically does not change in comparison with that of the corresponding local values, and the contamination levels between ET and LT galaxies are $\le$\,10\%. This picture starts to change when including fainter galaxies, down to F814W\,$\le$\,23.0, and even more at F814W\,$\le$\,24.0. Depending on the diagram the contamination rate changes between 2\,-\,25\% at F814\,$\le$\,23.0, and 6\,-\,35\% for ET and 20\,-\,70\% for LT galaxies at F814\,$\le$\,24.0. However, even at the highest analysed magnitude cut, which corresponds to even fainter magnitudes in ALHAMBRA and SXDS selected bands, we still don't observe the complete overlapping of the ET and LT regions like in the case of ground-based surveys, and the contamination levels are still below those obtained for the brightest ALHAMBRA/SXDS magnitudes. \\

\indent Although the impact from survey depth on the morphological parameters measured within COSMOS conditions is still present, it is significantly lower than in the case of deep ground-based surveys like ALHAMBRA and SXDS. We can assume that the spatial resolution is the main limitation that should be taken into account when dealing with the photometric data and morphological classification of galaxies. \\

\begin{table*}
\begin{center}
\caption{Expected level of contamination for densest regions (50\% of sources) of ET, LT\_et, and LT\_lt galaxies in some of the most used morphological diagrams in the ALHAMBRA (top), SXDS (middle), and COSMOS (bottom) surveys. 
\label{tab_cont_alh_sxds_cosmos}}
\begin{tabular}{| c | c | c | c | c | c | c | c | c | c | c |}
\hline
&&\multicolumn{3}{|c|}{\textbf{mag1\_cut\,$\le$\,20.0}}&\multicolumn{3}{|c|}{\textbf{mag2\_cut\,$\le$\,21.5}}&\multicolumn{3}{|c|}{\textbf{mag3\_cut\,$\le$\,23.0}}\\
\cline{3-11}
&\textbf{Diagram}&\textbf{cont. ET}&\textbf{cont. LT\_et}&\textbf{cont. LT\_lt}&\textbf{cont. ET}&\textbf{cont. LT\_et}&\textbf{cont. LT\_lt}&\textbf{cont. ET}&\textbf{cont. LT\_et}&\textbf{cont. LT\_lt}\\
&&\textbf{[\%]}&\textbf{[\%]}&\textbf{[\%]}&\textbf{[\%]}&\textbf{[\%]}&\textbf{[\%]}&\textbf{[\%]}&\textbf{[\%]}&\textbf{[\%]}\\
\cline{2-11}
&CCON&$-$  (LT\_et)&$4$ (ET)&$-$ (ET)&24 (LT\_et)&31 (ET)&2 (ET)&58 (LT\_et)&71 (ET)&92 (ET)\\
\cline{3-11}
A&vs. CABR&$3$  (LT\_lt)&11 (LT\_lt)&42 (LT\_et) &$-$ (LT\_lt)&20 (LT\_lt)&77 (LT\_et)&20 (LT\_lt)&35 (LT\_lt)&100 (LT\_et)\\
\cline{2-11}
L&CCON&11 (LT\_et)&13 (ET)&$-$ (ET)&31 (LT\_et)&38 (ET)&6 (ET)&56 (LT\_et)&78 (ET)&100 (ET)\\
\cline{3-11}
H&vs. GINI&$-$ (LT\_lt)&16 (LT\_lt)&58 (LT\_et) &1 (LT\_lt)&24 (LT\_lt)&86 (LT\_et)&22 (LT\_lt)&36 (LT\_lt)&100 (LT\_et)\\
\cline{2-11}
A&CCON&$-$ (LT\_et)&46 (ET)&$-$ (ET)&50 (LT\_et)&75 (ET)&13 (ET)&65 (LT\_et)&93 (ET)&100 (ET)\\
\cline{3-11}
M&vs. M20&34 (LT\_lt)&12 (LT\_lt)&40 (LT\_et) &2 (LT\_lt)&19 (LT\_lt)&88 (LT\_et)&21 (LT\_lt)&34 (LT\_lt)&100 (LT\_et)\\
\cline{2-11}
B&GINI&3 (LT\_et)&3 (ET)&16 (ET)&29 (LT\_et)&31 (ET)&49 (ET)&58 (LT\_et)&77 (ET)&100 (ET)\\
\cline{3-11}
R&vs. CABR&$-$ (LT\_lt)&22 (LT\_lt)&100 (LT\_et) &3 (LT\_lt)&36 (LT\_lt)&100 (LT\_et)&28 (LT\_lt)&42 (LT\_lt)&100 (LT\_et)\\
\cline{2-11}
A&M20&6 (LT\_et)&8 (ET)&$-$ (ET)&37 (LT\_et)&48 (ET)&23 (ET)&62 (LT\_et)&80 (ET)&100 (ET)\\
\cline{3-11}
&vs. CABR&$-$ (LT\_lt)&20 (LT\_lt)&64 (LT\_et) &5 (LT\_lt)&33 (LT\_lt)&100 (LT\_et)&27 (LT\_lt)&41 (LT\_lt)&100 (LT\_et)\\
\cline{2-11}
&M20&25 (LT\_et)&36 (ET)&$-$ (ET)&54 (LT\_et)&65 (ET)&85 (ET)&60 (LT\_et)&92 (ET)&100 (ET)\\
\cline{3-11}
&vs. GINI&2 (LT\_lt)&32 (LT\_lt)&100 (LT\_et) &21 (LT\_lt)&40 (LT\_lt)&100 (LT\_et)&30 (LT\_lt)&42 (LT\_lt)&100 (LT\_et)\\
\hline
\noalign{\vskip 2mm}    
\hline
&&\multicolumn{3}{|c|}{\textbf{mag1\_cut\,$\le$\,21.0}}&\multicolumn{3}{|c|}{\textbf{mag2\_cut\,$\le$\,23.0}}&\multicolumn{3}{|c|}{\textbf{mag3\_cut\,$\le$\,24.5}}\\
\cline{3-11}
&\textbf{Diagram}&\textbf{cont. ET}&\textbf{cont. LT\_et}&\textbf{cont. LT\_lt}&\textbf{cont. ET}&\textbf{cont. LT\_et}&\textbf{cont. LT\_lt}&\textbf{cont. ET}&\textbf{cont. LT\_et}&\textbf{cont. LT\_lt}\\
&&\textbf{[\%]}&\textbf{[\%]}&\textbf{[\%]}&\textbf{[\%]}&\textbf{[\%]}&\textbf{[\%]}&\textbf{[\%]}&\textbf{[\%]}&\textbf{[\%]}\\
\cline{2-11}
&CCON&31 (LT\_et)&29 (ET)&$-$ (ET)&45 (LT\_et)&41 (ET)&11 (ET)&51 (LT\_et)&42 (ET)&14 (ET)\\
\cline{3-11}
&vs. CABR&$-$ (LT\_lt)&15 (LT\_lt)&81 (LT\_et) &4 (LT\_lt)&19 (LT\_lt)&86 (LT\_et)&18 (LT\_lt)&22 (LT\_lt)&98 (LT\_et)\\
\cline{2-11}
&CCON&29 (LT\_et)&40 (ET)&2 (ET)&34 (LT\_et)&41 (ET)&8 (ET)&48 (LT\_et)&55 (ET)&28 (ET)\\
\cline{3-11}
S&vs. GINI&0.5 (LT\_lt)&15 (LT\_lt)&83 (LT\_et)&2 (LT\_lt)&22 (LT\_lt)&93 (LT\_et)&6 (LT\_lt)&26 (LT\_lt)&94 (LT\_et)\\
\cline{2-11}
X&CCON&52 (LT\_et)&69 (ET)&27 (ET)&56 (LT\_et)&77 (ET)&64 (ET)&71 (LT\_et)&77 (ET)&73 (ET)\\
\cline{3-11}
D&vs. M20&4 (LT\_lt)&20 (LT\_lt)&100 (LT\_et) &11 (LT\_lt)&23 (LT\_lt)&100 (LT\_et)&13 (LT\_lt)&26 (LT\_lt)&100 (LT\_et)\\
\cline{2-11}
S&GINI&24 (LT\_et)&36 (ET)&$-$ (ET)&30 (LT\_et)&33 (ET)&6 (ET)&47 (LT\_et)&42 (ET)&13 (ET)\\
\cline{3-11}
&vs. CABR&$-$ (LT\_lt)&18 (LT\_lt)&88 (LT\_et) &2 (LT\_lt)&21 (LT\_lt)&94 (LT\_et)&11 (LT\_lt)&25 (LT\_lt)&100 (LT\_et)\\
\cline{2-11}
&M20&38 (LT\_et)&46 (ET)&$-$ (ET)&48 (LT\_et)&56 (ET)&34 (ET)&56 (LT\_et)&49 (ET)&28 (ET)\\
\cline{3-11}
&vs. CABR&$-$ (LT\_lt)&18 (LT\_lt)&98 (LT\_et)&10 (LT\_lt)&25 (LT\_lt)&100 (LT\_et)&47 (LT\_lt)&24 (LT\_lt)&90 (LT\_et)\\
\cline{2-11}
&M20&35 (LT\_et)&55 (ET)&10 (ET)&38 (LT\_et)&63 (ET)&31 (ET)&60 (LT\_et)&67 (ET)&53 (ET)\\
\cline{3-11}
&vs. GINI&2 (LT\_lt)&27 (LT\_lt)&100 (LT\_et) &6 (LT\_lt)&24 (LT\_lt)&100 (LT\_et)&10 (LT\_lt)&30 (LT\_lt)&100 (LT\_et)\\
\hline
\noalign{\vskip 2mm}    
\hline
&&\multicolumn{3}{|c|}{\textbf{mag1\_cut\,$\le$\,21.0}}&\multicolumn{3}{|c|}{\textbf{mag2\_cut\,$\le$\,23.0}}&\multicolumn{3}{|c|}{\textbf{mag3\_cut\,$\le$\,24.0}}\\
\cline{3-11}
&\textbf{Diagram}&\textbf{cont. ET}&\textbf{cont. LT\_et}&\textbf{cont. LT\_lt}&\textbf{cont. ET}&\textbf{cont. LT\_et}&\textbf{cont. LT\_lt}&\textbf{cont. ET}&\textbf{cont. LT\_et}&\textbf{cont. LT\_lt}\\
&&\textbf{[\%]}&\textbf{[\%]}&\textbf{[\%]}&\textbf{[\%]}&\textbf{[\%]}&\textbf{[\%]}&\textbf{[\%]}&\textbf{[\%]}&\textbf{[\%]}\\
\cline{2-11}
&CCON&$-$  (LT\_et)&$-$ (ET)&$-$ (ET)&$-$ (LT\_et)&2 (ET)&$-$ (ET)&6 (LT\_et)&17 (ET)&$-$ (ET)\\
\cline{3-11}
&vs. CABR&$-$  (LT\_lt)&7 (LT\_lt)&36 (LT\_et) &$-$ (LT\_lt)&9 (LT\_lt)&31 (LT\_et)&$-$ (LT\_lt)&15 (LT\_lt)&62 (LT\_et)\\
\cline{2-11}
&CCON&$-$ (LT\_et)&$-$ (ET)&$-$ (ET)&4 (LT\_et)&11 (ET)&$-$ (ET)&11 (LT\_et)&30 (ET)&$-$ (ET)\\
\cline{3-11}
&vs. GINI&$-$ (LT\_lt)&11 (LT\_lt)&48 (LT\_et) &$-$ (LT\_lt)&12 (LT\_lt)&40 (LT\_et)&$-$ (LT\_lt)&19 (LT\_lt)&75 (LT\_et)\\
\cline{2-11}
C&CCON&$-$ (LT\_et)&$-$ (ET)&$-$ (ET)&7 (LT\_et)&15 (ET)&$-$ (ET)&22 (LT\_et)&38 (ET)&$-$ (ET)\\
\cline{3-11}
O&vs. M20&$-$ (LT\_lt)&3 (LT\_lt)&17 (LT\_et) &$-$ (LT\_lt)&6 (LT\_lt)&20 (LT\_et)&$-$ (LT\_lt)&12 (LT\_lt)&47 (LT\_et)\\
\cline{2-11}
S&GINI&$-$ (LT\_et)&$-$ (ET)&0 (ET)&2 (LT\_et)&4 (ET)&$-$ (ET)&13 (LT\_et)&30 (ET)&$-$ (ET)\\
\cline{3-11}
M&vs. CABR&$-$ (LT\_lt)&28 (LT\_lt)&100 (LT\_et) &$-$ (LT\_lt)&35 (LT\_lt)&100 (LT\_et)&$-$ (LT\_lt)&34 (LT\_lt)&100 (LT\_et)\\
\cline{2-11}
O&M20&$-$ (LT\_et)&$-$ (ET)&$-$ (ET)&2 (LT\_et)&5 (ET)&$-$ (ET)&15 (LT\_et)&32 (ET)&$-$ (ET)\\
\cline{3-11}
S&vs. CABR&$-$ (LT\_lt)&8 (LT\_lt)&30 (LT\_et) &$-$ (LT\_lt)&12 (LT\_lt)&47 (LT\_et)&$-$ (LT\_lt)&12 (LT\_lt)&77 (LT\_et)\\
\cline{2-11}
&M20&5 (LT\_et)&7 (ET)&$-$ (ET)&24 (LT\_et)&55 (ET)&$-$ (ET)&35 (LT\_et)&75 (ET)&$-$ (ET)\\
\cline{3-11}
&vs. GINI&$-$ (LT\_lt)&10.5 (LT\_lt)&43 (LT\_et) &$-$ (LT\_lt)&17 (LT\_lt)&63 (LT\_et)&7 (LT\_lt)&26 (LT\_lt)&100 (LT\_et)\\
\cline{2-11}
&logCABR&$-$  (LT\_et)&$-$ (ET)&$-$ (ET)&1.3 (LT\_et)&3 (ET)&$-$ (ET)&13 (LT\_et)&26 (ET)&$-$ (ET)\\
\cline{3-11}
&vs. logASYM&$-$  (LT\_lt)&13 (LT\_lt)&45 (LT\_et) &$-$ (LT\_lt)&18 (LT\_lt)&55 (LT\_et)&$-$ (LT\_lt)&24 (LT\_lt)&80 (LT\_et)\\
\cline{2-11}
&logCCON&$-$ (LT\_et)&$-$ (ET)&$-$ (ET)&5.5 (LT\_et)&11.3 (ET)&$-$ (ET)&11 (LT\_et)&17 (ET)&$-$ (ET)\\
\cline{3-11}
&vs. logSMOOTH&$-$ (LT\_lt)&6.3 (LT\_lt)&38 (LT\_et) &$-$ (LT\_lt)&10 (LT\_lt)&34 (LT\_et)&$-$ (LT\_lt)&16 (LT\_lt)&60 (LT\_et)\\
\cline{2-11}
&logSMOOTH&56 (LT\_et)&28 (ET)&64 (ET)&35 (LT\_et)&39 (ET)&$-$ (ET)&42 (LT\_et)&47 (ET)&46 (ET)\\
\cline{3-11}
&vs. logASYM&$-$ (LT\_lt)&11 (LT\_lt)&$-$ (LT\_et) &$-$ (LT\_lt)&18 (LT\_lt)&78 (LT\_et)&5 (LT\_lt)&22 (LT\_lt)&100 (LT\_et)\\
\cline{2-11}
&ASYM&7 (LT\_et)&6 (ET)&$-$ (ET)&16 (LT\_et)&31 (ET)&$-$ (ET)&28 (LT\_et)&60 (ET)&40 (ET)\\
\cline{3-11}
&vs. GINI&$-$ (LT\_lt)&20 (LT\_lt)&67 (LT\_et) &$-$ (LT\_lt)&25 (LT\_lt)&84 (LT\_et)&4 (LT\_lt)&30 (LT\_lt)&100 (LT\_et)\\
\hline
\end{tabular}
\end{center}
\begin{flushleft}
{\textbf{* Column description:}  \textbf{Diagram} - commonly used morphological diagrams (for the description of each parameter see Sec.~\ref{sec_method_param}); \textbf{mag\_cut} - for each survey, and in each diagram, the contamination levels are provided at three corresponding magnitude cuts (see Sec.~\ref{sec_method_highz}); \textbf{cont. ET}, \textbf{cont. LT\_et}, and \textbf{cont. LT\_lt} - contamination level of ET, LT\_et, and LT\_lt simulated local galaxies, respectively, with the other two morphological types at the corresponding magnitude cuts.}
\end{flushleft}
\end{table*}

\subsection{Diagnostic diagrams: implications for morphological classification of galaxies}
\label{sec_discussion_diagrams_implications}

\indent For intermediate- and high-redshift galaxy samples, and large data sets, typical for current (and future) deep photometric surveys, the application of automated non-parametric methods, that use the parameters analysed within this work, is one of the ways to obtain the information about the morphology. In some of the latest works, the classification based on non-parametric methods was done using a set of parameters simultaneously, therefore decreasing the impact of spatial resolution and depth \citep[e.g.][]{huertas09,huertas11,povic12,povic13,cibinel13,carollo13,cotini13,holwerda14}. However, for classifications based on only one, two, or three parameters, special attention should be taken in relation to the magnitude and redshift distribution of the analysed sample, and the data quality.\\

\indent The ASYM index was used in some of previous works to select mergers and disturbed galaxies \citep{lotz10a,lotz10b,lotz11,villforth14}. ASYM is especially sensitive to the survey noise, as showed in Sec.~\ref{sec_method_highz}, and to both, the spatial resolution and depth, as showed in Sec.~\ref{sec_results_morph_simul}. In surveys similar to COSMOS, for sources fainter than F814\,=\,23.0, we again do not recommend the use of a single criterion based on ASYM when classifying them morphologically, since the contamination levels of 50\% highest density regions of ET and LT rise to above 40\%. In the case of ground-based surveys, we do not recommend the use of this single criterion in any case (except for maybe the brightest galaxies), since the contamination levels are much higher, taking the noise effect especially into account; we suggest instead the simultaneous use of multiple diagnostic diagrams. \\  

\indent Over the past ten years, the M20\,-\,GINI diagram (Fig.~\ref{fig_morphdiag_m20_gini}) found wide applications. It was mainly used in identifying galaxy mergers and peculiar (disturbed) galaxies \citep[e.g.][]{lotz04,pierce07,lotz08,conselice08,mendez11,chen10,stott13,petty14,hung14}; often it was used in the analysis where different types of galaxies, e.g. ULIRGs, dust obscured galaxies, star-forming galaxies, or quiescent galaxies, were compared with control samples \citep{bussmann09,pentericci10,lanyon12,wang12,chung13,bohm13}; or for selecting ET and LT galaxies \citep{lotz04,lotz08}. \\
\indent From our results, in ALHAMBRA and SXDS surveys even for the brightest analysed sources (F613\,$\le$\,20.0 and $i'$\,$\le$\,21.0, respectively) the contamination levels between the ET and LT galaxies in M20\,-\,GINI diagram are about 30\%. In COSMOS however, at the lowest magnitude cut F814\,$\le$\,21.0 (z\,$\lesssim$\,1) the contamination is insignificant ($\lesssim$\,10\%, except for the youngest spiral/irregular galaxies), but already at the second magnitude cut F814\,$\le$\,23.0 (z\,$\lesssim$\,1.5) the contamination is $>$\,20\%. We represented in Fig.~\ref{fig_morphdiag_m20_gini} the limits established by \cite{lotz04,lotz08} to separate between normal ET and LT galaxies and major mergers, in the local universe (dotted lines), and at 0.2\,$<$\,z\,$<$\,0.4 (solid lines), using the All-wavelength Extended Groth strip International Survey (AEGIS) HST/ACS data. The solid-lines criterion (non-local sample) was later applied by \cite{lotz08,lotz10a,lotz10b} to z\,$\le$\,1.2 sources to select major mergers, reporting a population of $\sim$\,15\% of misclassified sources, and was used in other works, as mentioned above. From the established limits we can see again how sensitive the diagram is to spatial resolution and data depth, changing significantly between the ground- and space-based surveys. While in COSMOS the contamination by normal galaxies in the region assigned for major mergers, does not exceed the values measured by the authors (even at higher redshifts), in the analysed ground-based surveys the contamination can be $>$\,70\%.\\
\indent When applying the M20\,-\,GINI diagram to classify galaxies and select mergers, we highly recommend to use it always with additional criteria when dealing with ground-based data, among which: a combination with other morphological diagrams, a careful selection of the control sample and evaluation of the diagram effectiveness, and/or application of probabilistic approaches. In space-based surveys like COSMOS to distinguish between the ET and LT sources additional criteria should be taken into account at least for galaxies fainter than F814\,=\,23.0. \\

\indent Finally, different combinations of 2\,-\,3 parameters analysed in the previous sections, were used in many studies to distinguish between the ET and LT galaxies, and also to study the structure of galaxies with other properties, like SFR, mass, environment, nuclear activity, etc \citep[e.g.][]{pierce07,cassata07,bussmann09,povic09a,povic09b,pentericci10,chung13,bohm13}. The combination between the concentration parameters (CABR and/or CCON) and ASYM was also used for selecting ET and LT sources \citep{abraham94,abraham96,conselice06,huertas07,povic09a}, while in some cases the combination between the GINI and CCON was used to study the host properties of AGN and quasars \citep[e.g.][]{urrutia08}. The CCON\,-\,ASYM\,-\,SMOOTH diagram was used in classifying both galaxies and mergers \citep{conselice03,cassata07,conselice08}. We found SMOOTH to be especially sensitive to both spatial resolution and depth, and the most unstable parameter of all analysed. Moreover, together with ASYM, we also found it to be more sensitive to noise, as showed in Sec.~\ref{sec_method_highz}. As already mentioned, we do not recommend the use of these two parameters in the ground-based surveys if the noise effects are not taken into account carefully. In surveys similar to COSMOS, they can be used in combination with other parameters but for sources brighter than F814\,=\,24.0. The combination of only ASYM and SMOOTH gives high contamination levels even in the case of COSMOS (as showed in Fig.~\ref{fig_morphdiag_cosmos_3plots_twomagbins} and Table~\ref{tab_cont_alh_sxds_cosmos}). \\
\indent We suggest the consultation of Table~\ref{tab_cont_alh_sxds_cosmos} and Fig.~\ref{fig_morphdiag_ccon_cabr} to \ref{fig_morphdiag_cosmos_3plots_twomagbins} when using diagnostic diagrams in morphological studies, especially if they are based on 2\,-\,3 parameters. We discussed in Sec.~\ref{sec_results_morphdiagram} the effectiveness of a particular diagnostic diagram in the ground-based surveys and noted that in cases similar to both ALHAMBRA and SXDS, special care has to be taken into account when classifying galaxies at all magnitude cuts (redshifts). Our suggestion when using ground-based data is to avoid the diagnostic diagrams based on 2\,-\,3 parameters, and to apply instead all morphological parameters simultaneously, and to use the statistical approaches providing a probability for a galaxy to belong to a given morphological type (e.g., approaches similar to that used in the galSVM code and tested in Huertas-Company et al. 2008). Diagrams that include M20 moment of light showed to be more affected by spatial resolution, and less effective in separating galaxies. Moreover, the noise should be taken into account if using ASYM and SMOOTH parameters.\\
\indent In space-based surveys similar to COSMOS, even at the highest magnitude cuts (F814\,$\le$\,24.0), in all diagnostic diagrams the contamination levels are significantly lower than at the brightest magnitudes in the ground-based surveys. The observational bias is insignificant for all parameters at magnitudes brighter than F814\,=\,21.0 (except for the youngest spiral and irregular galaxies), as shown in Sec~\ref{sec_results_morphdiagram}; however when going to fainter magnitudes the contamination increases. Therefore, even when dealing with data sets similar to COSMOS, we remind that eventual contamination levels are present at fainter magnitudes; the diagnostic diagrams that show to be more effective in this case are CCON\,-\,CABR, CCON\,-\,GINI, CCON\,-\,SMOOTH, and/or CABR\,-\,ASYM. To apply again, when possible, different diagnostics will minimise the effect of the observational bias.\\

\subsection{Results from Linear Discriminant Analysis}
\label{sec_discussion_linear_discrimination}

In order to merge all the analysed morphological parameters, and to provide an additional and statistical test of our previous results, a Linear Discriminant Analysis (LDA; as implemented by Pedregosa et al. 2011) was performed. The LDA method allows us to find the linear combination of measured parameters that best separates two or more classes of objects or events. We tested the combination of the four morphological parameters (CABR, GINI, CCON, and M20) to distinguish between ET, LT\_et, and LT\_lt galaxies in the three analysed non-local surveys (ALHAMBRA, SXDS, and COSMOS). Knowing that ASYM and SMOOTH are found to be affected by noise, we performed the LDA without these two parameters. However we also performed a LDA including all the six parameters for comparison and to check the effect of noise since both ASYM and SMOOTH parameters resulted to be the ones more significantly affected by noise, mainly in the two ground-based surveys, as we described in Sec.~\ref{sec_method_highz}.

\begin{figure}
\centering
	\includegraphics[width=0.49\textwidth,angle=0]{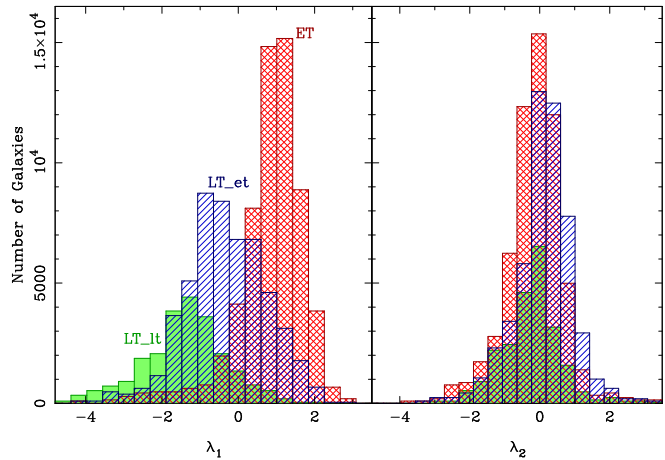}
	\caption{Histograms of the $\lambda_1$ (left panel) and $\lambda_2$ (right panel) eigenvectors resulting from the LDA of the local sample, without including the SMOOTH and ASYM parameters. The red cross-hatched, blue hatched, and  green solid histograms correspond to the ET, LT\_et, and LT\_lt galaxies, respectively. As it can be seen, $\lambda_1$ is the main morphology discriminator, whereas $\lambda_2$ only slightly improves the classification for the LT\_et galaxies.
		\label{his_lda}}
\end{figure}

The training phase of the LDA was obtained by fitting the coefficients over the measured parameters of the galaxies of the local sample. We found that all the possible discrimination is based on two eigenvectors ($\lambda_1$ and $\lambda_2$) both being linear combinations of the four parameters entering in the classification. The first one, $\lambda_1$, is dominated by GINI, CCON, and CABR, whereas the second one, $\lambda_2$, is based on CABR, GINI, and M20. Almost all the power of the discrimination arise from the $\lambda_1$ eigenvector whereas $\lambda_2$ only slightly helps to the classification of LT\_et galaxies, as it can be seen in Fig.~\ref{his_lda} where the distributions of both $\lambda_1$ and $\lambda_2$ for the three morphological types are shown. The three concentration indexes show to be more stable, discriminating better between different morphological types in comparison with M20. An attempt to improve the LDA was also carried out by doing a second LDA iteration including only those galaxies that show the same input and predicted output morphology. We did not find any significant change, and therefore did not proceed further with this additional step. As a result of the LDA of the local sample we found that about 30\% of the objects show a different input morphology than the one predicted by the resulting eigenvectors of the LDA. This fact could be caused by errors in the measured parameters or even in the original classification. To illustrate the results obtained by the LDA we present in Figure \ref{his_lda}, the histogram of the predicted morphologies after applying the eigenvector discrimination over the local sample, where for each output morphological type (ET, LT\_et, LT\_lt) we show the contribution of galaxies with different input morphology. As it clearly can be seen, whereas ET galaxies tend to be well classified, LT\_ET spread between adjacent input types, and LT\_LT are split between both input late-types. 

After obtaining the eigenvectors that better discriminate the morphology in the local sample, LDA transformations to the ($\lambda_1$ and $\lambda_2$) space were performed for each non local sample (ALHAMBRA, SXDX, COSMOS). The results so obtained at the three magnitude cuts and for each non-local sample are presented in Table \ref{tab_lda_4param}. In this table, for each survey, the first column represents the three corresponding magnitude cuts, the second column corresponds to the total number of galaxies (Ngal) at each magnitude cut and sample (i.e. galaxies with valid measured parameters after applying the filtering criteria explained in Sec.~\ref{sec_results_morphdiagram}) while the score column represents the ratio between the galaxies having the same LDA output morphology as the input one normalized to the total number of galaxies Ngal; a score of 1 would indicate a perfect LDA classification. We also tested the results when including ASYM and SMOOTH. In table \ref{tab_lda_all} we present the results obtained from the LDA taking into account all morphological parameters. In general, the scores obtained after removing these two parameters (see Table \ref{tab_lda_4param}) are found to be similar to those that would be obtained including also ASYM and SMOOTH, however the total number of valid galaxies increase significantly and this is more notable in the case of the ground base surveys (ALHAMBRA and SXDS). If ASYM and SMOOTH are excluded, the total number of galaxies with good morphology is found to be larger, thus the final classification get worst when including these two parameters as they are indeed affected by noise. Only in surveys similar to COSMOS, the inclusion of ASYM and SMOOTH would have a slightly positive net effect.

\begin{figure}
	\centering
	\includegraphics[width=0.45\textwidth,angle=0]{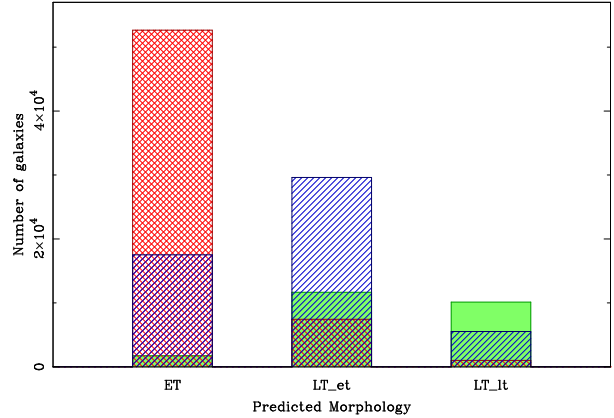}
	\caption{Histogram of the predicted morphology obtained from the LDA. For each predicted morphology, the number of galaxies of each different input morphological type are displayed. Colours and representations are the same as in Fig.~\ref{his_lda}. ET galaxies are found to be well classified, however LT\_ET spread between adjacent input types, and LT\_LT galaxies are split between both late-types. 
		\label{his_pred}}
\end{figure}

\begin{table}
\begin{center}
\caption{\textit{LDA without SMOOTH and ASYM.}
\label{tab_lda_4param}}
\scriptsize{\begin{tabular}{|ccc|}
\hline 
ALHAMBRA & Ngal & score\\
20.0 & 139470 & 0.583\\
21.5 & 139046 & 0.510\\
23.0 & 137463 & 0.356\\
\hline 
\end{tabular}}
\scriptsize{\begin{tabular}{|ccc|}
\hline 
SXDS & Ngal & score\\
21.0 & 13872 & 0.478\\
23.0 & 14579 & 0.533\\
24.5 & 14427 & 0.525\\
\hline 
\end{tabular}}
\scriptsize{\begin{tabular}{|ccc|}
\hline 
COSMOS & Ngal & score\\
21.0 & 49890 & 0.549\\
23.0 & 50150 & 0.601\\
24.0 & 48519 & 0.505\\
\hline 
\end{tabular}}
\end{center}
\end{table}

\begin{table}
\begin{center}
\caption{\textit{LDA using all parameters.}
\label{tab_lda_all}}
\scriptsize{\begin{tabular}{|ccc|}
		\hline 
		ALHAMBRA & Ngal & score\\
		20.0 & 114271 & 0.539\\
		21.5 & 88186 & 0.525\\
		23.0 & 66978 & 0.479\\
		\hline 
\end{tabular}}
\scriptsize{\begin{tabular}{|ccc|}
		\hline 
		SXDS & Ngal & score\\
		21.0 & 10752 & 0.491\\
		23.0 & 11067 & 0.512\\
		24.5 & 10077 & 0.489\\
		\hline 
	\end{tabular}}
\scriptsize{\begin{tabular}{|ccc|}
		\hline 
		COSMOS & Ngal & score\\
		21.0 & 49292 & 0.653\\
		23.0 & 47988 & 0.655\\
		24.0 & 43061 & 0.600\\
		\hline 
	\end{tabular}}
\end{center}
\end{table}

\begin{table}
\begin{center}
\caption{LDA results by morphological type.\label{tab_lda_types}}
\scriptsize{\begin{tabular}{|cccc|}
		\hline 
		ALHAMBRA & ET & LT\_ET & LT\_LT\\
		20.0 & 0.86 & 0.51  &0.02\\
		21.5 & 0.47 & 0.74 &0.10\\
		23.0 & 0.17 & 0.55 &0.41\\
		\hline 
\end{tabular}}
\scriptsize{\begin{tabular}{|cccc|}
		\hline 
		SXDS & ET & LT\_ET & LT\_LT\\
		21.0 & 0.73 & 0.33 &0.15\\
		23.0 & 0.72 & 0.51 &0.10\\
		24.5 & 0.55 & 0.67 &0.13\\
		\hline 
	\end{tabular}}
\scriptsize{\begin{tabular}{|cccc|}
		\hline 
		COSMOS & ET & LT\_ET & LT\_LT\\
		21.0 & 0.99 & 0.29 &0.00\\
		23.0 & 0.83 & 0.54 &0.15\\
		24.0 & 0.55 & 0.55 &0.15\\
		\hline 
	\end{tabular}}
\end{center}
\end{table}

Table \ref{tab_lda_types} corresponds to the obtained scores (ratio of galaxies with same output/input morphology over total galaxies) for each morphological type and magnitude cut in each survey, as resulting from the LDA with neither SMOOTH nor ASYM included. As can be seen, in all surveys, ET galaxies are better recognised at the brighter magnitude cuts, whereas LTs tend to be better recognised at fainter magnitude cuts.

\section[]{Summary and Conclusions}
\label{sec_summary_conclusions}

\indent In this paper we present for the first time a systematic study of the impact from spatial resolution and depth on the six morphological parameters commonly used for galaxy morphological classification: Abraham concentration index, Gini coefficient, Conselice-Bershady concentration index, M20 moment of light, asymmetry, and smoothness. We studied how strong the impact is for three data sets that have different spatial resolutions and depth: ALHAMBRA and SXDS, as an example of ground-based data, and COSMOS, as an example of deep space-based data. Our results correspond to maximum analysed redshift values of: 1.0, 2.2, and 1.6 in the three surveys, respectively, where the provided values describe the distribution of 95\% of each sample. We used a sample of 3000 early- and late-type local galaxies, with the available visual morphological classification, and we measured their parameters in 2 cases:\\
\indent - first, the reference values, that correspond to the real galaxy redshifts (magnitudes), and \\
\indent - second, the simulated values, obtained after moving the local galaxies to the observational conditions and magnitude/redshift distribution of three selected surveys.\\
\indent Comparing the results obtained in these two cases we showed how each parameter changes in each survey, at three particular magnitude cuts. Finally, we analysed and quantified how the impact from spatial resolution and depth affects some of the most used morphological diagrams, gave some suggestions for the galaxy/merger classification studies, and used the LDA statistical approach to analyse the most effective combination of parameters to distinguish between different types.\\

In the following, we summarise some of our main findings:\\
\indent - All six analysed morphological parameters suffer from significant biases related to the spatial resolution and data depth.  \\
\indent - The impact of the spatial resolution on the morphology is much stronger in comparison with the data depth, being therefore the most responsible for changing the parameters in the ground-based surveys, making in general the galaxies to appear less concentrated and more symmetric.\\
\indent - We stress that ASYM and SMOOTH are more sensitive to noise effects. Survey noise should be taken into account carefully when using these two parameters in morphological classification of galaxies, especially when dealing with ground-based data.\\
\indent - M20 results to be also significantly affected in all surveys, changing both the shape and range of its distribution at all brightness levels. However, M20\,-\,GINI is the most used diagram for selecting interacting (merging) systems. We highly recommend the use of this diagram simultaneously with other morphological parameters, when dealing with ground-based data sets, and in space-based surveys like COSMOS at least when studying sources fainter than F814\,=\,23.0.\\
\indent - CCON shows to be more sensitive to the spatial resolution in comparison to GINI and CABR. It works therefore better for space-based data, whereas the other two concentration indexes behave better for ground-based surveys.\\
\indent - In surveys similar to ALHAMBRA, when analysing the highest density regions of early- and late-type galaxies in the main morphological diagnostic diagrams, the impact from spatial resolution and data depth introduces contamination levels of 5\,-\,25\% for ET and 3\,-\,35\% for LT galaxies (depending on the diagram), for the brightest galaxies with F613\,$\le$\,20.0. The diagrams that seem to work the best in this case are those that do not include M20 moment of light. At the faintest analysed magnitudes (F613\,$\le$\,23.0) the contamination levels increase significantly, being as high as 60\,-\,100\%, making each of the diagnostics useless if used separately. Similar results are obtained in the case of SXDS, but at higher magnitude cuts (of order 1\,-\,2). Moreover, in both surveys, at all magnitude cuts, classification ot LT\_lt galaxies suffers very high levels of contamination. Taking all this into account, when dealing with ground-based data sets, we suggest to avoid the use of 2\,-\,3 parameter diagnostic diagrams in morphological classification, and to apply instead the use of all morphological parameters simultaneously, and to use statistical approaches based on probability distributions that the galaxy is ET or LT. \\  
\indent - In space-based surveys similar to COSMOS, even at the highest magnitude cuts (F814\,$\le$\,24.0), the contamination levels are significantly lower than those at the brightest magnitudes in the ground-based surveys. The observational bias is insignificant for all parameters at magnitudes brighter than F814\,=\,21.0 (except for the LT\_lt galaxies). However, when going to fainter magnitudes the contamination increases, and depending on the diagram it changes from 2\,-\,25\% at F814\,$\le$\,23.0, to 6\,-\,35\% for ET, and 20\,-\,70\% for LT galaxies at F814\,$\le$\,24.0. In surveys similar to COSMOS, we again suggest to use several diagnostics to classify galaxies when going to magnitudes fainter than F814\,=\,23.0.\\
\indent - Through LDA analysis we obtained that the combination of CABR, GINI, and CCON parameters, is the most effective to distinguish between different morphological types.\\ 
\indent The results presented in this paper can be directly applied to any survey similar to ALHAMBRA, SXDS and COSMOS, and also can serve as an upper/lower limit to take into account when classifying galaxies using shallower/deeper data sets.

\section*{Acknowledgments}

\indent We thank Chris Lintott for accepting to review this paper, giving us constructive comments that in our believe improved  the paper significantly. We also appreciate that he signed his report. We thank Kaz Sekiguchi and the SXDS team for informing us about the stage of the photometric redshift measurements. MP acknowledge financial support from JAE-Doc program of the Spanish National Research Council (CSIC), co-funded by the European Social Fund. This research was supported by the Junta de Andaluc\'ia through project TIC114, and the Spanish Ministry of Economy and Competitiveness (MINECO) through projects AYA2010-15169, AYA2013-42227-P, and AYA2013-43188-P. The work uses the observations collected at the Centro Astron\'omico Hispano Alem\'an (CAHA) at Calar Alto, operated jointly by the Max-Planck Institut fur Astronomie and the Instituto de Astrof\'isica de Andaluc\'ia (CSIC). The CEFCA is funded by the Fondo de Inversiones de Teruel, supported by both the Government of Spain (50\%) and the regional Government of Arag\'on (50\%). In this work, we made use of Virtual Observatory Tool for OPerations on Catalogues And Tables (TOPCAT). IRAF is distributed by the National Optical Astronomy Observatories, which are operated by the Association of Universities for Research in Astronomy, Inc., under cooperative agreement with the National Science Foundation

\label{lastpage}

\end{document}